\pdfoutput=1
\documentclass[a4paper,12pt]{article}

\usepackage[textwidth=16cm]{geometry}                		
\usepackage{graphicx}				
\usepackage{amssymb}

\usepackage{verbatim}

\usepackage[T1]{fontenc} 
\usepackage[section]{placeins}

\usepackage{siunitx}
\usepackage{graphicx}
\usepackage{mathtools}
\usepackage{amssymb}
\usepackage{amsfonts}
\usepackage[footnotesize]{caption}
\usepackage[font=scriptsize]{subcaption}
\usepackage{color}
\usepackage{braket}

\usepackage{hyperref}
\usepackage{url}
\usepackage{multirow}
\usepackage{relsize}

\usepackage{float}
\usepackage{placeins}

\usepackage{tikz}
\usepackage{cancel}
\RequirePackage{luatex85}
\usepackage[compat=1.1.0]{tikz-feynman}
\usetikzlibrary{shapes,arrows}
\usepackage{cite}

\usepackage[utf8]{inputenc}

\usepackage{tabularx}

\newcommand{\newc}{\newcommand}
\newc{\be}{\begin{equation}}
\newc{\ee}{\end{equation}}
\newc{\beq}{\begin{equation}}
\newc{\eeq}{\end{equation}}
\newc{\bea}{\begin{eqnarray}}
\newc{\eea}{\end{eqnarray}}
\newc{\simlt}{~\mbox{\smaller\(\lesssim\)}~}
\newc{\simgt}{~\mbox{\smaller\(\gtrsim\)}~}

\begin{document}

\begin{titlepage}
\begin{center}
{\bf\Large
\boldmath{
Muon g-2, Dark Matter and the Higgs mass in No-Scale Supergravity }
} 
\\[12mm]
Adam K. Forster$^{\star}$
\footnote{E-mail: \texttt{A.K.Forster@soton.ac.uk}} and Stephen~F.~King$^{\star}$
\footnote{E-mail: \texttt{king@soton.ac.uk}}
\\[-2mm]
\end{center}
\vspace*{0.50cm}
\centerline{$^{\star}$ \it
Department of Physics and Astronomy, University of Southampton,}
\centerline{\it
SO17 1BJ Southampton, United Kingdom }
\vspace*{1.20cm}

\begin{abstract}
{\noindent
We discuss the phenomenology of no-scale supergravity (SUGRA), in which the universal scalar mass is zero at the high scale, focussing on the recently updated muon g-2 measurement, and including dark matter and the correct Higgs boson mass. Such no-scale supergravity scenarios arise naturally from string theory and are also
inspired by the successful Starobinsky inflation, with 
a class of minimal models leading to a strict upper bound on the 
gravitino mass $m_{3/2}<10^3$ TeV.
We perform a Monte Carlo scan over the allowed parameter space, assuming a mixture of pure gravity mediated 
and universal gaugino masses,
using the SPheno package linked to FeynHiggs, MicrOmegas and CheckMate,
displaying the results in terms of a Likelihood function.
We present results for zero and non-zero trilinear soft parameters, and for different signs of gaugino masses, giving
a representative set of benchmark points for each viable region
of parameter space. We find that, while
no-scale SUGRA can readily satisfy the dark matter 
and Higgs boson mass requirements, consistent with all other phenomenological constraints,
the muon g-2 measurement may be accommodated only in certain regions of parameter space, 
close to the LHC excluded regions for light sleptons and charginos.
}
\end{abstract}
\end{titlepage}

\tableofcontents

\section{Introduction}

Recent experimental results have propelled the anomalous muon magnetic moment $g-2$ back into the centre stage of particle physics. Following the original Brookhaven National Laboratory measurements \cite{Muong-2:2006rrc}, 
the Fermilab Muon collaboration has recently affirmed these findings \cite{Muong-2:2021ojo}, with the 
combined results now showing a $4.2\sigma$ discrepancy with the Standard Model (SM) 
calculation \cite{Aoyama:2020ynm,Aoyama:2012wk,Aoyama:2019ryr,Czarnecki:2002nt,Gnendiger:2013pva,Davier:2017zfy,Keshavarzi:2018mgv,Colangelo:2018mtw,Hoferichter:2019gzf,Davier:2019can,Keshavarzi:2019abf,Kurz:2014wya,Melnikov:2003xd,Masjuan:2017tvw,Colangelo:2017fiz,Hoferichter:2018kwz,Gerardin:2019vio,Bijnens:2019ghy,Colangelo:2019uex,Blum:2019ugy,Colangelo:2014qya},
\begin{equation}
a_\mu^{EXP} - a_\mu^{SM} = (2.51\pm 0.59)\times 10^{-9}
\end{equation}
where  $a_\mu = \frac{g-2}{2}$.
One of the classic explanations of the discrepancy is provided by supersymmetry (SUSY), due to loops of light sleptons, $\widetilde \mu$, $\widetilde \nu_\mu$, and light electroweak gauginos \cite{Czarnecki:2001pv}.
These days such an explanation must be made consistent with the measurement of the Higgs boson mass,
and LHC constraints on superpartners, both of which point towards rather heavy squarks and gluinos, and such studies have been recently performed in various SUSY models
\cite{Shafi:2021jcg,Zhao:2021eaa,Li:2021pnt,Wang:2021bcx,Gu:2021mjd,Han:2021ify,Endo:2020mqz,Cao:2021tuh}.

Here we shall be interested in the phenomenology of 
no-scale supergravity (SUGRA) (for a review and early references see \cite{Lahanas:1986uc}),
focussing on the 
interplay between the muon $g-2$, dark matter and the Higgs boson mass in particular.
No-scale SUGRA is an ultraviolet (UV) completion of the minimal supersymmetric standard model (MSSM), in which the scalar masses are zero at the high scale
and are subsequently generated by renormalisation group (RG) running via couplings 
to the non-zero gaugino masses. 
Apart from its minimality, 
it is motivated by string theory and more recently by the fact that the hidden sector contains the ingredients for 
inflation. However, its phenomenological viability is not straightforward, given the fact the Higgs mass requires large
stop masses, while the muon g-2 requires light slepton masses, and dark matter is known to be non-trivial to achieve in the general MSSM, 
and all this must be achieved in no-scale SUGRA starting from zero scalar masses at the high scale.
This provides the main motivation for a detailed phenomenological study of no-scale SUGRA.
In our approach, there will be an important constraint coming from inflation, to which we now turn.

The idea of inflation is one of the key concepts in modern cosmology  \cite{Guth:1980zm,Linde:1981mu,Mukhanov:1981xt,Albrecht:1982wi,Linde:1983gd,Linde:2007fr}. Not only does it explain the vast size of the universe (the flatness problem), it also explains the extreme homogeneity and isotropy of the universe on cosmological scales (the horizon problem), as well as diluting cosmological relics (the monopole problem). Furthermore, the slow rolling inflaton provides small quantum fluctuations which eventually lead to large scale structure \cite{Linde:1990flp, Lyth:1998xn}. 
However, despite its successes, the precise mechanism that causes inflation is unknown.
Clues to the theory behind inflation, may come from current observational data \cite{Ade:2015lrj},
where the spectral index is measured to be $n_s\approx 0.96 \pm  0.007$ with a low tensor-to-scalar ratio $r< 0.08$.
These data exclude the simplest chaotic models with inflaton potentials $\phi^2$ or $\phi^4$  \cite{Martin:2013tda}.
Amongst the successful models which are consistent with these data is Starobinsky inflation
\cite{Mukhanov:1981xt,Starobinsky:1980te}, which may have a link to SUSY models
\cite{Ellis:1982dg,Ellis:1982ed,Ellis:1982ws}.
Since inflation is sensitive to ultraviolet (UV) scales, one must also consider SUGRA when dealing with inflation, and 
the no-scale SUGRA models \cite{Ellis:1984bf} in particular are well suited for maintaining the flatness of the inflaton potential (thereby solving the $\eta$ problem), although other approaches have also been discussed
\cite{Antusch:2009ty,Antusch:2013eca,Nakayama:2013jka,Davis:2008fv,Kallosh:2011qk}.
Also the Lyth bound \cite{Lyth:1984yz} on the tensor-to-scalar ratio also
suggests a scale of inflation below the Planck scale,
leading to testable at collider tests of inflation.

Ellis, Nanopoulos, Olive (ENO) have shown 
that no-scale SUGRA can behave like the successful Starobinsky inflationary model \cite{Ellis:2013xoa,Ellis:2013nxa,Ellis:2013nka}.
However, in the ENO approach, a term with constant modular weight 
is used to break SUSY, and there is no connection between inflation and SUSY breaking.
Subsequently one of us has considered the above ENO model, but 
with a linear Polonyi term added to the superpotential \cite{Romao:2017uwa}. 
The purpose of adding this term was to provide
an explicit mechanism for breaking SUSY in order to provide a link between inflation and SUSY breaking. 
Indeed we showed that inflation requires a strict upper bound for the gravitino mass $m_{3/2}< 10^3$ TeV \cite{Romao:2017uwa}.
It was subsequently shown \cite{King:2019omb} how the model in \cite{Romao:2017uwa} may be 
generalised to include the fields in the 
visible sector of the minimal supersymmetric standard model (MSSM).
In such a framework the 
soft-SUSY breaking parameters depend on the modular weights in the superpotential and lead to new phenomenological possibilities for supersymmetry (SUSY) breaking, based on generalisations of no-scale SUSY breaking and pure gravity mediated SUSY breaking.
The strict upper limit on the gravitino mass $m_{3/2}< 10^3$ TeV provides an important phenomenological constraint, bearing in mind that the gaugino masses are typically suppressed by a loop factor $1/(16\pi^2)$ in 
gravity mediated scenarios.

In the present paper, motivated by the desire to relate inflation to collider physics, including dark matter and the muon $g-2$ measurement, we study the phenomenology of the no-scale SUGRA inflation model in  \cite{King:2019omb}. Although related phenomenological studies of similar works have been undertaken \cite{Ellis:2013xoa, Ellis:2013nxa, Ellis:2013nka} a full phenomenological study of the model in \cite{King:2019omb}, which is subject to the upper bound on the gravitino mass, has not yet been undertaken.
Here we shall perform a phenomenological study of two of the simplest cases suggested in \cite{King:2019omb}: the first case consisting of no-scale SUGRA with zero soft scalar masses $m_0=0$ and zero trilinear soft parameter $A_0=0$, where the only source of SUSY breaking in the visible sector is via the gaugino masses $M_i$; the second case switches on a small soft trilinear mass $A_0\neq 0$, while maintaining zero soft scalar masses $m_0=0$. In both cases we assume that the gaugino masses $M_i$ (which are not fixed by the Kahler potential and are therefore independent of the details of inflation) to arise from a hybrid of anomaly mediated and universal  sources
\cite{Jeong:2021qey}.
Such gaugino masses $M_i$ at the high scale will act as the seed of all soft squark and slepton masses at low energy via renormalisation group (RG) running effects.
Given the small number of input parameters, we shall use a Monte Carlo scan over parameter space,
using the SPheno \cite{Porod:2003um,Porod:2011nf} package linked to FeynHiggs \cite{Heinemeyer:1998yj}, MicrOmegas \cite{Belanger:2004yn,Belanger:2006is,Belanger:2008sj,Belanger:2010pz,Belanger:2013oya} and CheckMate \cite{Drees:2013wra,Dercks:2016npn}. For each case above, we find viable regions of 
parameter space displaying the results in terms of a Likelihood function, including the requirement of successful dark matter relic density.
We then consider a set of representative benchmark points from viable regions of parameter space, and discuss the 
prospects for discovering the resulting SUSY spectrum at colliders. In particular we show that the 
no-scale SUGRA case allows for dark matter while satisfying all phenomenological constraints, including the correct Higgs mass and dark matter relic density, leading to the prospect that 
SUSY may be discovered at the LHC or FCC.

The layout of the remainder of the paper is as follows.
In section \ref{parameters}, we summarise the basic parameters of the no-scale SUGRA models that we analyse.
In section \ref{method} we describe our calculational approach and numerical tools and algorithms that we employ in the analysis.
In section \ref{caseI} we present our results for non-scale SUGRA case I, with all soft parameters equal to zero at the high scale apart from the gaugino masses. 
In section \ref{caseII} we present our results for non-scale SUGRA case II, where we allow in addition (small) non-zero values of $A_0$, which relaxes the collider constraints somewhat.
Section \ref{conclusion} concludes the paper.

\section{The no-scale SUGRA model parameters}
\label{parameters}

In the considered model \cite{King:2019omb},
the soft supersymmetry breaking parameters are well approximated by 
\begin{equation}
	\begin{aligned}
		m_{0} &= 0 \\ 
		\frac{A_{0}}{m_{\frac{3}{2}}} &= -6\alpha  ,\\
		\frac{B_{0}}{m_{\frac{3}{2}}} &= 2(1-\beta)
	\end{aligned}
	\label{eq:c1 soft term equations}
\end{equation} 
where $\alpha,\beta$ are undetermined modular weights in the K{\"a}hler potential.

Turning to the gauginos, the model does not constrain the gaugino mass parameters. We therefore examine a general breaking scenario where the gaugino mass parameters are partly derived from "anomaly mediated" susy breaking giving three mass parameters. Each one is determined partly by loop corrections given below \cite{Evans:2019oyw} and partly by a universal term 
parameterised below by the dimensionless coefficient $k$ \cite{2106.04238},
\begin{equation}
	\begin{aligned}
		M_{1} &= (\frac{33}{5}\frac{g_{1}^{2}}{16\pi^2}+k)m_\frac{3}{2},\\
		M_{2} &= (\frac{g_{2}^{2}}{16\pi^2}+k)m_\frac{3}{2},\\
		M_{3} &= (-3\frac{g_{3}^{2}}{16\pi^2}+k)m_\frac{3}{2}
	\end{aligned}
	\label{eq: gaugino soft mass terms}
\end{equation} 
Note that the sign of $k$ plays an important phenomenological role in determining the gaugino mass spectrum.
For small and positive $k$ the electroweak gaugino masses $M_{1,2}$ are enhanced while the 
magnitude of the gluino mass $M_3$ is reduced due to the partial cancellation against the negative anomaly mediated contribution.
This yields a spectrum with a relatively light gluino and heavy winos and binos. On the other hand, for  
small and negative $k$, the electroweak gaugino masses $M_{1,2}$ are reduced due to a partial cancellation,
while the 
magnitude of the gluino mass $M_3$ is increased.
This yields a spectrum with relatively light winos and binos which is more suited to explaining the muon $g-2$.

The supersymmetric theory produced by this model gives a very small set of high scale parameters; the scalar mass scale, $m_{0}$, the bilinear coupling, $B_{0}$, and the trilinear coupling, $A_{0}$. In turn the two unified coupling parameters are determined by the gravitino scale $m_{\frac{3}{2}}$ and our choice of the modular weights $\alpha$ and $\beta$. Furthermore, $m_{0}$ is determined by our choice of K{\"a}hler potential \cite{King:2019omb}. In addition we choose a suitable value of $tan(\beta)$ and $\mu$ to minimise the broken higgs potential and thus satisfy the higgs potential minimisation conditions given in \cite{Ellis:2004qe}.
We are therefore left with just five input parameters; $\alpha$, $\beta$, $tan(\beta)$, $m_{\frac{3}{2}}$ and $\mu$, that must be selected in order to fully characterise the high scale model and its spontaneously broken characteristics. 
We shall mainly focus on the simplest cases $\alpha=0$ and $\beta=1$ which lead to zero soft parameters
$A_0=0$ and $B_0=0$ at the high scale.

\section{Method}
\label{method}

The aim of this project is to Monte-Carlo scan over the small set of input parameters for the two cases given above. We do not use the Metropolis-Hastings algorithm \cite{Hastings:1970aa}. This is simply because the input parameter set is so small that a random scan on the IRIDIS computer cluster will be suffice to cover the parameter landscape. We then calculate various experimental outcomes including collider phenomenology, mass spectrum, and dark matter relic density to calculate a likelihood associated with each parameter point.

We hope that this will give us a greater understanding of the model and its physical viability. We aim to place some lower bound on the gravitino mass scale thus constraining the model from above, via inflation constraints, and below, via Collider phenomenolgy. Furthermore, we hope to find some best fit parameter points that might satisfy the latest $g-2$ results.

In order to understand the parameter space, we start with the simplest possible case, and subsequently increase its complexity. Therefore, initially we fix $\alpha=0$ such that $A_0=0$. As $B_0$ is determined by electroweak symmetry breaking, we leave $\beta$ and $tan(\beta)$ as free parameters. SPheno also requires the $sign(\mu)$ to be prescribed. Although previous analyses suggest positive $\mu$ is more in-keeping with modern results, we chose to allow for both signs.

\FloatBarrier
\begin{figure}[h!]
\centering
\includegraphics[width=.99\linewidth]{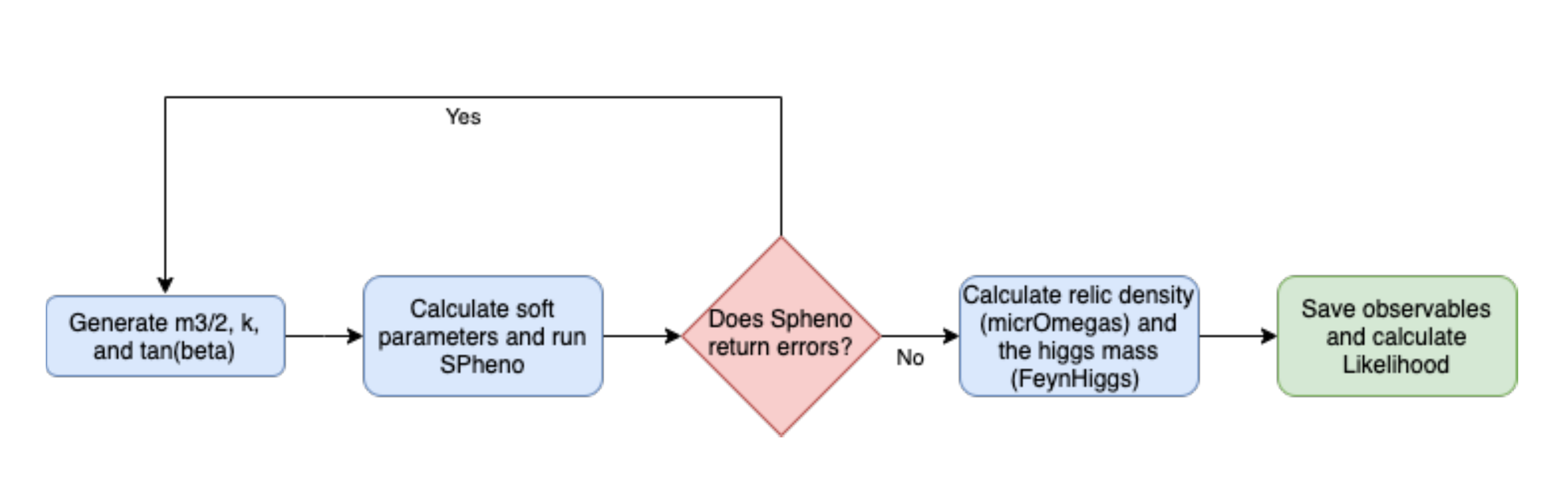}
\caption{Shows the algorithm flow used for this analyses. }
\label{method}
\end{figure}
\FloatBarrier

Generally speaking, the algorithm works as shown in Fig.~\ref{method}.
Four primary open source softwares were used to perform the calculations. For calculating the SUSY spectrum, $g-2$, and a number of other , we chose to use a modified version of SPheno that printed the value of $B_0$. Although for the most part our use of this software was very standard, we encountered a technical problem regarding the $\overline{DR}$ renormalisation scheme. Large values of the gluino mass induce large logs that must be re-summed. To solve this we  calculate the Higgs mass on-shell using FeynHiggs. In order to calculate the relic density, we use MicrOmegas.

For post data collection analysis we simulate collider effects using CheckMate \cite{Dercks:2016npn}. This software combines a number of subsidiary packages to simulate the events in a collider and exclude any given model to a 95\% confidence level. We chose to use $13TeV$ and $8TeV$ $pp$ collision data with the ATLAS detector in a signal region that focused on the search for squarks and gluinos in final states with jets and missing transverse momentum using $36fb$. This parameter space is commonly used in SUSY searches. However, further information could be garnered by broadening the analysis.

Instead of predefining $\beta$, as we did with $\alpha$, and therefore fixing a value of $B_{0}$, we decided to leave $tan(\beta)$ as a free parameter to be scanned over. This affords us some flexibility in understanding the parameter space. Furthermore, with the results in hand it is simple to solve Equation \ref{eq:c1 soft term equations} and find the relevant $\beta$ parameter. 

\begin{table}[h!]
\centering
\begin{tabular}{|l|l|l|l|l|l|l|l|l|}
\hline
case                     	& $m_{0}$           & $A_{0}$                       & $B_{0}$  & Section              \\ \hline
I          	& 0 & 0                             & $2(1-\beta) m_\frac{3}{2}$   & \ref{sec:c1}  \\ \hline
II           	& 0 & $-6\alpha m_\frac{3}{2}$                             & $2(1-\beta) m_\frac{3}{2}$   &   \ref{sec:c2}  \\ \hline
\end{tabular}

\caption {Table showing the two cases considered in this work. Gaugino mass terms (with a value for $k$), $tan(\beta)$, the sign of $\mu$, and the gravitino mass scale are also generated depending on the model being considered.  \label{tab:scenarios table}} 
\end{table}

In order to fully explore the parameter space, a number of scans (and accordant parameter limits) were initialised. Each scan is assigned to a subsection and summarised in Table \ref{tab:scans}. Note that each scan is performed twice; once for $\mu>0$ and again for $\mu<0$.
\begin{table}[h!]
\centering
\begin{tabular}{| c | c | c | c | c | c | }
\hline
  & $m_{\frac{3}{2}}\ (\rm{TeV})$           & $\alpha$                       & $tan(\beta)$ & $k$  & Subsection              \\ \hline
  Scan 1 & $[1, 1000]$           & $0$                       & $[1.5,30]$ & $[0,0.1]$  & \ref{sec:s1}              \\ 
  Scan 2 & $[1, 400]$           & $0$                       & $[1.5,50]$ & $[-0.01,-0.04]$  &  \ref{sec:s2}               \\ 
   Scan 3 & $[1, 1000]$           & $[-0.166,0.166]$                       & $[1.5,50]$ & $[0,0.1]$  &  \ref{sec:s3}               \\ 
  Scan 4 & $[1, 200]$           & $[-0.005, 0.005]$                       & $[1.5,50]$ & $[-0.014,-0.035]$  &  \ref{sec:s4}               \\ \hline  
\end{tabular}

\caption {Show the parameters for the 4 primary scans conducted in this paper. Links to the subsections in which each scan is presented are also included. In general, parameter ranges were chosen by trial an error so as to be representative of the parameter scape without losing excessive efficiency. \label{tab:scans}} 
\end{table}

\section{No-scale SUGRA with $A_0=0 $ (Case I) \label{sec:c1}}
\label{caseI}

Completely no-scale SUGRA with $A_0=0$ represents a fascinating scheme for this model. 
Previously thought to be ruled out, such scale-less supersymmetry has seen something of a resurrection
motivated by inflationary model building \cite{Ellis:2020mno}. As shown in Table \ref{tab:scenarios table}, $\alpha=0$, and therefore $A_0=0$, precludes trilinear terms from the model. Although we allow variance in $\beta$ in order to understand the parameter space, we shall see a strong preference for $\beta \approx 1$ implying $B_0=0$ bilinear coupling too. With $m_0 = 0$ 
the scalar masses are all zero at the high scale, 
however, the gauginos get mass terms by Equation \ref{eq: gaugino soft mass terms} and generate non-zero 
low energy scalar masses
via the RGEs. Remarkably, it turns out that such a scheme is phenomenologically viable, as we shall see. 

We begin by analysing the results of Scan 1 in Table \ref{tab:scans}, which assumes positive universal gaugino masses $k$,
and allows the gravitino mass $m_{\frac{3}{2}}$ to vary up to its upper bound from inflation of $1000$ TeV.

\subsection{Positive $k$ (Scan 1) \label{sec:s1}}
\begin{figure}[h!]
\begin{subfigure}{.5\textwidth}
\includegraphics[width=.9\linewidth]{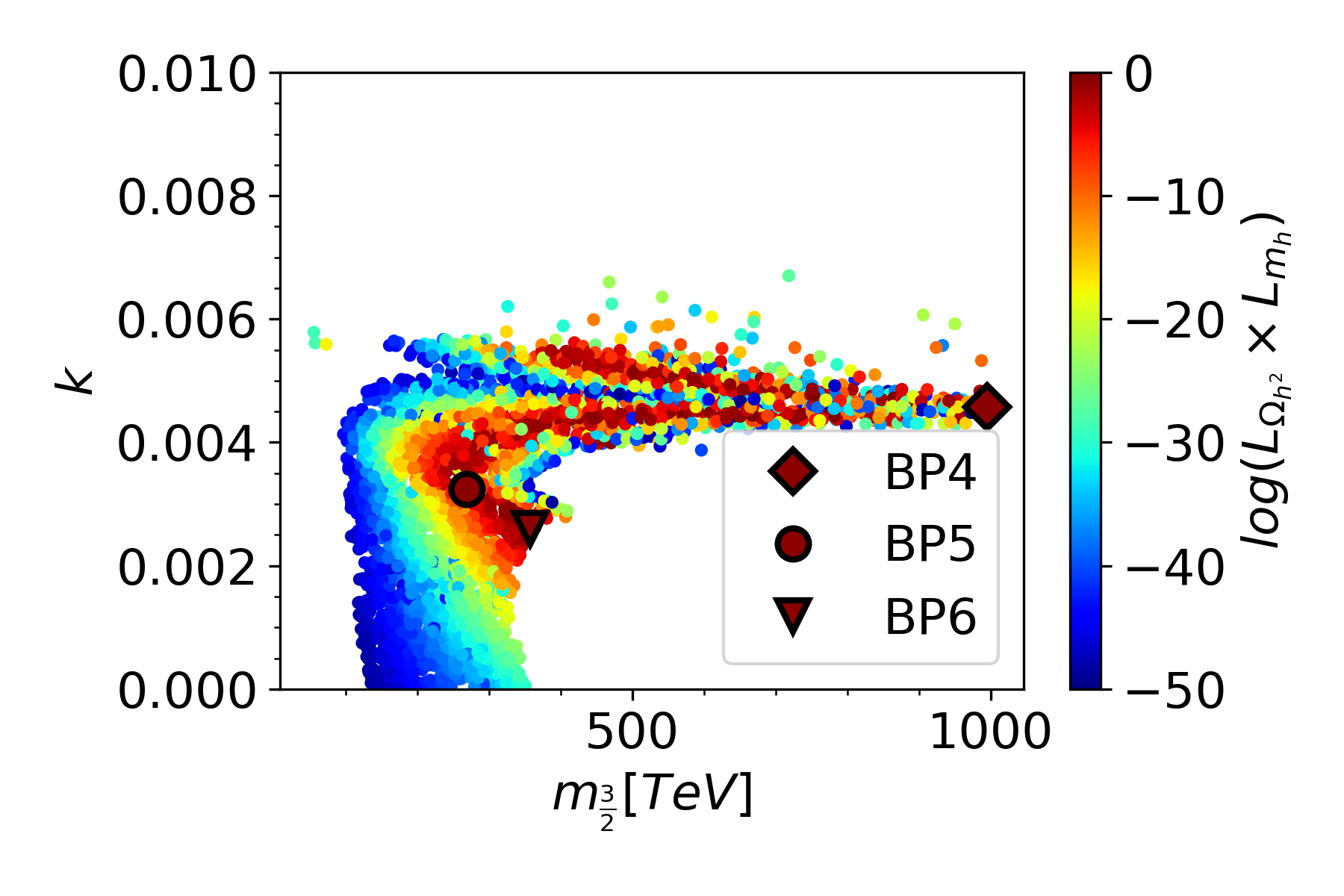}
\caption{$\mu<0$, case I }
\label{fig:scaleless/mu-/m32_vs_k_mupos}
\end{subfigure}
\begin{subfigure}{.5\textwidth}
\includegraphics[width=.9\linewidth]{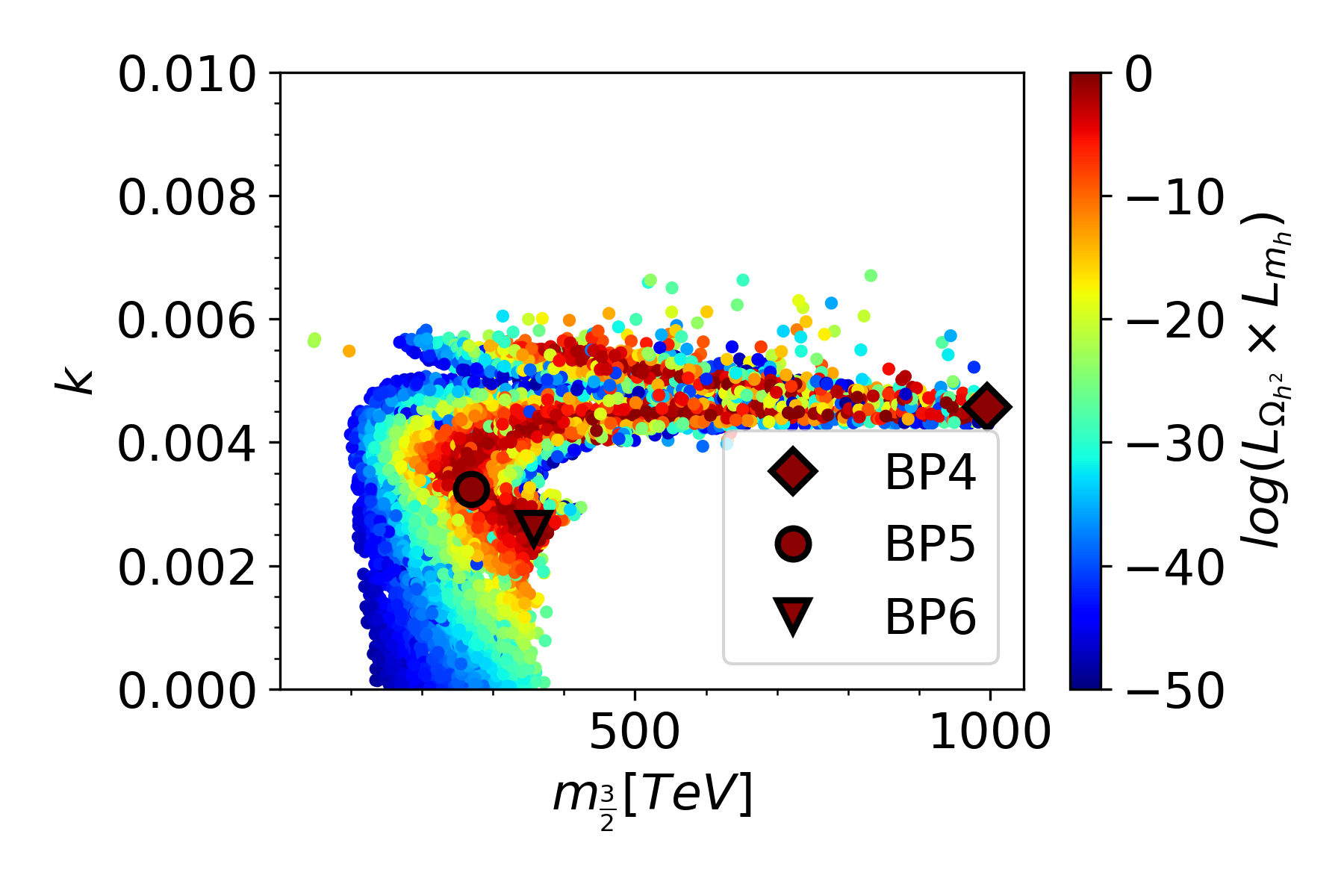}
\caption{$\mu>0$, case I }
\label{fig:scaleless_k/mu+/m32_vs_k_mupos}
\end{subfigure}

\caption{Shows $k$ against $m_\frac{3}{2}$ for case I, $\alpha=0$ data from a Monte-Carlo scan with parameter ranges ranges $tan(\beta) \in [1.5,30], k \in [0,0.1]$, and $m_\frac{3}{2} \in [10^{3}GeV,10^{6}GeV]$. Colour denotes likelihood,
with hotter colours corresponding with high likelihoods. The likelihood is dominated by the relic density calculation. The range of $k$ is naturally restricted to $k\lesssim 0.006$ due to $|\mu|^2<0$. Two bands of high likelihood points; one from above, one from below, converge at $(1000TeV, 0.0044)$ in $(m_\frac{3}{2}, k)$ space. The 6 benchmarks that are presented in Table \ref{tab:input-table} are also marked. The central colour of the benchmark denotes its likelihood.}
\label{fig:scaleless_k/m32_vs_k_mupos}
\end{figure}

As discussed earlier, positive universal gaugino mass parameter $k$ will tend to yield a 
spectrum with a relatively light gluino and relatively heavy winos and binos, so we do not expect this choice to explain the muon
$g-2$, so we will focus mainly on the Higgs mass and dark matter in this case.
Figure \ref{fig:scaleless_k/m32_vs_k_mupos} shows a scatter plot between the two most influential scan variables, $k$ and $m_\frac{3}{2}$. It should be noted that, although $tan(\beta)$ is a scan parameter, its influence on the overall likelihood is limited.

\begin{figure}[h!]
\begin{subfigure}{.5\textwidth}
\includegraphics[width=.9\linewidth]{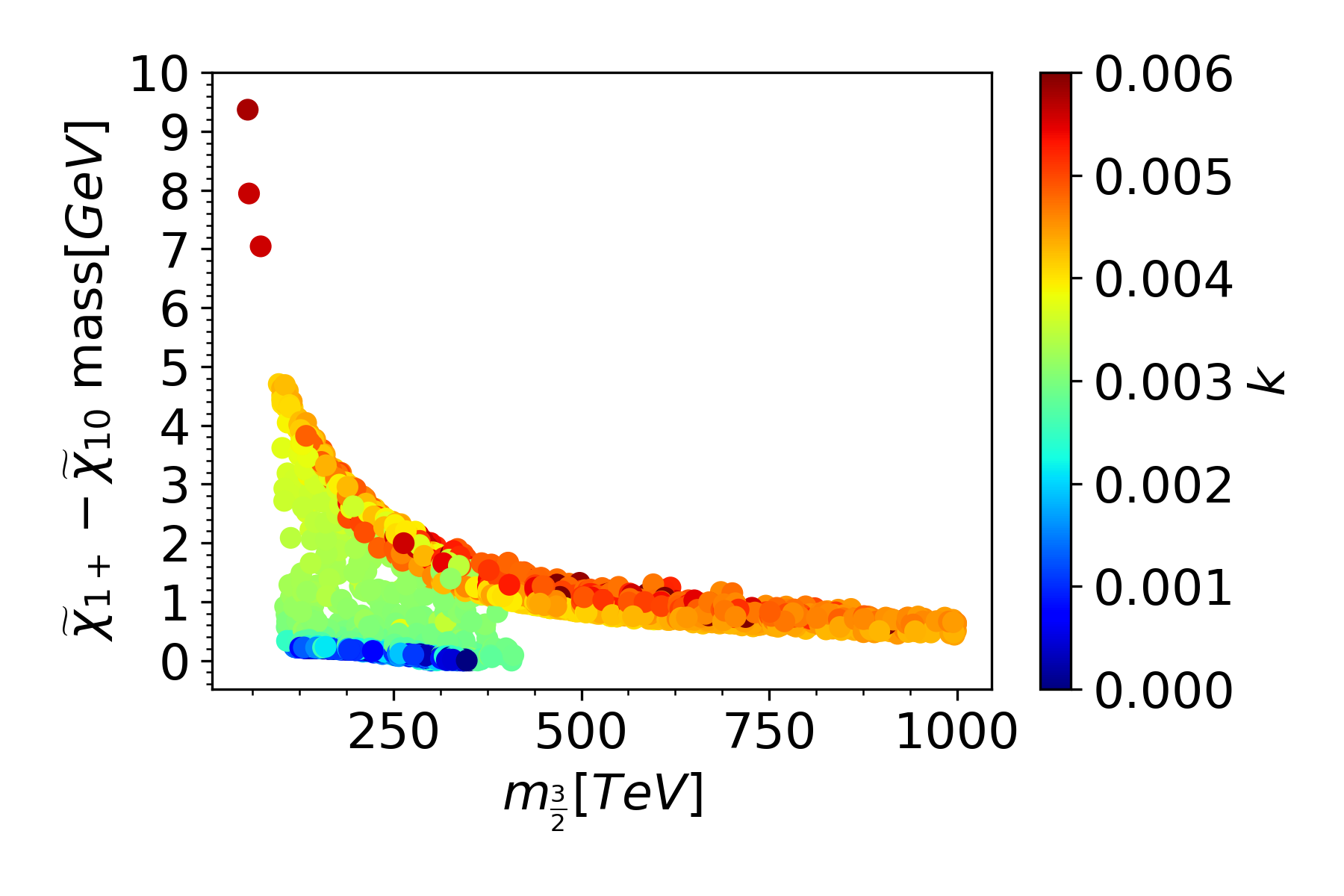}
\caption{$\mu<0$, case I,}
\label{fig:scaleless_k/high_k/mu-/m32_vs_mass_diff_mupos}
\end{subfigure}
\begin{subfigure}{.5\textwidth}
\includegraphics[width=.9\linewidth]{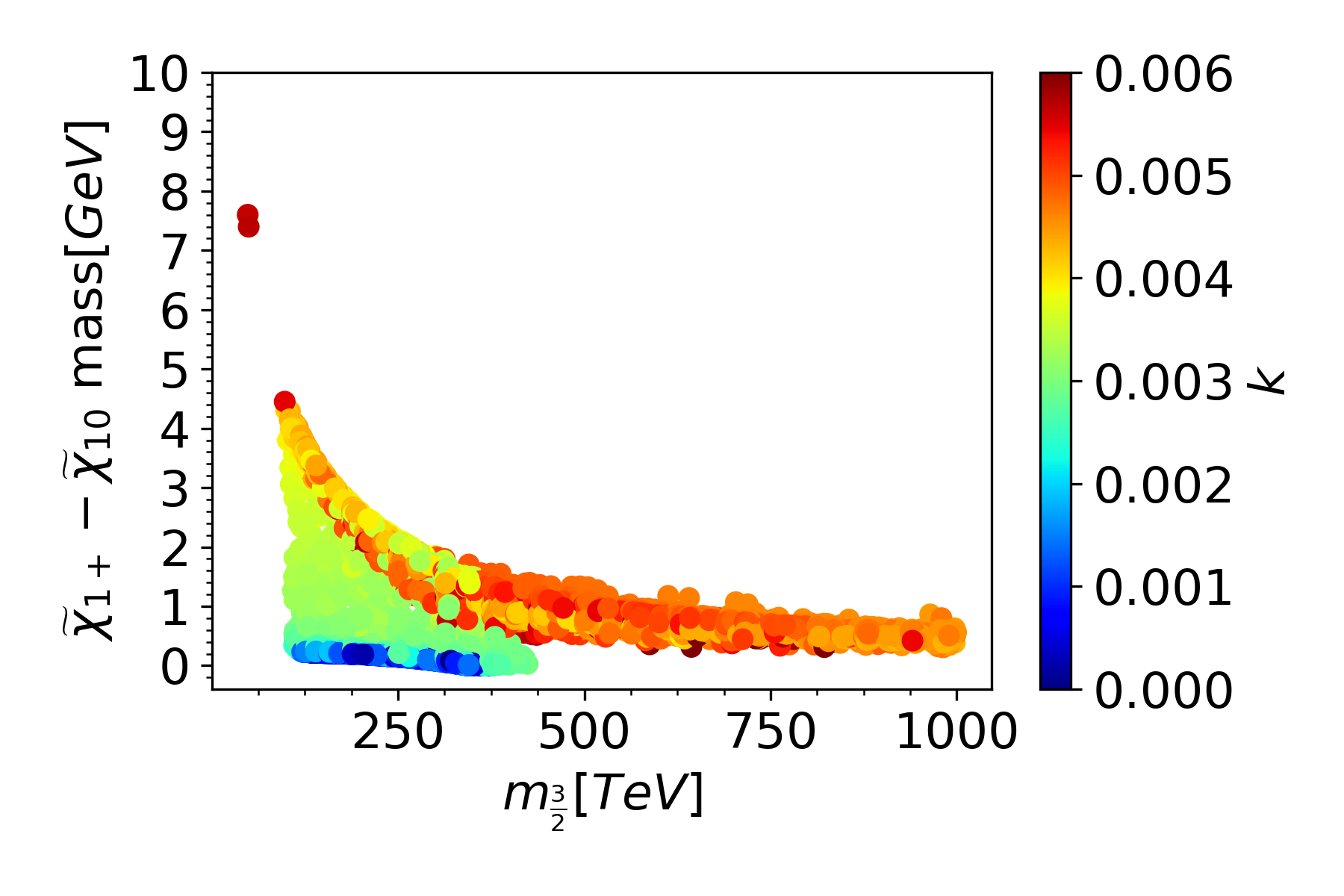}
\caption{$\mu>0$, case I, }
\label{fig:scaleless_k/high_k/mu+/m32_vs_mass_diff_mupos}
\end{subfigure}
\caption{Shows the mass difference between the LSP and the nLSP for case I, $\alpha=0$. The colour denotes the value of $k$. Small values of k lead to a negative difference between the $\widetilde{\chi}_{1+}$ and $\widetilde{\chi}_{10}$ at $400TeV$ implying a charged LSP.}
\label{fig:scaleless_k/m32_vs_mass_diff_mupos}
\end{figure}

As points that failed to produce correct electroweak symmetry breaking or a suitable dark matter candidate have been excluded from these plots, a fascinating structure of allowed, disallowed, and high likelihood points emerges. Firstly, for values of $k \gtrsim 0.006$ electroweak symmetry breaking cannot be satisfied (except for some anomalous points that, due to numeric instability, achieve electroweak symmetry breaking). Secondly, for values of $k\lesssim 0.004$ with $m_\frac{3}{2} \gtrsim 400$ TeV the LSP becomes charged. This can be seen in Figure \ref{fig:scaleless_k/m32_vs_mass_diff_mupos} where "cold" points tend to 0 mass difference and therefore, a charged LSP for large $m_{\frac{3}{2}}$ values.

In addition to the structure of excluded points, a band of hot points in the ($m_{3/2},k)$ plane
from (350 TeV, 0.0025) up to (250 TeV, 0.004) and across to (1000 TeV, 0.0045) where the relic density likelihood is maximised. An additional strip of high likelihood points begins at (400 TeV, 0.0055) and tracks down to (1000 TeV, 0.0046). 
We speculate that there could be some symmetry about $k\approx 0.0045$ that is broken above $k\approx 0.006$ due to electroweak symmetry breaking. 

The masses of the first two neutralinos are given by

\begin{equation}
	\begin{aligned}
		m_{\widetilde{\chi}_{10}} = M_1 - \frac{m_W^2(M_1+\mu sin(2\beta))}{\mu^2 - M_1^2}+...\\
		m_{\widetilde{\chi}_{20}} = M_2 - \frac{m_W^2(M_2+\mu sin(2\beta))}{\mu^2 - M_2^2}+...\\
	\end{aligned}
	\label{eq:neutralino masses}
\end{equation}

And the first chargino by,

\begin{equation}
	\begin{aligned}
		m_{\widetilde{\chi}_{1+}} = M_2 - \frac{m_W^2(M_2+\mu sin(2\beta))}{\mu^2 - M_2^2}+...\\
	\end{aligned}
	\label{eq:chargino masses}
\end{equation}

From Table \ref{tab:table-omega} we can see that as $k$ reduces, the LSP becomes increasingly "wino-like". Therefore, the mass of the lightest neutralino is dominated by the mass of $m_{\widetilde{\chi}_{20}}$ (Equation \ref{eq:neutralino masses}). In turn, the mass of lightest chargino is given, to leading order, by the same expression (Equation \ref{eq:chargino masses}). Therefore, their masses are exceptionally close. Although this can be helpful in reducing the relic density by some co-annihilation processes, for high $m_\frac{3}{2}$ the contributing factors can lead to a switch in hierarchy between the two mass states. This is further confirmed in Figure \ref{fig:scaleless_k/m32_vs_mass_diff_mupos} where the mass difference between the LSP and the nLSP tends to 0 for small $k$. 

In this case $m_0 = 0$ and thus the susy scale is relatively low. Therefore, it is important to consider the spectrum masses, their potential collider signature, and the effect new SUSY diagrams could have on certain branching ratios. We begin by looking at the mass distributions of the SUSY spectrum.

\begin{figure}[h!]
\begin{subfigure}{.5\textwidth}
\includegraphics[width=.99\linewidth]{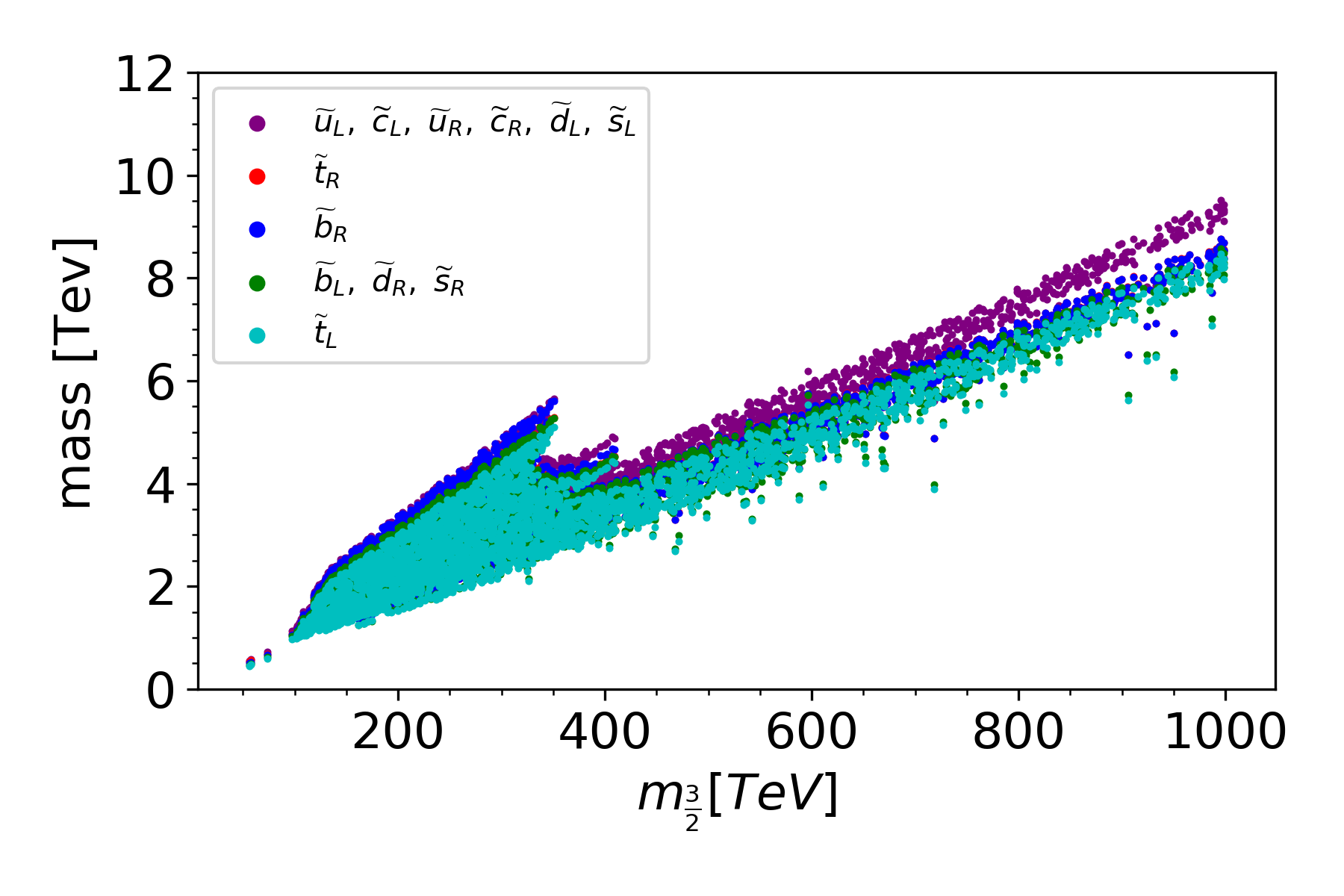}
\caption{$\mu<0$, case I  }
\label{fig:k/sparticles_vs_mgrav}
\end{subfigure}
\begin{subfigure}{.5\textwidth}
\includegraphics[width=.99\linewidth]{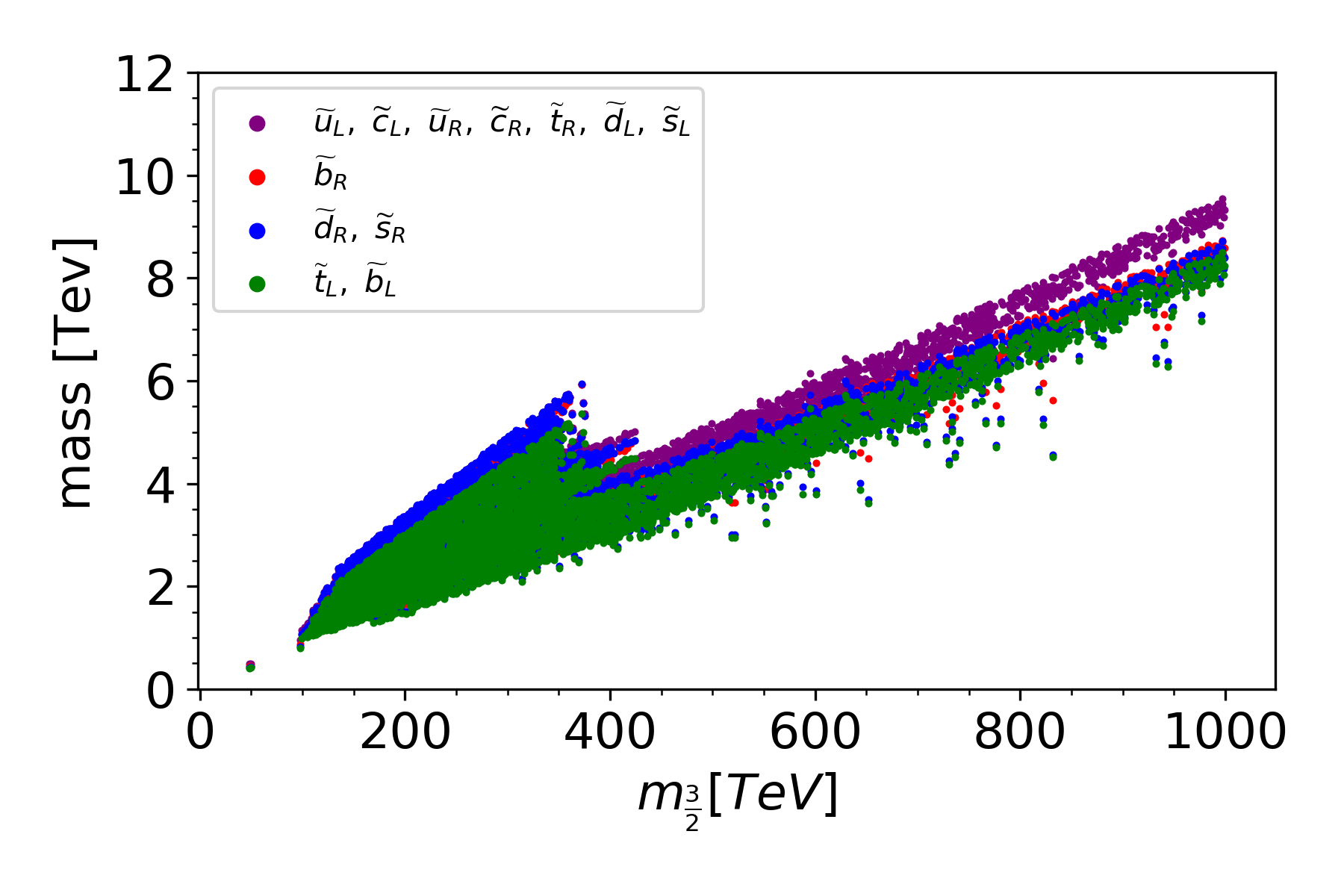}
\caption{$\mu>0$, case I }
\label{fig:k/sparticles_vs_mgrav}
\end{subfigure}
\caption{Shows the squark mass spectrum for case I, $\alpha=0$. Particles are plotted in the same colour if they are sufficiently mass degenerate as to be indistinguishable in this plot. Typically, this would entail masses within 15GeV of each other. Increasing $k$ increases the mass of the squarks. Low $k$ points cut-off just below $400TeV$.}
\label{fig:squark mass spectrum}
\end{figure}

\begin{figure}[h!]
\begin{subfigure}{.5\textwidth}
\includegraphics[width=.99\linewidth]{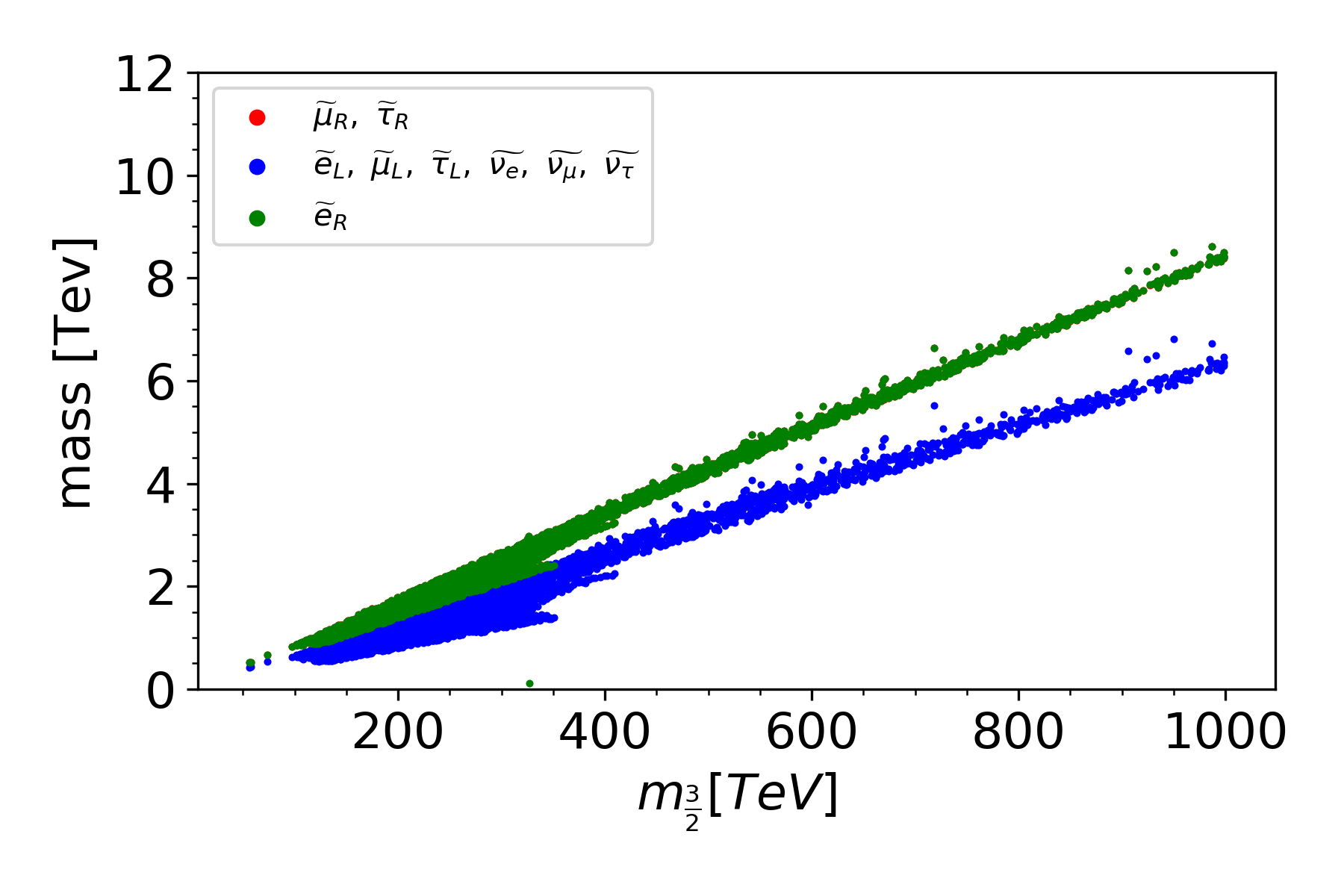}
\caption{$\mu<0$, case I}
\label{fig:k/sparticles_vs_mgrav}
\end{subfigure}
\begin{subfigure}{.5\textwidth}
\includegraphics[width=.99\linewidth]{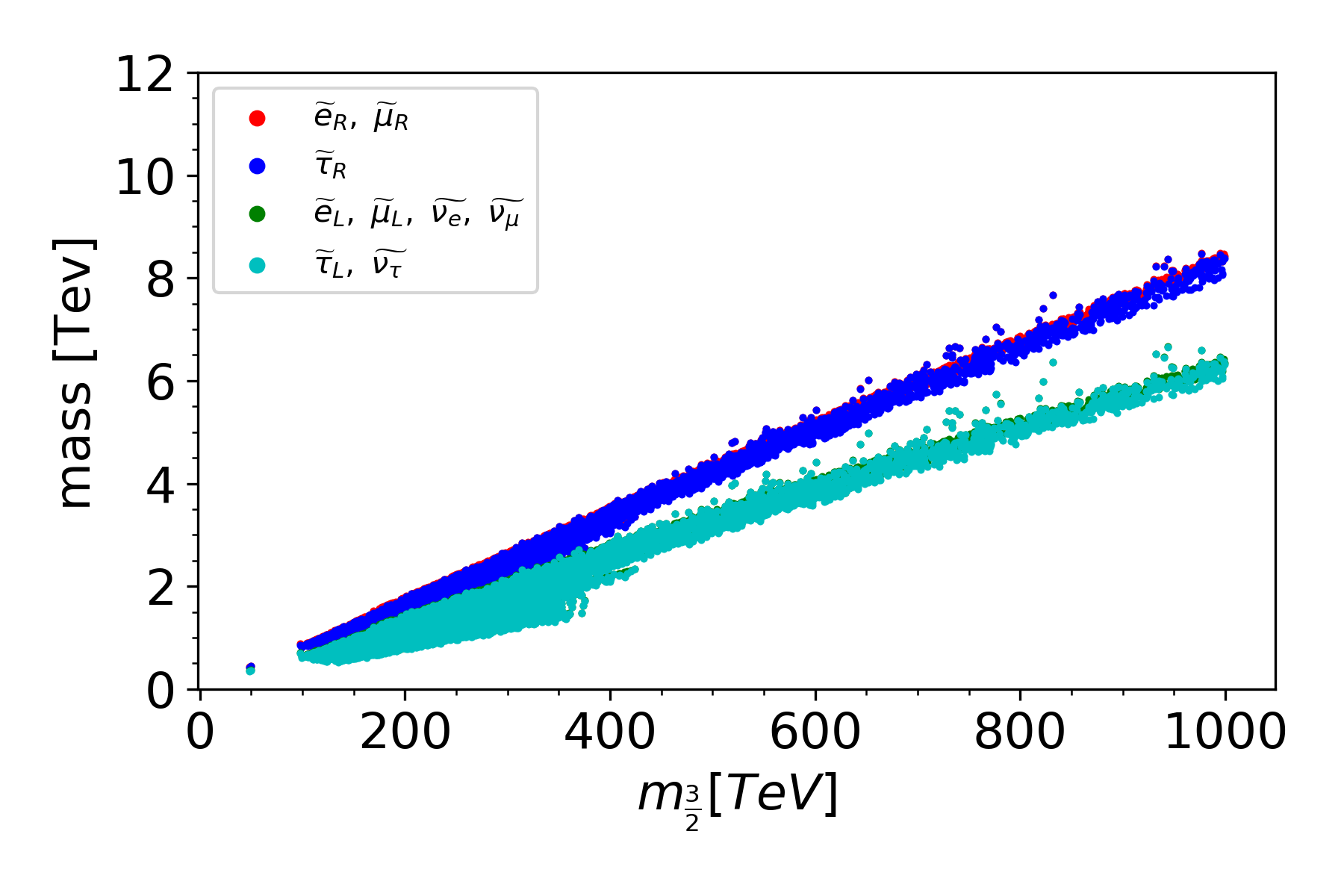}
\caption{$\mu>0$, case I }
\label{fig:k/sparticles_vs_mgrav}
\end{subfigure}
\caption{As in Figure \ref{fig:squark mass spectrum} but showing the slepton masses. Increasing $k$ decreases the mass of the sleptons. Low $k$ points cut-off just below $400TeV$.}
\label{fig:mass spectrum leptons}
\end{figure}

\begin{figure}[h!]
\begin{subfigure}{.5\textwidth}
\includegraphics[width=.99\linewidth]{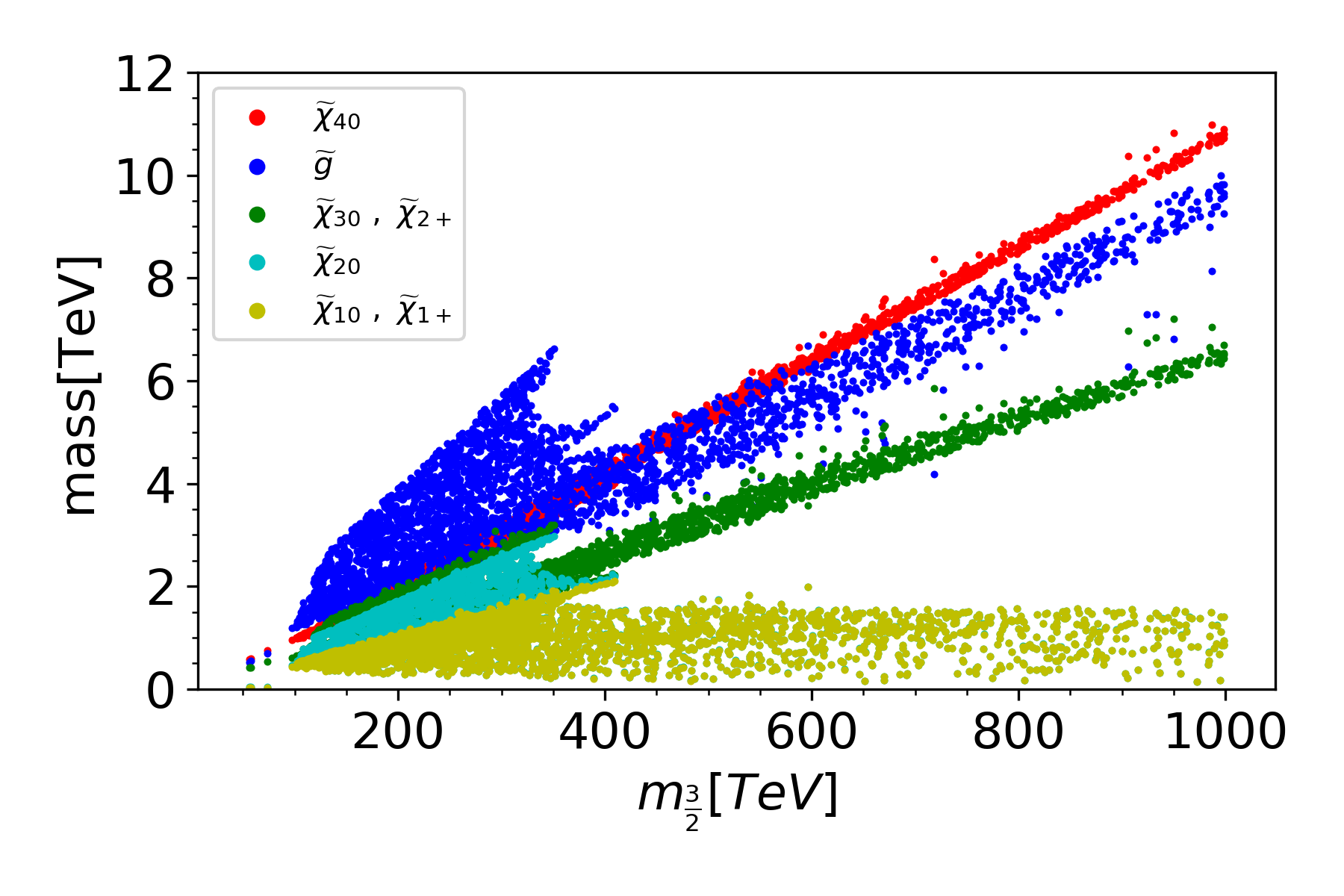}
\caption{$\mu<0$, case I }
\label{fig:k/sparticles_vs_mgrav}
\end{subfigure}
\begin{subfigure}{.5\textwidth}
\includegraphics[width=.99\linewidth]{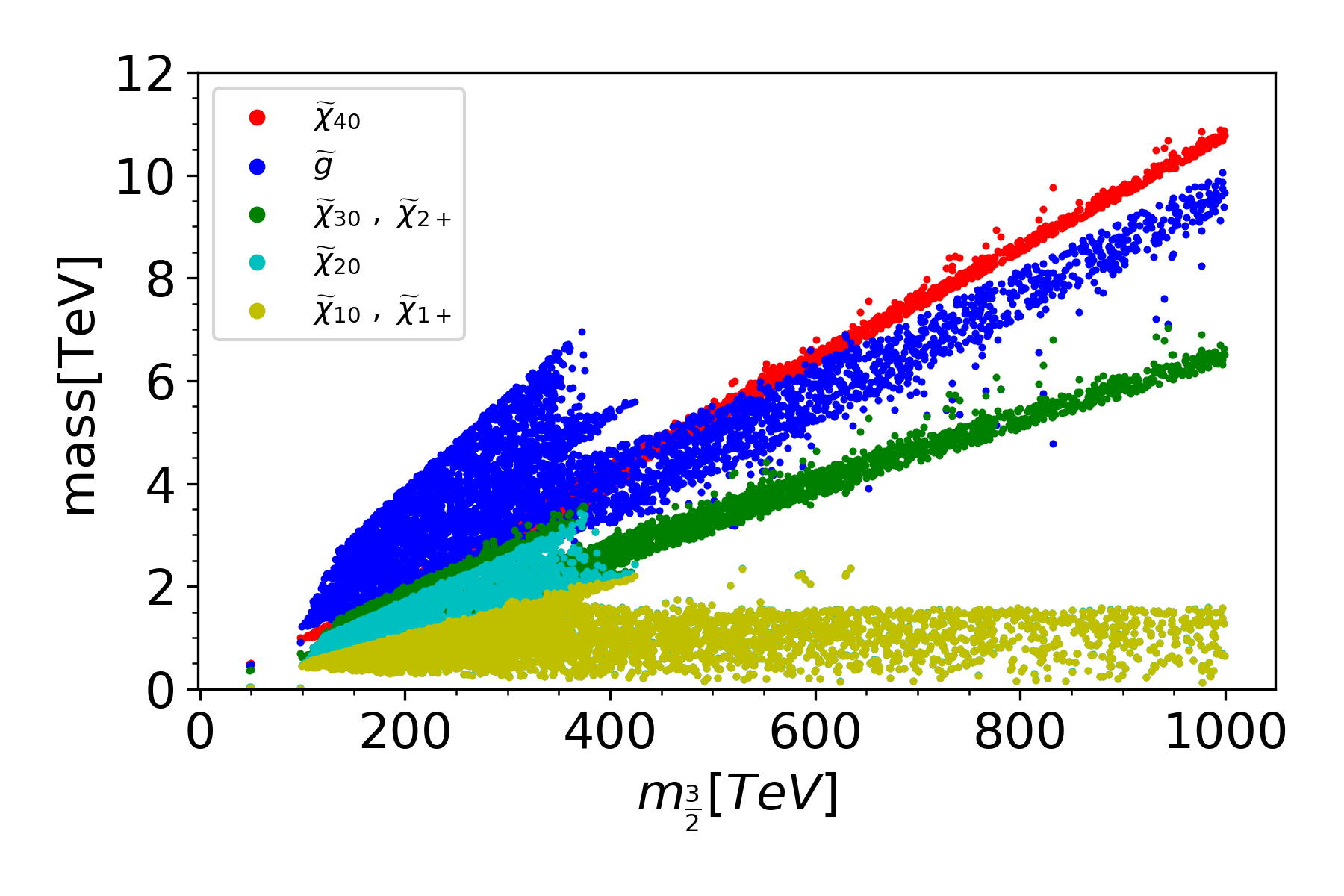}
\caption{$\mu>0$, case I }
\label{fig:k/sparticles_vs_mgrav}
\end{subfigure}
\caption{As in  Figure \ref{fig:squark mass spectrum} but showing the gaugino masses. Increasing $k$ decreases the mass of the gauginos. Low $k$ points cut-off just below $400TeV$.}
\label{fig:mass spectrum}
\end{figure}

From Fig \ref{fig:squark mass spectrum}, we see the mass distribution of the squarks with respect to $m_\frac{3}{2}$. Although it is not explicitly plotted we can see the effect of $k$ variance as a part of the mass distribution seems to cut-off just below $400TeV$. This is a reflection of the point demonstrated in Figure \ref{fig:scaleless_k/m32_vs_mass_diff_mupos}, where low $k$ points tend to having a charged LSP for large $m_\frac{3}{2}$. This same dependance of $k$ can also be seen in Figures \ref{fig:mass spectrum leptons} and \ref{fig:mass spectrum}.

We see that a reduction in $k$ increases the overall scale of the squarks. Their mass is given predominantly by $M_3$ contributions. Therefore, a reduction of $k$ will lead to an increase in the absolute value of $M_3$ as, by Equation \ref{eq: gaugino soft mass terms}, the anomaly mediated term is negative. This increase in mass scale leads to an increase in the mass scale of the squarks. We also note that the left handed $\widetilde{t}$ tends to be the the lightest squark with the first generation squarks making up the heaviest squarks as is typical in such models without flavour mixing.

Unlike the squarks, a reduction in $k$ decreases the mass scale of the sleptons. As they are uncharged under $SU(3)$, they receive their mass from contribution from $M_1$ and $M_2$. Furthermore, their anomaly mediated terms are positive, so a reduction in $k$ leads to a reduction in their absolute scale and thus a reduction in the slepton masses. Analogously to the squarks, the left handed $\widetilde{\tau}$ states tend to be the lightest sleptons.

Perhaps predictably, given the scaleless nature of the model, Figure \ref{fig:mass spectrum} shows the most interesting structure with regard to the mass spectrum. Firstly, as $m_\frac{3}{2}$ increases, the hierarchy of the heaviest neutralino and the gluino reverse. Furthermore, as $k$ increases, the gluino mass increases significantly. This can be understood by a similar argument to that of the squarks; an increase in $k$ leads to a decrease in the absolute scale of $M_3$ and thus a decrease in the gluino mass. 

Furthermore, for large values of $k$ the first neutralino mass, $\chi_{10}$, does not depend on $m_\frac{3}{2}$. The large values of $k$ increase the proportion of the first neutralino that is "higgsino-like". As this mass depends predominantly on $\mu$, this stabilises the mass of the neutralino with respect to $m_\frac{3}{2}$. Conversely, small $k$ creates a "wino-like" neutralino whose mass is proportional to $M_2$ and thus to $m_\frac{3}{2}$. 

\begin{figure}[h!]
\begin{subfigure}{.5\textwidth}
\includegraphics[width=.9\linewidth]{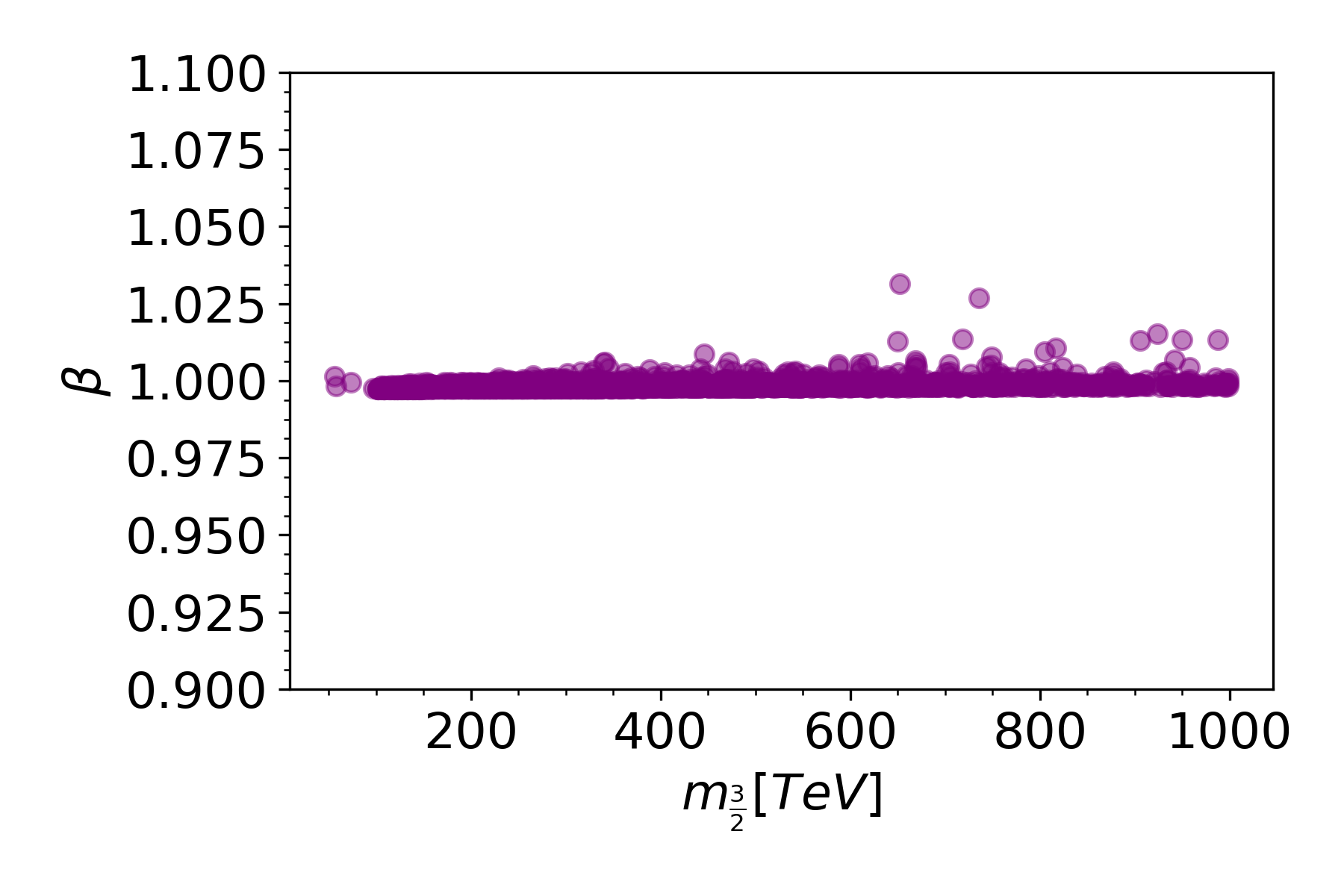}
\caption{$\mu<0$, case I }
\label{fig:scaleless/mu-/m32_vs_k_mupos}
\end{subfigure}
\begin{subfigure}{.5\textwidth}
\includegraphics[width=.9\linewidth]{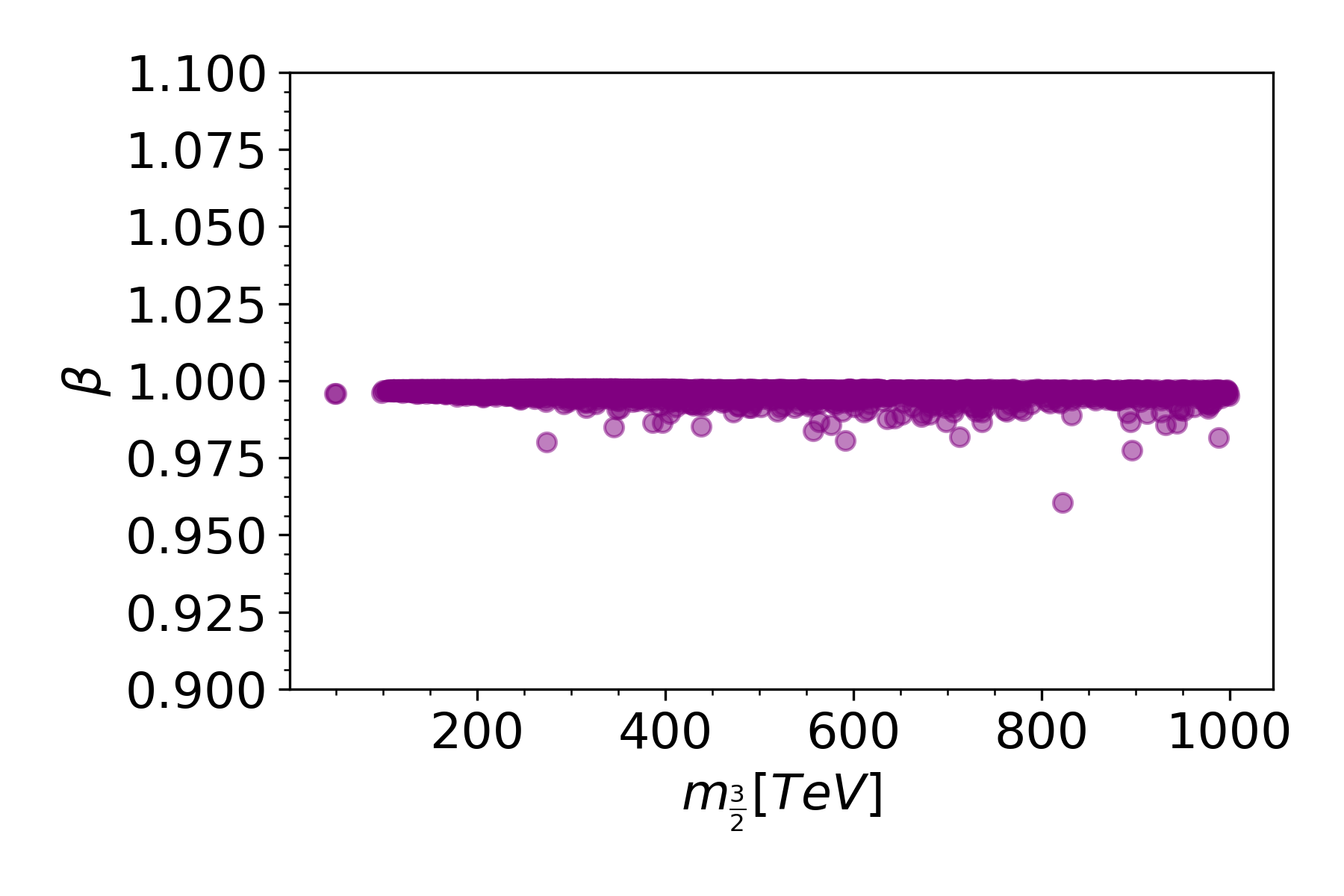}
\caption{$\mu>0$, case I}
\label{fig:scaleless_k/mu+/m32_vs_k_mupos}
\end{subfigure}

\caption{Shows the values of the modular weight, $\beta$, produced for case I, $\alpha=0$. There is a natural tendency for values very close to $1$. According to Eq.~\ref{eq:c1 soft term equations}, this suggests $B_0\approx 0$, in-keeping with scale-less supersymmetry or no-scale SUGRA. 
\label{fig:scaleless_k/m32_vs_beta_mupos}}
\end{figure}

Due to the different natures of the model weights, we decided to allow $tan(\beta)$, and therefore $B_0$ and $\beta$\footnote{$\beta$ is used both as the inverse tangent of the ratio of the higgs vevs, and as the modular weight. From context it should be clear which is being referred to.}, to vary as a free parameter but keep $\alpha$ fixed. Therefore, it is interesting to see the implications of the scan for the free $\beta$ parameter. Looking at Fig \ref{fig:scaleless_k/m32_vs_beta_mupos} we find a striking prediction for the model. For both signs of $\mu$ this model predicts that $\beta \approx 0.998$ (excluding some anomalous points). Recall that, by Eq.~\ref{eq:c1 soft term equations}, $\beta$ is connected to the bilinear coupling such that if $\beta=1\implies B_0=0$. As can be seen, the model clearly favours a bilinear coupling very close to 0 making a fully scale-less model.

\FloatBarrier
\begin{figure}[h!]
\begin{subfigure}{.5\textwidth}
\includegraphics[width=.9\linewidth]{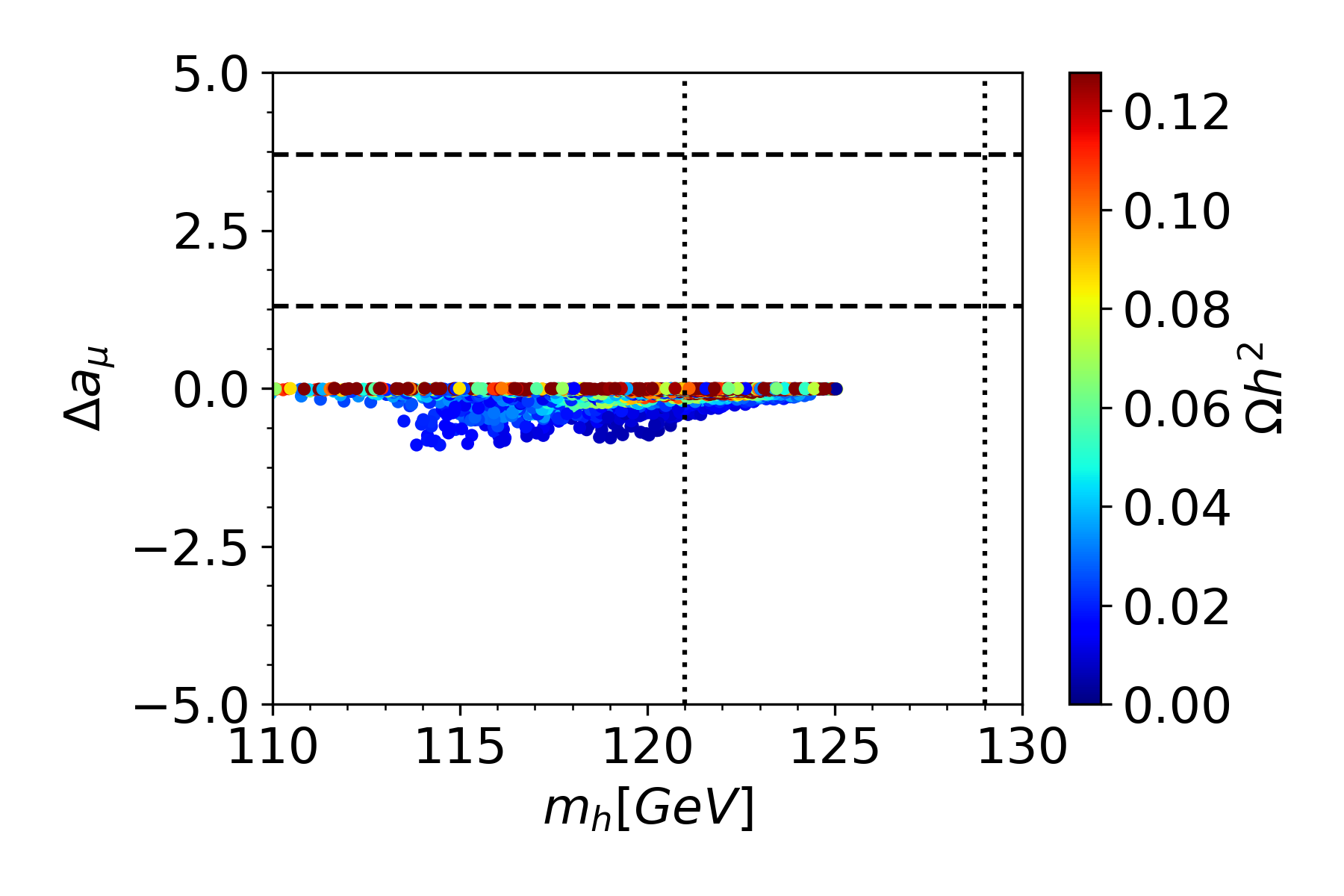}
\caption{$\mu<0$, case I}
\label{fig:A0/gmin2/mu+/mh_vs_amu_mupos_constrained}
\end{subfigure}
\begin{subfigure}{.5\textwidth}
\includegraphics[width=.9\linewidth]{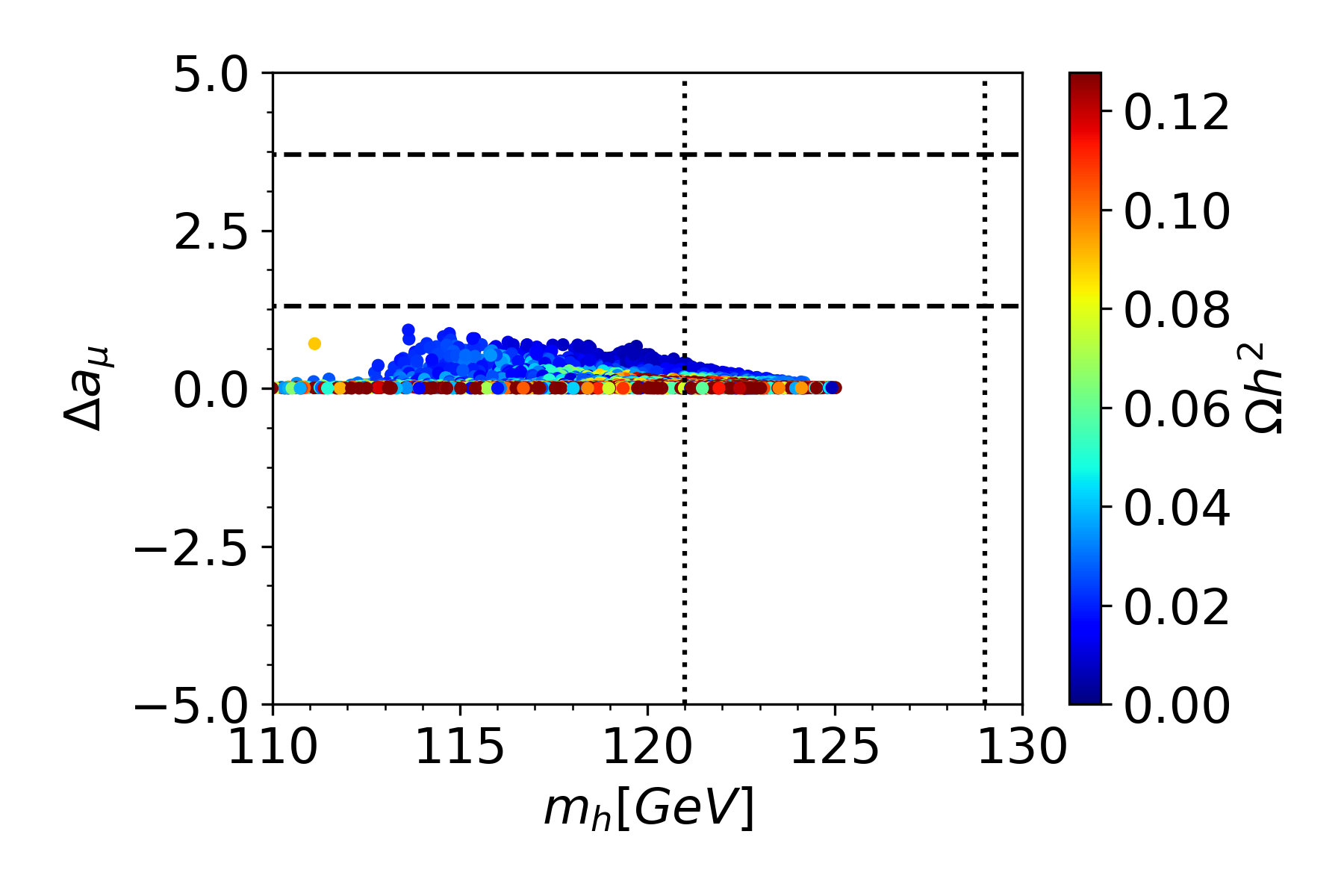}
\caption{$\mu>0$, case I }
\label{fig:A0/gmin2/mu-/mh_vs_amu_mupos_constrained}
\end{subfigure}
\caption{Shows a scatter plot of the higgs mass against the anomalous muon magnetic moment for a scan with ranges $tan(\beta) \in [1.5,30], k \in [0,0.1]$, and $m_\frac{3}{2} \in [10^{3}GeV,10^{6}GeV]$. The $2\sigma$ region of $a_\mu$ is marked with dotted lines while the $2\sigma$ region of $m_h$ is marked with dashed lines. The colour denotes the relic density. No point satisfies the observed values of the muon $g-2$.}
\label{fig:Final_Plots/scaleless_k/mh_vs_amu}
\end{figure}
\FloatBarrier

From Figure \ref{fig:Final_Plots/scaleless_k/mh_vs_amu} we see that this model cannot satisfy this condition for either sign of $\mu$. In the $\mu<0$ case, we see that the contributions are in fact negative. The contributions generally take the form $\Delta a_\mu \propto {\rm sign} [\mu] \frac{1}{(m_{\widetilde{\mu}, \widetilde \nu_\mu})^2}$. Therefore, a change in the sign of $\mu$ reverses the sign of these contributions. In general, large $m_\frac{3}{2}$ leads to a large mass spectrum which in turn suppresses the $g-2$ contributions. From Figure \ref{fig:mass spectrum leptons} we can see that the masses of the $\widetilde \mu$s and $\widetilde \nu_\mu$s grow linearly with $m_\frac{3}{2}$ leading to quadratic suppression of $\Delta a_\mu$. Although in the positive $\mu$ case, the contributions are themselves positive, no point can satisfy the measured discrepancy.

To gain further insight into this model, a set of benchmark points is presented. Table \ref{tab:input-table} shows the input parameters for each benchmark point selected. The benchmark points were selected to reflect a variety of viable regions in the model. 

Benchmark points 1,2, and 3 all have their sign of $\mu$ positive while the reverse is true for points 4,5, and 6. BP1 and BP4 are the highest likelihood points of the model for their respective signs of $\mu$. BP4 comes from the second band of high likelihood points at $k>0.0045$. They both have very high $m_\frac{3}{2}$ values existing just below the cut-off set by the CMB in the context of inflation \cite{King:2019omb}. BP2 and BP5 have the lowest values of $m_\frac{3}{2}$ whilst staying in a high likelihood range and BP3 and BP6 have the lowest values of $k$ whilst staying in a high likelihood range.

\begin{table}

\begin{tabularx}{\textwidth}{  m{0.11\textwidth}    m{0.12\textwidth} m{0.12\textwidth}   m{0.12\textwidth}  m{0.12\textwidth}   m{0.12\textwidth} m{0.12\textwidth}  }

\hline
Quantity  &  BP1 & BP2 & BP3 & BP4 & BP5 & BP6  \\
\hline
$\alpha $ &    			  				0		&	0		&	0		&	0		&	0		&	0				\\                  
$\beta $ &     							0.996	&	0.997	&	0.997	&	0.998	&	0.998	& 	0.998				\\                             
$ m_{\frac{3}{2}} $ [TeV] &  				903		&	268		&	359		&	995		&	269		&	358T				\\	
 $k$ &   				 				0.00443	&	0.00323	&	0.00251	&	0.00457	&	0.00324	& 	0.0026				\\              
\hline
\underline{SPheno:} &          							&			&			&		 	&			&					\\ 
\hline
$m_{0}$ [GeV] &      							0		&	0		&	0		&	0		&	0		&	0				\\		                   
$tan(\beta)$ &          						18.1		&	23.9		&	19.6		&	29.3		&	20.0		&	29.6				\\		     
 $sign(\mu)$ &       						1		&	1		&	1		&	-1		&	-1		&	-1				\\		          
 $A_{0}$ [GeV] &     							0		&	0		&	0		&	0		&	0		&	0				\\		                      
 $M_{1}$ [GeV] &   							20800	&	5700		&	7600		&	23000	&	5900		&	7600				\\		               
 $M_{2}$ [GeV] &     							6800		&	1700		&	2000		&	7600		&	1700		&	2000				\\		               
 $M_{3}$ [GeV] &      							-4400	&	-1600	&	-2400	&	-4700	&	-1600	&	-2400				\\	
\hline
\end{tabularx}

\caption{Shows six benchmark points representing six different areas of interest in the parameter where case I, $\alpha=0$. We present the model parameters and the resultant SPheno input parameters. BP1 and BP4 show the highest likelihood points for $\mu>0$ and $\mu<0$ respectively. BP2 and BP5 show points where $m_\frac{3}{2}$ is minimised whilst still satisfying our main constraints for $\mu>0$ and $\mu<0$ respectively. BP3 and BP6 show points for minimal values of $k$ whilst still satisfying our main constraints for $\mu>0$ and $\mu<0$ respectively. Dimensions of the parameters are given where "-" means dimensionless. $ m_{\frac{3}{2}} $ is given in units of TeV, and $m_{0}, A_{0}, M_i $ are given in GeV.\label{tab:input-table}}
\end{table}

\begin{table}

\begin{tabularx}{\textwidth}{  m{0.11\textwidth}  m{0.12\textwidth} m{0.12\textwidth}   m{0.12\textwidth}  m{0.12\textwidth}   m{0.12\textwidth} m{0.12\textwidth}  }

\hline
Masses &  BP1 & BP2 & BP3 & BP4 & BP5 & BP6  \\
\hline
$\widetilde e_{L} $&   					5690		&	1540		&	1900		&	6340		&	1540		&	1910				\\
$\widetilde e_{R} $&  	 				7590		&	2160		&	2790		&	8400		&	2170		&	2790				\\
$\widetilde \nu_{e L} $&      				5690		&	1530		&	1890		&	6340		&	1540		& 	1910				\\
$\widetilde \mu_{L} $&         				5690		&	1540		&	1900		&	6340		&	1540		&	1910				\\
$\widetilde \mu_{R} $&          				7590		&	2160		&	2790		&	8400		&	2170		&	2790				\\
$\widetilde \nu_{\mu L} $&   				5690		&	1530		&	1890		&	6330		&	1540		& 	1910				\\
$\widetilde \tau_{1} $&        				5690		&	1490		&	1860		&	6100		&	1510		&	1820				 \\
$\widetilde \tau_{2} $&       				7460		&	2100		&	2740		&	8040		&	2130		& 	2680				\\
$\widetilde \nu_{\tau L} $&       				5610		&	1490		&	1860		&	6100		&	1510		& 	1820				\\
\hline
$\widetilde{d}_{L} $ &        				8580         &	3190		&	4500		&	9260		&	3190		&	4450				\\
$\widetilde d_{R} $&        					7780		&	3060		&	4390		&	8310		&	3060		&	4330				\\
$\widetilde u_{L} $&       					8580		&	3190		&	4500		&	9260		&	3190		&	4450				 \\
$\widetilde u_{R} $ &       					8920		&	3300		&	4680		&	9620		&	3310	 	&	4630					\\
$\widetilde s_{L} $&        					8580		&	3190		&	4500		&	9260		&	3190 	&  	4450					\\
$\widetilde s_{R} $&       					7780		&	3060		&	4390		&	8310		&	3060	  	&	4330				\\
$\widetilde c_{L} $&          					8580		&	3190		&	4500		&	9260		&	3190		& 	4450				 \\
$\widetilde c_{R} $&        					8920		&	3300		&	4680		&	9620		&	3310	 	&	4630				\\  
$\widetilde b_{1} $ &         					7670		&	2910		&	4160		&	8040		&	2940		&  	4090					\\
$\widetilde b_{2} $ &   					7870		&	2980		&	4320		&	8420		&	3020		& 	4230					\\
$\widetilde t_{1} $&         					7600		&	2840		&	4050		&	8190		&	2840		&	4010				 \\
$\widetilde t_{2} $&     					7880		&	2940		&	4170		&	8420		&	2960	  	&	4110				\\
\hline
$\widetilde g $&                  				8940 	&	3510		&	5080		&	9540		&	3510		&	5010				 \\
$\widetilde \chi_{10} $&          				1110		&	1370		&	1750		&	1110		&	1370	 	&	1770				\\
$\widetilde \chi_{20} $&               			1111		&	1410		&	2170		&	1110		&	1400		& 	2110				\\
$\widetilde \chi_{30} $&              			5810		&	1500		&	2170		&	6520		&	1500	   	&	2110				 \\  
$\widetilde \chi_{40} $&             				9650		&	2620		&	3440		&	10800	&	2640		&	3440    				\\  
$\widetilde \chi_{1+} $&            				1110		&	1370		&	1750		&	1110		&	1370	    	& 	1770				\\
$\widetilde \chi_{2+} $&             			5810		&	1500		&	2180		&	6520		&	1500		&   	2110				 \\  
\hline
${h_{0}}$ &         						124.5	&	120		&	122		&	125		&	121		& 	123				\\                  
${H_{0}}$ &      						5510		&	1860		&	2690		&	5410		&	1970		&  	2520				\\               
$A_{0}$ &     						5510		&	1860		&	2690		&	5410		&	1970	 	&	2520				\\                   
${H_{\pm}}$ &  						5520		&	1870		&	2700		&	5410		&	1970		&	2520				 \\
$\mu$  &  								1100		&	1400		&	2100		&	-1070	&	-1390	&	-2090				\\
$B_{0}$ &      							7610		&	1510		&	1970		&	3890		&	1240		&	1700				\\                               
\hline

\end{tabularx}

\caption{Shows the spectrum of SUSY masses for the benchmark points given in Table \ref{tab:input-table}. The difference between the mass of $\widetilde \chi_{1+}$ and $\widetilde \chi_{10}$ is also given as this pertains to the production of dark matter. We also include the highscale bilinear coupling value $B_0$ for its relevance to the high scale parameters of the model. All massses are given in GeV.  \label{tab:table-spectrum} }
\end{table}

Table \ref{tab:table-spectrum} shows the resultant SUSY mass spectrum for the 6 benchmark points presented so far. As changing the sign of $\mu$ plays little role, we will refer to the combination of BP1 and BP4 as BP14, and similarly for the other benchmark points. Many points do not achieve a the precise value for the higgs mass. However, there is a relatively large theoretical uncertainty on this parameter leaving all these points with relatively high likelihoods. 

BP14 shows a hugely inflated mass spectrum for the squarks and sleptons. Furthermore, we see huge masses for the heaviest gauginos. However, although much of the spectrum is high in mass, the lighter gauginos are only about 1TeV. These could leave a tell tale signature in current or future colliders. We also see a relatively low $\mu$ value. This causes the high "higgsino-like" proportions of the LSP. $\widetilde \nu_{\tau L} $ is the smallest supersymmetric scalar (excluding the higgs boson).

BP25 shows a significantly reduced SUSY spectrum due to the reduced scale of $m_\frac{3}{2}$ and therefore the reduction in $M_{1,2,3}$. Although the majority of the spectrum is still mostly out of range of modern detectors, the model still produces low mass light gauginos. Although the mass gap between $\widetilde \chi_{10} $ and $\widetilde \chi_{1+} $ is still small, the gap between the first two charged states has increased due to a change in the mixing of the gauginos. Again, $\widetilde \nu_{\tau L} $ is the smallest supersymmetric scalar.

Finally, BP36 shows a slight increase in the overall SUSY scale in comparison to the BP25. Although the mass gap between the heavy and light gaugino states has increased, the overall scale for the states has increased significantly as the LSP becomes almost exclusively "wino-like". In general, a high wino-like state can lead to excessive co-annihilation with the first chargino; especially when, as can be seen in Table \ref{tab:table-omega}, the mass gap between said states is so low. However, for sufficiently high mass neutralinos, the freeze out temperature is high thus ending these problematic processes early in the universes cosmological past and thereby preventing the excessive annihilation of the candidate. Although $\widetilde \chi_{1+}$ is still the second lowest mass particle, the $\widetilde \nu_{\tau L} $ is lighter than the $\widetilde \chi_{20}$ unlike in th other benchmark points where a clear hierarchy existed between the fermionic and scalar states.

\begin{table}

\begin{tabularx}{\textwidth}{  m{0.2\textwidth}    m{0.1\textwidth} m{0.1\textwidth}   m{0.1\textwidth}  m{0.1\textwidth}   m{0.1\textwidth} m{0.1\textwidth}  }

\hline
Quantity &  BP1 & BP2 & BP3 & BP4 & BP5 & BP6  \\
\hline     
$\Omega_{DM} h^2$  & 					0.119 	& 	0.124	& 	0.115 	&	0.119	&	0.119	&	0.124			\\         
\hline
$\widetilde \chi_{10} $ [GeV]& 				1109.5	&	1372.7	&	1748.3	&	1110.5	&	1372.1  	&	1767.0  			\\
$\widetilde \chi_{1+} -\widetilde \chi_{10} $ [GeV] & 	0.48072	&	1.5		&	0.0329	&	0.59656	&	1.699	&	0.03567					\\     
\hline
$|\alpha_1|^2$ &							0.000016	&	0.000441	&	0.000004	&	0.000009	&	0.0004	&	0.000004				\\
$|\alpha_2|^2$ &							0.000144	&	0.312481	&	0.982081	&	0.0001	&	0.228484	&	0.976144				\\
$|\alpha_3|^2$ &							0.499849	&	0.352836	&	0.011025	&	0.499849	&	0.394384	&	0.014161				\\
$|\alpha_4|^2$ &							0.499849	&	0.335241	&	0.007396	&	0.499849	&	0.375769	&	0.009604				\\
\hline
\end{tabularx}

\caption{Shows the relic density of the LSP for each benchmark point. The difference between the LSP and the nLSP is also given. Finally, we give the probability of finding the given LSP in a particular flavour state. That is to say; we give $|\alpha_i|^2$  where $\widetilde \chi_{10} = \alpha_1 \widetilde B + \alpha_2 \widetilde W + \alpha_3 \widetilde H_1 + \alpha_4 \widetilde H_2$ and $\sum |\alpha_i|^2 = 1$. Dimensionful parameters are given in GeV. \label{tab:table-omega}  }
\end{table}

Table \ref{tab:table-omega} shows information specifically regarding the relic density for the benchmarks points. In all cases we see a very compressed gaugino mass spectrum inducing the requisite co-annihilarion processes. BP36 shows an exceptionally compressed mass gap. As was previously argued, the high mass leads to an early freeze out temperature preventing these co-annihilartion processes eradicating the dark matter too efficiently. We also present the proportion of bino, wino, and higgsino for the given particle. We see a very interesting shift between the respective benchmark points with BP14 being mostly higgsino, BP25 being evenly split between higgsino and wino, and BP36 being majority wino.

Tables \ref{tab:decay-BP14}, \ref{tab:decay-BP25}, \ref{tab:decay-BP36} show phenomenological information regarding potential constraints and collider physics. BP1 and BP4, BP2 and BP5 , and BP3 and BP6 are paired together as they belong to similar regions of parameter space. We implemented a checkmate analysis for both 8, 13, and 14 TeV ATLAS and CMS analyses. However, we chose not to include 14TeV analysis in our tables as they are not based on LHC runs but rather Monte-Carlo simulations. Instead, where appropriate, we have simply noted the $r_{max}$ value (defined below) produced as suggestive of the type of the effects future colliders could have. We use MadGraph\_2.6.7 \cite{Frederix:2018nkq, Alwall:2014hca} to generate events with SUSY final states. Pythia\_8.2.45 \cite{Sjostrand:2014zea, Sjostrand:2007gs} is then used to shower and hadronise the events. Finally, Delphes\_3.4.2 \cite{deFavereau:2013fsa} and some subsidiary tools \cite{Cacciari:2005hq, Cacciari:2011ma, Cacciari:2008gp, Read:2002hq} are used to perform event and detector analysis. This approach allows us to assess a given benchmark points viability in comparison with experimental data. We find that all presented benchmark points cannot be ruled out by the LHC at $\sqrt{s}\leq13TeV$. We include a quantity $r_{max}$ defined by

\begin{equation}
r_{max} = \frac{S-1.64 \cdot \Delta S}{S95}
\end{equation}

where $S$ is the number of signal events, $\Delta S$ is its uncertainty, and $S95$ is the experimental upper limit on the number of signal events. This quantity indicates whether a point is ruled out by the analyses or not. Points with $r_{max}\geq1 $ are ruled out; those with $r_{max}<1 $ are not. We also include the total proton to proton collision cross section, $\sigma_{LO}$ at the given centre of mass energy. Finally, the most important signal regions analysed by CheckMATE are given. We find a variety of different analyses are important due to the changing mixing matrices and SUSY spectrum. 

We include a number important beyond the standard model constraints; $\Delta a_\mu$, BR $(B_s \rightarrow \mu^+ \: \mu^-)$, the relic density, and BR $(b \rightarrow s \: \gamma)$. We find that the two branching ratios agree with experimental data very well. However, the lack of light $\widetilde \mu$s and $\widetilde \nu_{\mu}$s lead to insufficient contributions to the $\Delta a_\mu$ loop diagrams as these are suppressed by the propagator mass squared. Indeed, BP4, 5, and 6 have negative contributions to the anomalous muon magnetic moment. This is due to the reversal of sign of the $\mu$ parameter, and the dependancy on $\mu$ in the gaugino mediated diagrams that contribute to $\Delta a_\mu$. As the relic density is our strongest constraint, all points have been chosen to satisfy modern cosmological constraints on the production of dark matter. 

Finally, we include the decay width and branching ratios of the two lightest particles (excluding the LSP who is stable). In BP14 and BP25 these are light gaugino states. However, in BP36, the $\widetilde \nu_{\tau}$ is lighter than the $\chi_{20}$. Only branching ratios that contribute by more than 1\% are included. 

Although BP14 shows a long lifetime it is drastically insufficient for the particle to escape the detector. Therefore, the particles decay products are very important in assessing the signature of this model. The analysis region of most significance is one that focuses on mono-jets. This analysis has a very large luminosity. It should be noted that, unlike BP25 and BP36, 14TeV analysis yields an $r_{max}$ value of $0$, perhaps due to the lack of significance at such high energy scales. The branching ratios for both the two lightest non-LSP particles are dominated by $\widetilde \chi_{10}$ decays. This would suggest large corresponding missing momenta. We find an excellent fit for the BSM constraints as well as dark matter. However, $\Delta a_\mu$ cannot be satisfied.

\begin{table}
\begin{tabularx}{\textwidth}{  m{0.3\textwidth}     m{0.35\textwidth} m{0.35\textwidth}   }
\hline
Quantity &     BP1 &  BP4\\
\hline
$\Gamma \chi_{1+}$ [GeV]&  									$8.4\times10^{-14}$ 		&	$1.7\times10^{-13}$			\\
BR $(\chi_{1+} \rightarrow \chi_{10} \: \pi^+)$ [\%]  &  				92.9					&	89.3	 					\\
BR $(\chi_{1+} \rightarrow \chi_{10} \: e^+ \: \nu_e)$ [\%]   &  			4.00					&	5.71						\\
BR $(\chi_{1+} \rightarrow \chi_{10} \: \mu^+ \: \nu_\mu)$ [\%]   &  		3.14 					&	4.90						\\
\hline
$\Gamma \chi_{20}$ [GeV] &  									$5.12\times10^{-12}$ 	&	$2.9\times10^{-12}$			\\
BR $(\chi_{20} \rightarrow \chi_{10} \: \pi^0)$ [\%]  & 					63.8					&	74		 				\\
BR $(\chi_{20} \rightarrow \chi_{1+} \: \pi^-)$ [\%]  &  					6.1 					&	2.64						\\
BR $(\chi_{20} \rightarrow \chi_{1-} \: \pi^+)$ [\%]  &					6.1 					&	2.64						\\
BR $(\chi_{20} \rightarrow \chi_{10} \: \gamma)$ [\%]  &  				4.9 					&	5.79						\\

BR $(\chi_{20} \rightarrow \chi_{10} \: e^- \: e^+)$ [\%]  &  				2.0 					&	1.81						\\
BR $(\chi_{20} \rightarrow \chi_{10} \: \mu^- \: \mu^+)$ [\%]  &  			2.0 					&	1.72						\\
BR $(\chi_{20} \rightarrow \chi_{10} \: \nu_e \: \overline{\nu_e})$ [\%]  &  	2.0 					&	10.8						\\
\hline
BR $(b \rightarrow s \: \gamma)$ [\%]  & 							0.032 				&	0.032					\\ 
BR $(B_s \rightarrow \mu^+ \: \mu^-)$ [\%]  &  						$2.9\times 10^{-7}$		&	$2.9\times10^{-7}$			\\
$\Delta a_\mu$ & 										$8.39\times 10^{-12}$ 	&	$-1.15\times10^{-11}$		\\
$\Omega_{DM} h^2$  & 									0.119				&	0.119					\\
$\chi_{10}$ [GeV] & 											1110					&	1110						\\
$\sigma_{q \overline{q} \rightarrow  \chi_{10} \chi_{10}}$ [pb] &  				$0$					&	$0$						\\
\hline
$r_{max}$ &												$3.72\times10^{-4}$		&	$2.03\times10^{-4}$			\\
$\sqrt{s}$ [TeV]&											13				&	13						\\
Analysis	&												atlas\_conf\_2017\_060	&	atlas\_conf\_2017\_060		\\
Signal Region &											EM7					&	IM6						\\
Ref.			&										\cite{ATLAS:2017bfj}		&	\cite{ATLAS:2017bfj}							\\
$\sigma_{LO}$ [pb]	&										$4.126\times10^{-4}$	&	$4.036\times10^{-4}$		\\
\end{tabularx}

\caption{Shows branching ratios for lightest supersymmetric particles in the spectrum for the benchmarks points with highest likelihood, BP1 and BP4. Only branching ratios greater than $1\%$ are included. We also include some the the beyond the standard model observables BR $(b \rightarrow s \: \gamma)$, BR $(B_s \rightarrow \mu^+ \: \mu^-)$, $\Delta a_\mu$, and $\Omega_{DM} h^2$, where $\Delta a_\mu$ is a calculation of the SUSY contribution beyond the standard model. The model successfully predicts the b decays discrepancy as well as the relic density. However, the anomalous muon magnetic moment cannot be satisfied. CheckMATE runs using 13TeV and 8TeV analyses cannot rule out these points. The cross section $\sigma_{LO}$ is calculated by pythia as a summation of subprocesses giving the total p+ p+ cross section. Decay widths and masses are given in GeV, branching ratios are given in \%, and cross sections are given in pb. \label{tab:decay-BP14} }
\end{table}

\begin{table}
\begin{tabularx}{\textwidth}{  m{0.3\textwidth}     m{0.35\textwidth} m{0.35\textwidth}   }
\hline
Quantity &     BP2 &  BP5\\
\hline
$\Gamma \chi_{1+}$ [GeV] & 										$8.9\times10^{-12}$		&	$1.4\times10^{-11}$			\\
BR $(\chi_{1+} \rightarrow \chi_{10} \:  \overline{d} \: u)$ [\%] & 				60.0					&	58.9						\\
BR $(\chi_{1+} \rightarrow \chi_{10} \: \overline{s} \: c)$ [\%]& 				0.34					&	2.02						\\
BR $(\chi_{1+} \rightarrow \chi_{10} \:  e^+ \nu_e)$ [\%]& 					20.1					&	19.7						\\
BR $(\chi_{1+} \rightarrow \chi_{10} \:  \mu^+ \nu_\mu)$ [\%]& 				19.6					&	19.4						\\
\hline
$\Gamma \chi_{20}$ [GeV] & 										$2.45\times10^{-4} $ 	&	$6.1\times10^{-5}$				\\
BR $(\chi_{20}  \rightarrow  \chi_{10} \: u \:  \overline{u}) $ [\%]					&	3.92	 		&	4.22								\\
BR $(\chi_{20}  \rightarrow  \chi_{10} \: c \:  \overline{c}) $ [\%]					&	3.90	 		&	4.18								\\
BR $(\chi_{20}  \rightarrow  \chi_{10} \: d \:  \overline{d}) $ [\%]					&	5.09	 		&	5.49								\\
BR $(\chi_{20}  \rightarrow  \chi_{10} \: s \:  \overline{s}) $ [\%]					&	5.09	 		&	5.49								\\
BR $(\chi_{20}  \rightarrow  \chi_{10} \: b \:  \overline{b}) $	[\%]				&	4.68	 		&	4.74								\\
BR $(\chi_{20}  \rightarrow  \chi_{10} \: e^- \: e^+) $ [\%]						&	1.17	 		&	1.26								\\
BR $(\chi_{20}  \rightarrow  \chi_{10} \: \mu^- \: \mu^+) $ [\%]					&	1.17	 		&	1.26								\\
BR $(\chi_{20}  \rightarrow  \chi_{10} \: \tau^- \: \tau^+) $ [\%]					&	1.16	 		&	1.24								\\
BR $(\chi_{20}  \rightarrow  \chi_{10} \: \nu_e \:  \overline{\nu_e}) $ [\%]			&	7.03	 		&	7.57								\\
BR $(\chi_{20}  \rightarrow  \chi_{1+} \: d \:  \overline{u}) $ [\%]					&	11.2	 		&	10.8								\\
BR $(\chi_{20}  \rightarrow  \chi_{1-} \: \overline{d} \: u) $ [\%]					&	11.2	 		&	10.8								\\
BR $(\chi_{20}  \rightarrow  \chi_{1+} \: s \:  \overline{c}) $ [\%]					&	11.1	 		&	10.7								\\
BR $(\chi_{20}  \rightarrow  \chi_{1-} \:  \overline{s} \: c) $ [\%]					&	11.1	 		&	10.7								\\
BR $(\chi_{20}  \rightarrow  \chi_{1+} \: e^- \: \overline{\nu_e}) $ [\%]				&	3.72	 		&	3.60								\\
BR $(\chi_{20}  \rightarrow  \chi_{1-} \: e^+ \: \nu_e) $ [\%]					&	3.72	 		&	3.60								\\
BR $(\chi_{20}  \rightarrow  \chi_{1+} \: \mu^- \:  \overline{\nu_\mu}) $ [\%]		&	3.72	 		&	3.60								\\ 
BR $(\chi_{20}  \rightarrow  \chi_{1-} \: \mu^+ \: \nu_\mu) $ [\%]					&	3.72	 		&	3.60								\\
BR $(\chi_{20}  \rightarrow  \chi_{1+} \: \tau^- \:  \overline{\nu_\tau}) $ [\%]		&	3.68	 		&	3.54								\\
BR $(\chi_{20}  \rightarrow  \chi_{1-} \: \tau^+ \: \nu_\tau) $ [\%]					&	3.68	 		&	3.54								\\
\hline
BR $(b \rightarrow s \: \gamma)$ [\%] &								0.033 				&	0.033							\\ 
BR $(B_s \rightarrow \mu^+ \: \mu^-)$ [\%] & 							$2.9\times 10^{-7}$		&	$3.0\times10^{-7}$					\\
$\Delta a_\mu$ &  											$1.15\times 10^{-10}$ 	&	$-9.6\times10^{-11}$					\\
$\Omega_{DM} h^2$  & 										0.124				&	0.119							\\
$\chi_{10}$ [GeV] & 												1370					&	1370								\\
$\sigma_{q \overline{q} \rightarrow  \chi_{10} \chi_{10}}$ [pb] &  			$8.629\times10^{-8}$		&	$2.718\times10^{-8}$			\\
\hline
$r_{max}$ &												$1.12\times10^{-2}$		&	$8.10\times10^{-3}$					\\
$\sqrt{s}$ [TeV] &												13TeV				&	13TeV								\\
Analysis	&												atlas\_1712\_02332		&	atlas\_1712\_02332					\\
Signal Region &											2j-3600				&	2j-3600							\\
Ref.			&											\cite{ATLAS:2020syg}	&	\cite{ATLAS:2020syg}								\\
$\sigma_{LO}$ [pb]	& 											$1.611\times10^{-4}$	&	$1.630\times10^{-4}$				\\
\end{tabularx}

\caption{As in Table \ref{tab:decay-BP14} but for BP2 and BP5; points with low $m_\frac{3}{2}$.  \label{tab:decay-BP25}}
\end{table}

The most constraining region for BP25 focuses on squarks and gluinos, with 0 leptons, and 2-6 jets at 13Tev. This is due to the relatively low mass of the strongly coupled particles. We see the highest value of $r_{max}$ of all the benchmark points due to the abundance of these lighter particles. $\widetilde \chi_{20}$ shows a huge variety in decay channels as the mass gap is too small for the squark pairs to hadronise. We find that 14TeV analysis yields very high $r_{max}$ values of 0.276 in the positive $\mu$ case. This suggests that future colliders could probe regions of interest in this model. Again, the relic and BSM branching ratios can be fitted well, but $\Delta a_\mu$ cannot.

As previously eluded to, BP36 shows a change in the spectrum hierarchy. From Table \ref{tab:table-omega}, the mass difference between $\widetilde \chi_{10}$ and $\widetilde \chi_{1+}$ is the smallest. Therefore, a smaller phase space is available leading to fewer decay channels. In contrast to $\widetilde{\nu_\tau}$, whose lifetime is very small, the lifetime of $\chi_{1+}$ is sufficiently long that the particle could escape the detector. As it is a charged particle, this would appear as a charge track in the calorimeter.  As was the case for BP25, 14TeV analysis gives a whole order of magnitude increase in $r_{max}$ hinting at the exciting prospects for physics to come. Again, the BSM constraints and the relic are satisfied; however, $\Delta a_\mu$ is not.

\begin{table}
\begin{tabularx}{\textwidth}{  m{0.3\textwidth}     m{0.35\textwidth} m{0.35\textwidth}   }
\hline
Quantity &     BP3 &  BP6\\
\hline
$\Gamma \chi_{1+}$ [GeV] &									$7.2\times10^{-20}$		&	$1.0\times10^{-19}$			\\
BR $(\chi_{1+} \rightarrow \chi_{10} \: \overline{d} \: u)$ [\%] &			72					&	72						\\
BR $(\chi_{1+} \rightarrow \chi_{10} \: e^+ \: \nu_e)$ [\%] & 			28					&	28						\\
\hline
$\Gamma \widetilde{\nu_\tau}$ [GeV] &  							$2.9\times10^{-1}$		&	$7.7\times10^{-2}$			\\
BR $(\widetilde{\nu_\tau} \rightarrow \chi_{10} \: \nu_\tau)$ [\%]	&	33.3					&	33.0						\\
BR $(\widetilde{\nu_\tau} \rightarrow \chi_{1+} \: \tau^-)$ [\%]	& 		66.7					&	67.0						\\
\hline
BR $(b \rightarrow s \: \gamma)$ [\%] & 							0.032 				&	0.032					\\ 
BR $(B_s \rightarrow \mu^+ \: \mu^-)$ [\%] & 						$2.9\times 10^{-7}$		&	$3.0\times10^{-7}$			\\
$\Delta a_\mu$ & 										$5.50\times 10^{-11}$ 	&	$-8.30\times10^{-11}$		\\
$\Omega_{DM} h^2$  &									0.115				&	0.124					\\
$\chi_{10}$ [GeV] & 											1750					&	1750						\\
$\sigma_{q \overline{q} \rightarrow  \chi_{10} \chi_{10}}$ [pb] &  		$7.805\times10^{-8}$		&	$6.467\times10^{-8}$			\\
\hline
$r_{max}$ &											$3.458\times10^{-3}$	&	$3.189\times10^{-3}$				\\
$\sqrt{s}$ [TeV] &											13TeV					&	13TeV								\\
Analysis	&											atlas\_conf\_2017\_060	&	atlas\_conf\_2017\_060				\\
Signal Region &										EM10				&	EM10							\\
Ref.			&										\cite{ATLAS:2017bfj}		&	\cite{ATLAS:2017bfj}						\\
$\sigma_{LO}$ [pb]	&										$1.520\times10^{-5}$	&	$1.422\times10^{-5}$				\\
\end{tabularx}

\caption{As in Table \ref{tab:decay-BP14} but for BP3 and BP6; points with low $k$.  \label{tab:decay-BP36}}
\end{table}
\FloatBarrier

\subsection{Negative $k$ (Scan 2) \label{sec:s2}}

In this case we expect a negative universal gaugino mass parameter $k$ to tend to yield a 
spectrum with a heavy gluino and relatively light winos and binos, possibly suitable to explain the muon
$g-2$, as well as the Higgs mass and dark matter.
Moreover, in order to explain the measured $\Delta a_\mu$, small values of the $\widetilde \mu$ mass are required. As seen in Figures \ref{fig:squark mass spectrum}, \ref{fig:mass spectrum leptons}, and \ref{fig:mass spectrum}, a reduction in $k$ reduces the slepton masses, whilst increasing the gluon and squark masses. Further reductions in $k$ reduce the $\widetilde \mu$ mass sufficiently but keep the mass of the higgs boson high as this is dependant on the SU(3) charged squarks and sleptons. 

In Scan 2, the $k$ parameter is varied from $0$ to $-0.04$. As noted, negative values of $k$ will increase the absolute scale of $M_3$ but reduce the scale of $M_2$. This will further contribute to the effect described above, increasing squark masses whilst decreasing slepton masses.
As previously argued, the $\Delta a_\mu$ contributions depend on $m_{\widetilde \mu}^{-2}$ and therefore large values of $m_\frac{3}{2}$ will lead to a suppression of the contributions. We therefore scan for lower values of $m_{\frac{3}{2}}$ between 0 and 400 TeV.

\begin{figure}[h!]
\begin{subfigure}{.5\textwidth}
\includegraphics[width=.9\linewidth]{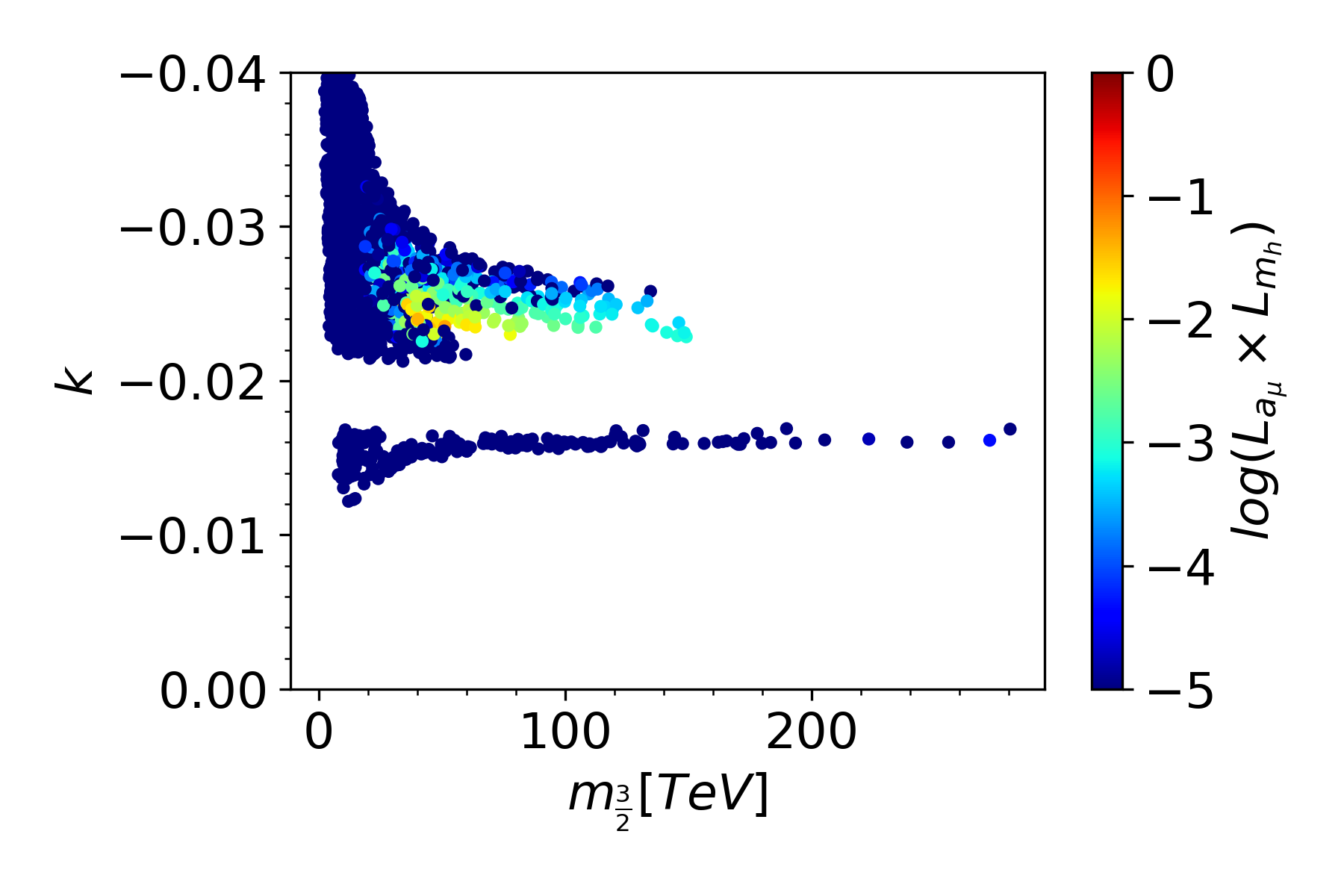}
\caption{$\mu<0$, case I}
\label{fig:scaleless_k_neg_k/mu-/m32_vs_k_mupos_amu_mh_like_constrained.}
\end{subfigure}
\begin{subfigure}{.5\textwidth}
\includegraphics[width=.9\linewidth]{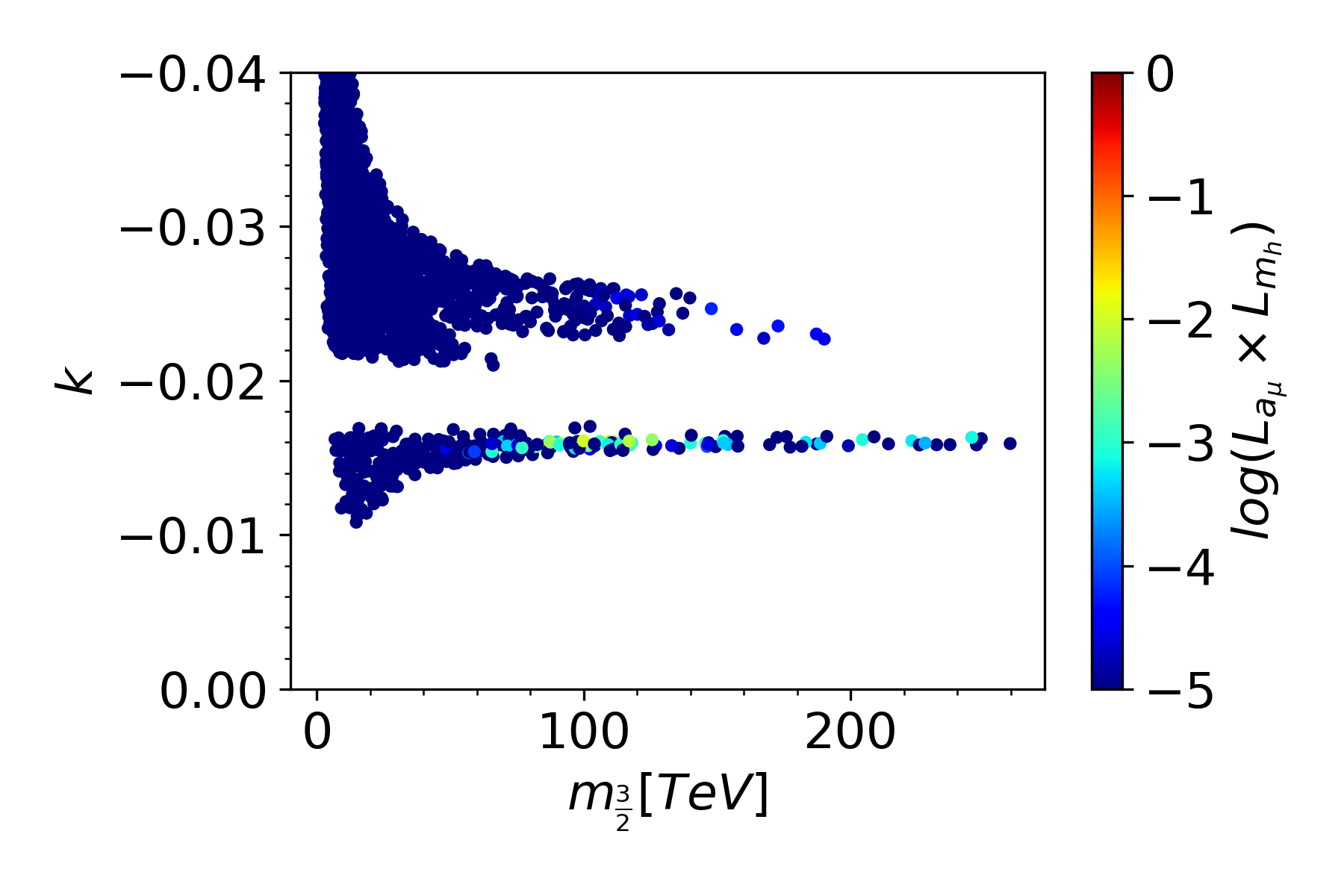}
\caption{$\mu>0$, case I}
\label{fig:scaleless_k_neg_k/mu+/m32_vs_k_mupos_amu_mh_like_constrained}
\end{subfigure}
\caption{Shows the distribution of $k$ against $m_\frac{3}{2}$ with $a_\mu$ and $m_h$ log likelihood for $m_{\frac{3}{2}}\in[0TeV,400TeV]$, $\alpha =0$ and $k\in[-0.035,-0.014]$. A region with $\mu<0$, $k\approx-0.023$ and $m_\frac{3}{2} \in [40TeV, 60Tev]$ is preferred. }
\label{fig:scaleless_k_neg_k/m32_vs_k_mupos_amu_mh_like_constrained}
\end{figure}

Fig \ref{fig:scaleless_k_neg_k/m32_vs_k_mupos_amu_mh_like_constrained} shows the distribution of $k$ against $m_\frac{3}{2}$ for both signs of $\mu$. As we are now focussing on $a_\mu$ we redefine the likelihood as $L=L_{a_\mu}\times L_{m_h}$. However, we do not want relic particles whose abundance would rule out the model entirely. Therefore, we impose the condition that $\Omega h^2 < 0.12+2\times0.0078$ such that no point can be ruled out by leaving a non-phenomenologically large relic abundance. 

From Fig \ref{fig:scaleless_k_neg_k/m32_vs_k_mupos_amu_mh_like_constrained}, we see two distinct areas of points that give viable results; $k\approx-0.016$ and $k\approx-0.023$. However we find that $k\approx -0.023$ with $\mu<0$ is phenomenologically preferred. In  this region, we also find $30TeV<m_\frac{3}{2}<100TeV$ to be preferred.

\begin{figure}[h!]
\begin{subfigure}{.5\textwidth}
\includegraphics[width=.9\linewidth]{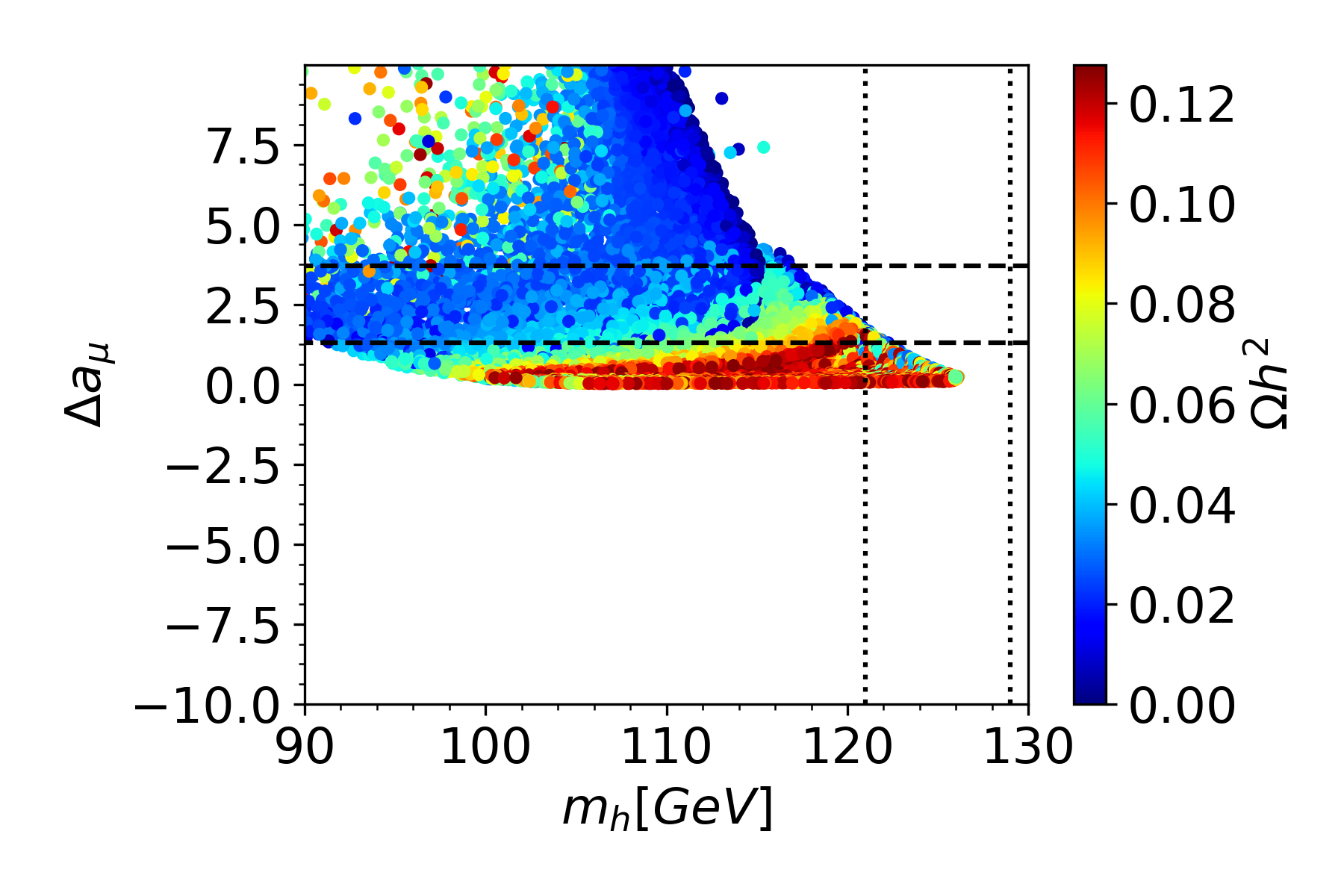}
\caption{$\mu<0$, case I }
\label{fig:scaleless_k_neg_k_sss/mu-/mh_vs_amu_mupos_constrained}
\end{subfigure}
\begin{subfigure}{.5\textwidth}
\includegraphics[width=.9\linewidth]{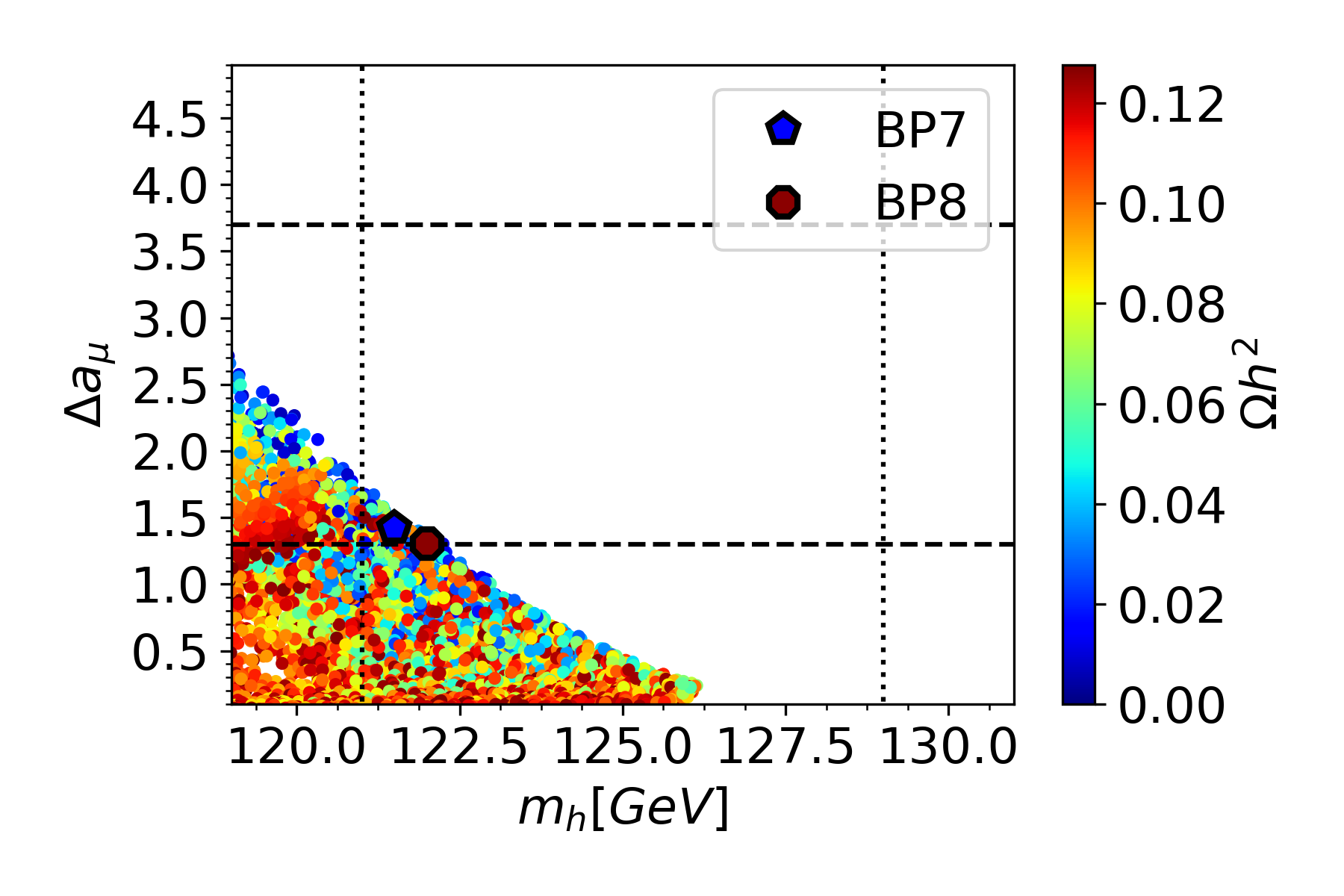}
\caption{$\mu<0$, case I, zoomed }
\label{fig:scaleless_k_neg_k_sss/mu-/mh_vs_amu_mupos_constrained_zoomed}
\end{subfigure}
\caption{Shows a scatter plot of $a_\mu$ and $m_h$ for a focussed scan with ranges $m_{\frac{3}{2}}\in[0TeV,160TeV]$, $k\in[-0.022,-0.027]$ and $tan(\beta)\in[2,12]$. The $2\sigma$ region of $a_\mu$ is marked with dotted lines while the $2\sigma$ region of $m_h$ is marked with dashed lines. The colour denotes the relic density.}
\label{fig:scaleless_k_neg_k_sss/mh_vs_amu_mupos_constrained}
\end{figure}

Finally, we present the results of a scan focussing on the good region as seen in Figure \ref{fig:scaleless_k_neg_k_sss/mh_vs_amu_mupos_constrained} where k varies between -0.022 and -0.027 and the sign of $\mu$ is fixed as negative. Negative results for $\Delta a_\mu$ are now ruled out as the absolute value of the gaugino mass parameters is large, suppressing some key SUSY contributions to $(g-2)$ that involve these particles. This leaves us with one first order loop diagram that contributes to this result whose sign is fixed by the sign of $\mu$.

Figure \ref{fig:scaleless_k_neg_k_sss/mh_vs_amu_mupos_constrained} shows the distribution of $\Delta a_\mu$ against the higgs mass. We see many points sit within the $2\sigma$ region. Furthermore, we find that many of these points have excellent dark matter relic densities. Two benchmark points are marked on Figure \ref{fig:scaleless_k_neg_k_sss/mh_vs_amu_mupos_constrained} who are presented later. 

\begin{figure}[h!]
\begin{subfigure}{.5\textwidth}
\includegraphics[width=.9\linewidth]{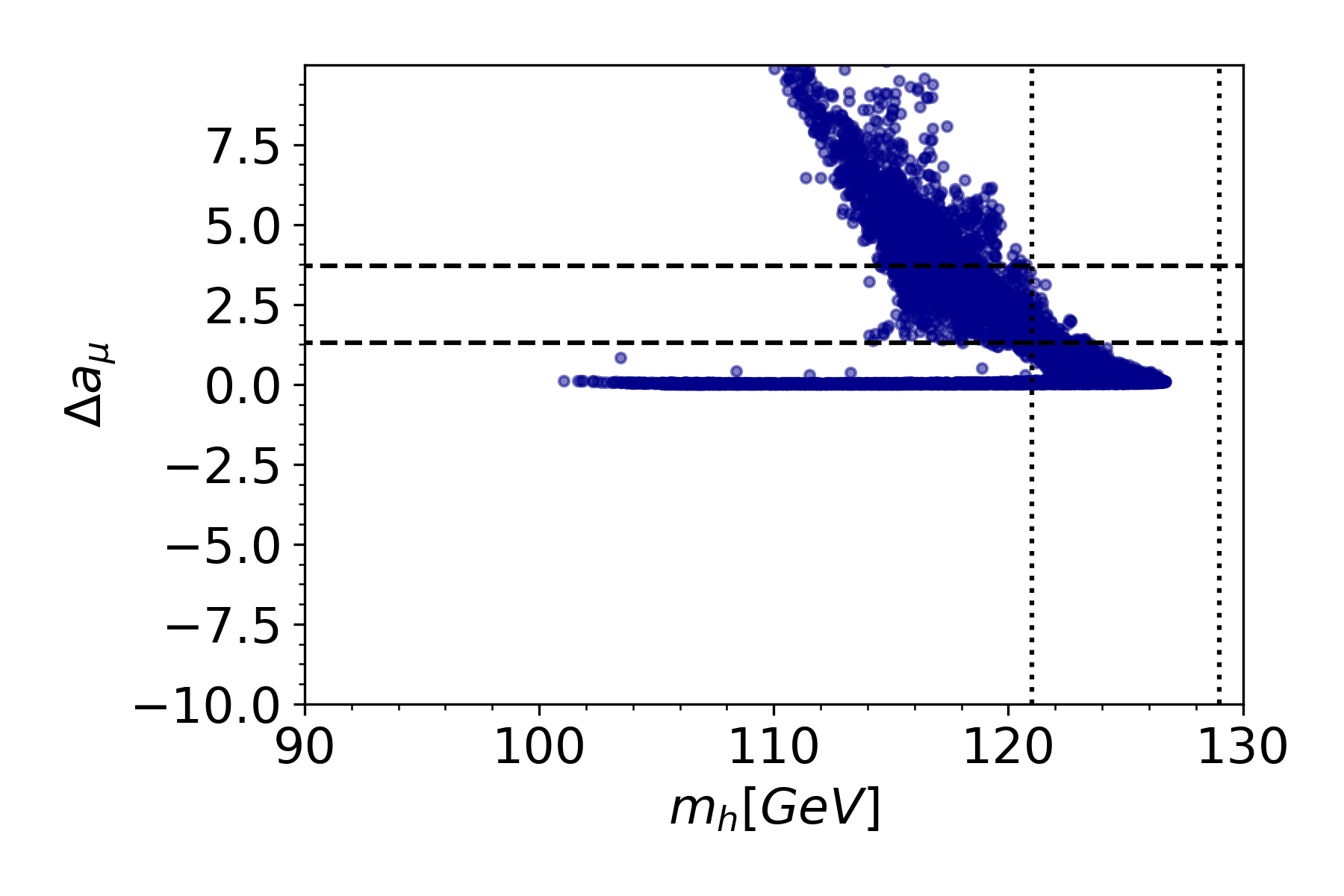}
\caption{$\mu<0$, case I }
\label{fig:scaleless_k_neg_k_sss/mu-/RPV/mh_vs_amu_mupos_constrained_combined}
\end{subfigure}
\begin{subfigure}{.5\textwidth}
\includegraphics[width=.9\linewidth]{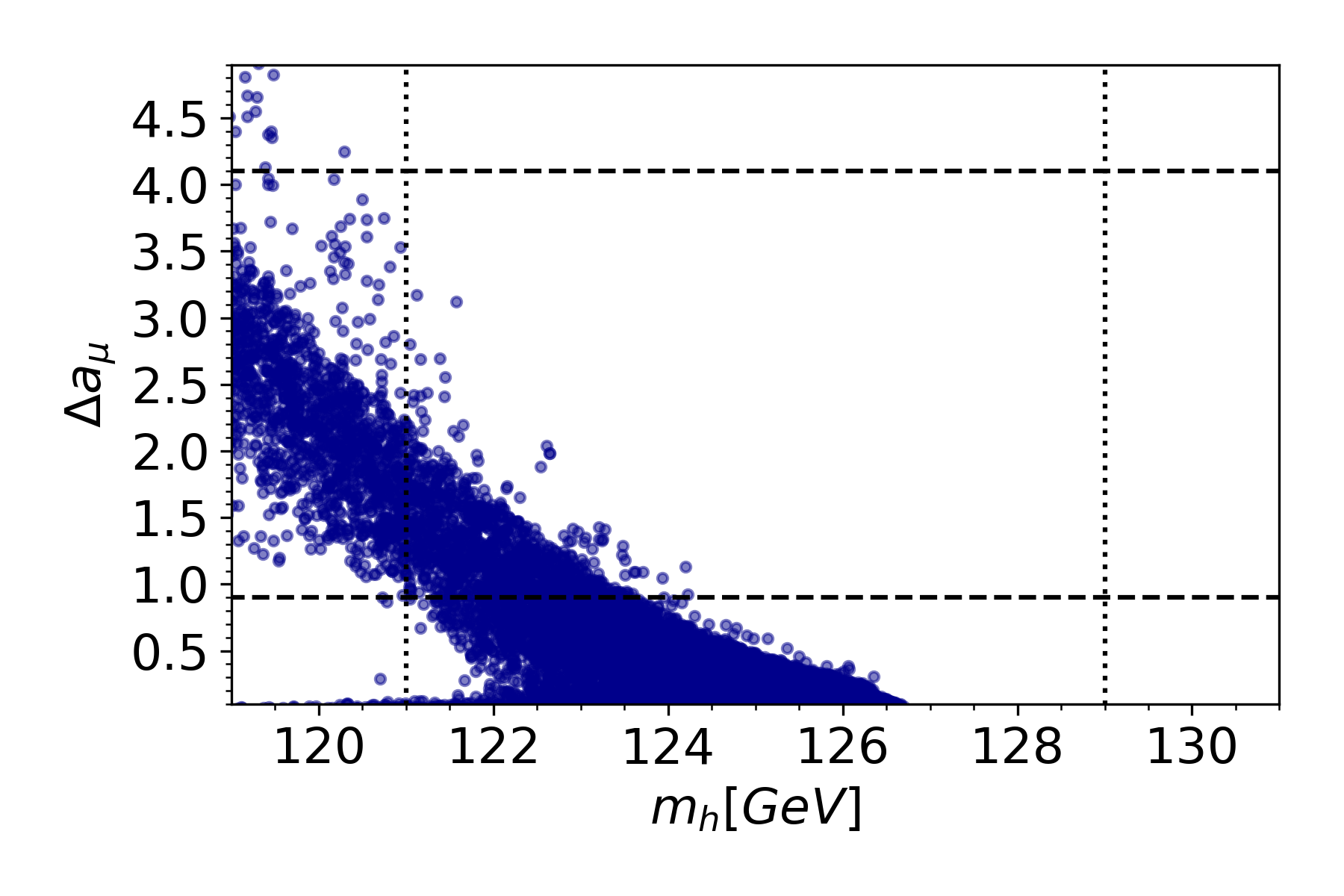}
\caption{$\mu<0$, case I, zoomed }
\label{fig:scaleless_k_neg_k_sss/mu-/RPV/mh_vs_amu_mupos_constrained_combined_zoomed}
\end{subfigure}
\caption{As for Fig \ref{fig:scaleless_k_neg_k_sss/mh_vs_amu_mupos_constrained} but where points that have charged LSP states are shown. Such states are marked in blue to indicate that no relic density could be calculated.}
\label{fig:scaleless_k_neg_k_sss/RPV/mh_vs_amu_mupos_constrained_combined_zoomed}
\end{figure}

Until now, the relic abundance of dark matter has played a pivotal role in assessing the veracity of any given point. However, we find large parts of the parameter space give the right handed stau state as the lightest sparticle. In all preceding plots and in Figure \ref{fig:scaleless_k_neg_k_sss/mh_vs_amu_mupos_constrained} we present results where the neutralino is the LSP. However, it is also interesting to consider potential R-parity violating models in which these light, charged particles decay into standard model particles. Although a detailed discussion of the viability of such points is beyond the scope of this paper, we include Figure \ref{fig:scaleless_k_neg_k_sss/RPV/mh_vs_amu_mupos_constrained_combined_zoomed} as these states can lead to excellent fits for $a_\mu$ and $m_h$.

\FloatBarrier
\FloatBarrier

Continuing our analysis, we took a number of points from the allowed region and ran them through CheckMATE. Finding that a sample of points were excluded we wrote a code to systematically check points in the allowed region. We found that all points were either excluded by CheckMATE or in a region near exclusion (allowing for uncertainties in the analysis). We include some of the more promising benchmark points below. It should be noted that we ran CheckMATE using the default 5000 points. However, if a warning was presented speculating that more data could exclude the point in question we increased the number of events by the specified amount. 

In Table \ref{tab:input-table-78} we present the input parameters for two benchmarks points from Case I with negative $k$ values. One point was selected for its relatively high values of $\Delta a_\mu$ whilst the other was selected for it's relatively high value of $\Omega h^2$. Both points are $k=-0.023$ in accordance with the highest likelihood region. Indeed, both points are very similar in a variety of ways suggesting a small input deviations can give large variance in output; particularly for the LSP relic abundance.

\begin{table}
\begin{tabularx}{\textwidth}{  m{0.11\textwidth}    m{0.35\textwidth} m{0.35\textwidth}     }
\hline
Quantity &    BP7 & BP8    \\
\hline
$\alpha $ &    			  				0		&	0						\\                  
$\beta $ &   							1		&	1						\\                             
$ m_{\frac{3}{2}} $ [TeV] & 				51		&	57						\\	
 $k$ &    				 				-0.023	&	-0.023					\\              
\hline
\underline{SPheno:} &          							&							\\ 
\hline
$m_{0}$ [GeV] &       					0		&	0						\\		                   
$tan(\beta)$ &          						9.72		&	9.37						\\		     
 $sign(\mu)$ &        						-1		&	-1						\\		          
 $A_{0}$ [GeV] &     						0		&	0					\\		                      
 $M_{1}$ [GeV] &   						-263		&	-263 				\\		               
 $M_{2}$ [GeV] &     						-1058	&	-1152						\\		               
 $M_{3}$ [GeV] &     						-1694	&	-1863					\\	
\hline
\end{tabularx}
\caption{Shows two benchmark points representing two different areas of interest in the parameter space. We present the model parameters and the resultant SPheno input parameters. BP7 shows a point with high $a_\mu$ and BP8 shows a point with high $\Omega$. $ m_{\frac{3}{2}} $ is given in TeV and all other dimensionful parameters are given in GeV.  \label{tab:input-table-78}}
\end{table}

Table \ref{tab:table-spectrum-78} shows the SUSY spectrum for benchmark points 7 and 8. Strikingly, we see that the lightest stau state is the nLSP behind the usual neutralino LSP. This ofcourse has implications for the most powerful diagrams for the relic density. Furthermore, the nnLSP is given by $\widetilde \mu_{R} $ as anticipated by our efforts to generate a high $\Delta a_\mu$ value using diagrams involving this particle. Unlike the sleptons, the squarks are very high in mass contributing to the phenomenologically viable higgs mass. In both cases $\widetilde \tau_1 -\widetilde \chi_{10} \approx10 GeV$. This is important for critical co-annihilation diagrams for the relic density. Again we see that $B_0$ is very small in comparison with $m_\frac{3}{2}$ suggesting a fully scale-less model. 

\begin{table}
\vspace{-1em}
\begin{tabularx}{\textwidth}{  m{0.11\textwidth}   m{0.35\textwidth} m{0.35\textwidth}     }
\hline
Masses &    BP7 & BP8 \\
\hline
$\widetilde e_{L} $&   							661		&	717				\\
$\widetilde e_{R} $&    	 					133  	&	138				\\
$\widetilde \nu_{e L} $&   						656		&	712				\\
$\widetilde \mu_{L} $&        					661		&	717				\\
$\widetilde \mu_{R} $&          				133		&	137				\\
$\widetilde \nu_{\mu L} $&  					656		&	712				\\
$\widetilde \tau_{1} $&        					104		&	109				 \\
$\widetilde \tau_{2} $&       					661		&	717				\\
$\widetilde \nu_{\tau L} $&       				654		&	710				\\
\hline
$\widetilde{d}_{L} $ &      				    3110	&	3390				\\
$\widetilde d_{R} $&          					3050	&	3330				\\
$\widetilde u_{L} $&    						3110	&	3390				 \\
$\widetilde u_{R} $ &        					3050	&	3320				\\
$\widetilde s_{L} $&        					3110	&	3390				\\
$\widetilde s_{R} $&         					3050	&	3330				\\
$\widetilde c_{L} $&         					3110	&	3390				 \\
$\widetilde c_{R} $&           					3050	&	3320				\\  
$\widetilde b_{1} $ &         					2890	&	3150				\\
$\widetilde b_{2} $ &     						3040	&	3310				\\
$\widetilde t_{1} $&        					2580	&	2820				 \\
$\widetilde t_{2} $&      						2910	&	3170				\\
\hline
$\widetilde g $&                   				3560	&	3890				\\
$\widetilde \chi_{10} $&          				100		&	99			\\
$\widetilde \chi_{20} $&               			855		&	933			\\
$\widetilde \chi_{30} $&             			1890	&	2060				 \\  
$\widetilde \chi_{40} $&             			1890	&	2060					\\  
$\widetilde \chi_{1+} $&           				855		&	933			\\
$\widetilde \chi_{2+} $&            			1890	&	2060				 \\  
\hline
${h_{0}}$ &          							121.5	&	122				\\                  
${H_{0}}$ &    									1990	&	2170				\\               
${A_{0}}$ &       								1990	&	2170				\\                   
${H_{\pm}}$ &  									2000	&	2170				 \\
$\mu$  & 									-1777	&	-1931				\\
$B_{0}$ &      								-12		&	-26.5				\\                               
\hline
$\widetilde \tau_1 -\widetilde \chi_{10} $ &  		4.2		&	9.6								\\      
\hline
\end{tabularx}
\caption{Shows the spectrum of SUSY masses for the benchmark points given in Table \ref{tab:input-table-78}. The difference between the mass of $\widetilde \tau_1$ and $\widetilde \chi_{10}$ is also given as this pertains to the production of dark matter. We also include the high scale bilinear coupling value $B_0$ for its relevance to the high scale parameters of the model. All parameters are given in GeV.   \label{tab:table-spectrum-78}}
\end{table}

Both benchmarks produce a completely bino-like LSP state caused by the large scale difference between $M_1$ and $M_2$ and have very similar dark matter physics in general. Although BP7 has a dark matter relic density within the allowed region, BP8 gives an excellent value. We see a much larger mass gap between the LSP and the nLSP than for Benchmark points 1 to 6. This is because the points produce a much smaller mass LSP. Therefore, the freeze out temperature is low and as such, there is more cosmological time for co-annihilation to occur. A larger mass gap reduces the strength of these co-annihilation channels allowing for the phenomenological Higgs boson mass. 

\begin{table}
\begin{tabularx}{\textwidth}{  m{0.3\textwidth}     m{0.35\textwidth} m{0.35\textwidth}     }
\hline
Quantity &    BP7 & BP8  \\
\hline     
$\Omega_{DM} h^2$  & 									0.020	&	0.112			\\           
\hline
$\widetilde \chi_{10} $ [GeV]&          								100		&	99				\\
$\widetilde \tau_1 -\widetilde \chi_{10} $ [GeV] & 					4.2		&	9.6				\\     
\hline
$|\alpha_1|^2$ &							 	1		&	1						\\
$|\alpha_2|^2$ &								0		&	0						\\
$|\alpha_3|^2$ &								0		&	0						\\
$|\alpha_4|^2$ &								0		&	0						\\
\hline
\end{tabularx}
\caption{Shows the relic density of the LSP for each benchmark point. The difference between the LSP and the nLSP is also given. Finally, we give the probability of finding the given LSP in a particular flavour state. That is to say; we give $|\alpha_i|^2$  where $\widetilde \chi_{10} = \alpha_1 \widetilde B + \alpha_2 \widetilde W + \alpha_3 \widetilde H_1 + \alpha_4 \widetilde H_2$ and $\sum |\alpha_i|^2 = 1$. Dimensionful parameters are given in GeV.\label{tab:table-omega-78} }
\end{table}

Table \ref{tab:decay-BP78} shows some key observables for the given benchmark points. As previously stated, both points are at the borderline of exclusion by Checkmate analysis, since, both points still have relatively low $r_{max}\sim 1$ values, suggesting that both points might be viable within the uncertainties of the analysis, as we discuss further below.

We see short life times for $\widetilde \tau_1$ with only one decay channel therefore potentially leaving a strong collider signature. Both points get their strongest constraints from the CMS analysis \cite{CMS:2017moi} that focuses on direct electroweak production of charginos and neutralinos leading to final state leptons, little hadronic activity, and a large missing momentum.  As the colour charged particles are far more massive than the leptons and gauginos, the latter will represent the dominant production mechanism for the sparticles. The signal region focuses on a final state with three light leptons where two of the three are either $e$ or $\mu$ and as we have a light gaugino LSP with light sleptons for $g-2$ this is likely to be constraining. Furthermore this region focuses on states where the leptonic pair have invariant mass greater than $105GeV$ and transverse mass of the third lepton greater than $160GeV$.  $\widetilde \chi_{20}$ has a $32\%$ branching ratio into $\widetilde \mu^\pm \mu^\mp$ / $\widetilde e^\pm e^\mp$ combined with $\widetilde \chi_{1+}$'s tendency to decay into $\widetilde \tau \nu_{\tau}$ pairs creates a strong signal in this region placing the parameter point on the edge of exclusion. Furthermore, the relatively large mass of $\widetilde \chi_{20}$ means that the resultant lepton pair will exceed the required $105$ GeV invariant mass and the production of the $\tau$ are on the edge of exclusion of the transverse mass limit. The b-type decays satisfy the experimental constraints for both points.

\begin{table}
\begin{tabularx}{\textwidth}{  m{0.3\textwidth}    m{0.35\textwidth} m{0.35\textwidth}   }
\hline
Quantity &    BP7 &  BP8\\
\hline
$\Gamma \widetilde \tau_1$ [GeV] &  										$3.30\times10^{-3}$		&	$1.63\times10^{-2}$						\\
BR $(\widetilde \tau^-_1 \rightarrow \chi_{10} \tau^-)$ [\%]	&						100					&	100   						\\
\hline
$\Gamma \widetilde \mu_{R}$ [GeV] & 										$1.32\times10^{-1}$ 		&	$1.67\times10^{-1}$				\\
BR $(\widetilde \mu^-_R -\rightarrow \chi_{10} \mu^-)$ [\%]& 						100					&	100							\\
\hline
BR $(b \rightarrow s \: \gamma)$ [\%] &  										0.032 				&	0.032					\\ 
BR $(B_s \rightarrow \mu^+ \: \mu^-)$ [\%] &									$2.97\times 10^{-7}$		&	$2.96\times10^{-7}$			\\
$\Delta \frac{(g-2)_\mu}{2}$ & 												$1.42\times10^{-9}$ 		&	$1.30\times10^{-9}$			\\
$\Omega_{DM} h^2$  & 													0.020				&	0.112					\\
$\chi_{10}$ [GeV]& 														100					&	99						\\
$\sigma_{q \overline{q} \rightarrow  \chi_{10} \chi_{10}}$ [pb] &  					$1.796\times10^{-14}$	&	$3.562\times10^{-14}$		\\

\hline
$r_{max}$ &															$1.22$				&	$1.04$					\\
$\sqrt{s}$ [TeV] &														13					&	13						\\
Analysis	&															cms\_sus\_16\_039		&	cms\_sus\_16\_039			\\
Signal Region &														SR\_A44				&	SR\_A44					\\
Ref.			&														\cite{CMS:2017moi}		&	\cite{CMS:2017moi}			\\
$\sigma_{LO}$ [pb]	& 													$7.127\times10^{-11}$	&	$5.744\times10^{-11}$		\\
\end{tabularx}

\caption{Shows branching ratios for lightest supersymmetric particles in the spectrum for BP7 and BP8. Only branching ratio greater than $1\%$ are included. We also include some beyond the standard model observables BR $(b \rightarrow s \: \gamma)$, BR $(B_s \rightarrow \mu^+ \: \mu^-)$, $\Delta \frac{(g-2)_\mu}{2}$, and $\Omega_{DM} h^2$, where $\Delta \frac{(g-2)_\mu}{2}$ is a calculation of the SUSY contribution beyond the standard model. The model successfully predicts the b decays discrepancy and satisfies the anomalous muon magnetic moment to $2\sigma$. The relic density is too small and is therefore not ruled out phenomenologically. CheckMATE runs using 13 TeV and 8 TeV analyses rule out these points. Decay widths and masses are given in GeV, branching ratios are given in \%, and cross sections are given in pb.   \label{tab:decay-BP78} }
\end{table}

\section{No-scale SUGRA with non-zero $A_0$ (Case II)\label{sec:c2} } 
\label{caseII}

In order to fully explore the parameter space we can also allow for non-zero values of $A_0$. Indeed, non-zero $A_0$ will serve to increase the Higgs boson mass, encouraging a better fit to the current experimental results. We therefore present a subsequent scan where we vary $\alpha$ between $-0.016$ and $0.016$. These values were chosen by trial and error such that computer time would not be wasted by producing many points with large $A_0$ values breaking colour charge symmetry. 

Analogously to the previous, we start by analysing the positive $k$ values. Further to this analogy, we revert to the previous definition of the likelihood to emphasise the relic density as key constraint in this paradigm and highlight the fact that $g-2$ will not be satisfied for positive values of $k$.

\subsection{Positive $k$ (Scan 3) \label{sec:s3}}

\begin{figure}[h!]
\begin{subfigure}{.5\textwidth}
\includegraphics[width=.9\linewidth]{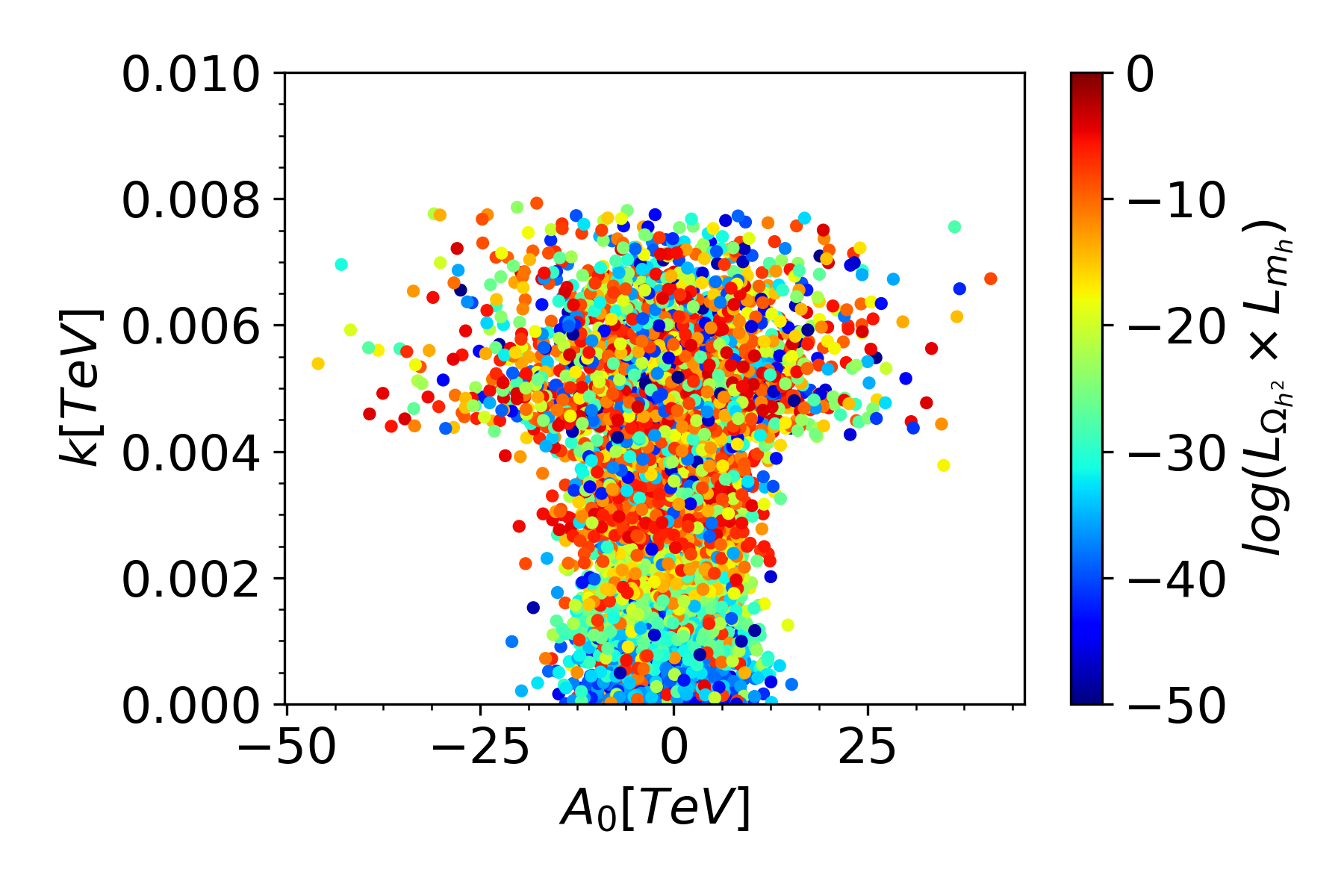}
\caption{$\mu<0$, case II }
\label{fig:A0_posk/mu-/A0_vs_tanB_mupos}
\end{subfigure}
\begin{subfigure}{.5\textwidth}
\includegraphics[width=.9\linewidth]{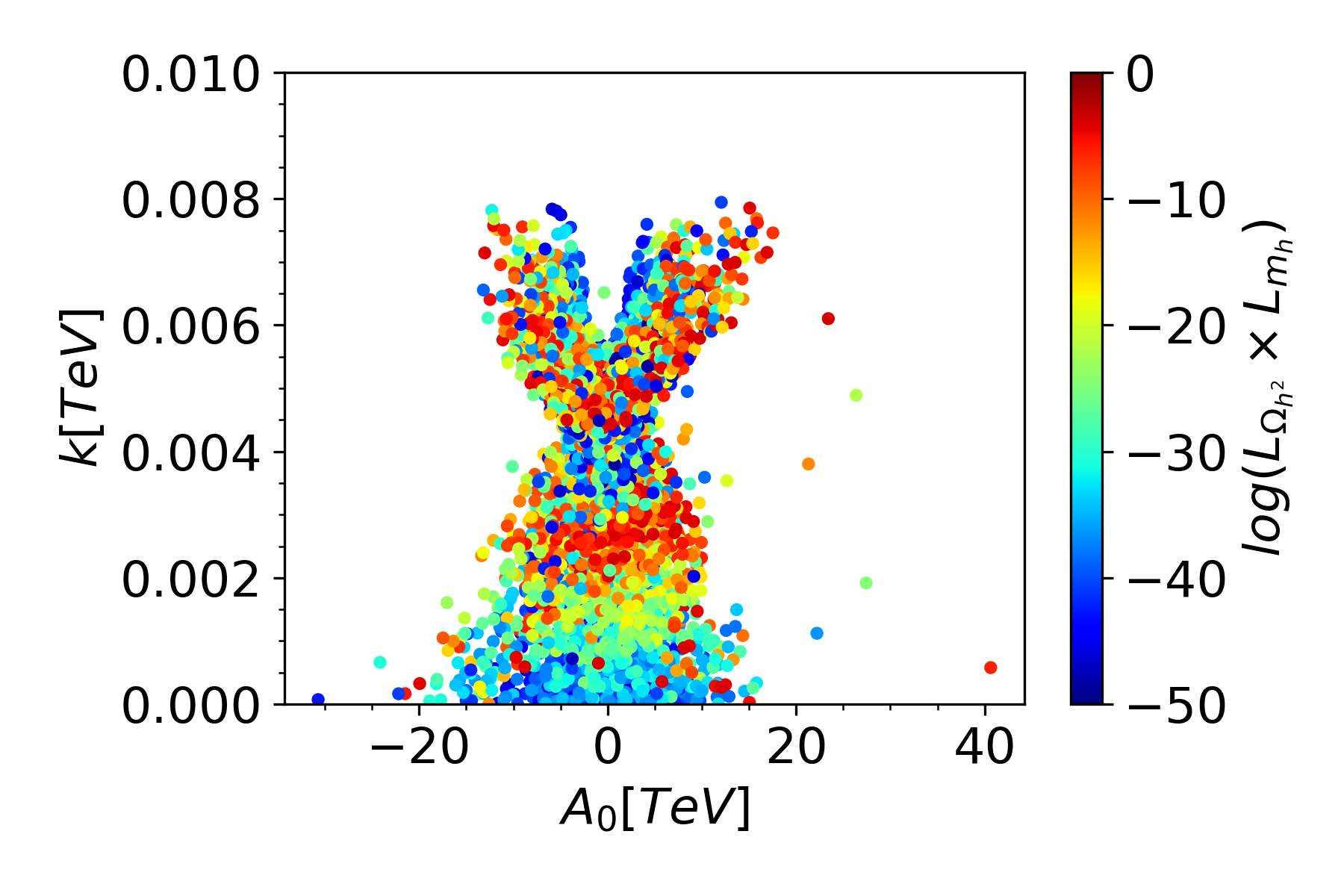}
\caption{$\mu>0$, case II  }
\label{fig:A0_posk/mu+/A0_vs_k_mupos}
\end{subfigure}
\caption{Shows the distribution of $A_0$ where case I data from a Monte-Carlo scan with parameter ranges $tan(\beta) \in [1.5,50], k \in [0,0.1]$, $\alpha \in [-0.166, 0.166]$, and $m_\frac{3}{2} \in [10^{3}GeV,10^{6}GeV]$. Colour denotes likelihood, with hotter colours corresponding with high likelihoods. As before, the likelihood is dominated by the relic density calculation.  }
\label{fig:A0_posk/A0}
\end{figure}

\begin{figure}[h!]
\begin{subfigure}{.5\textwidth}
\includegraphics[width=.9\linewidth]{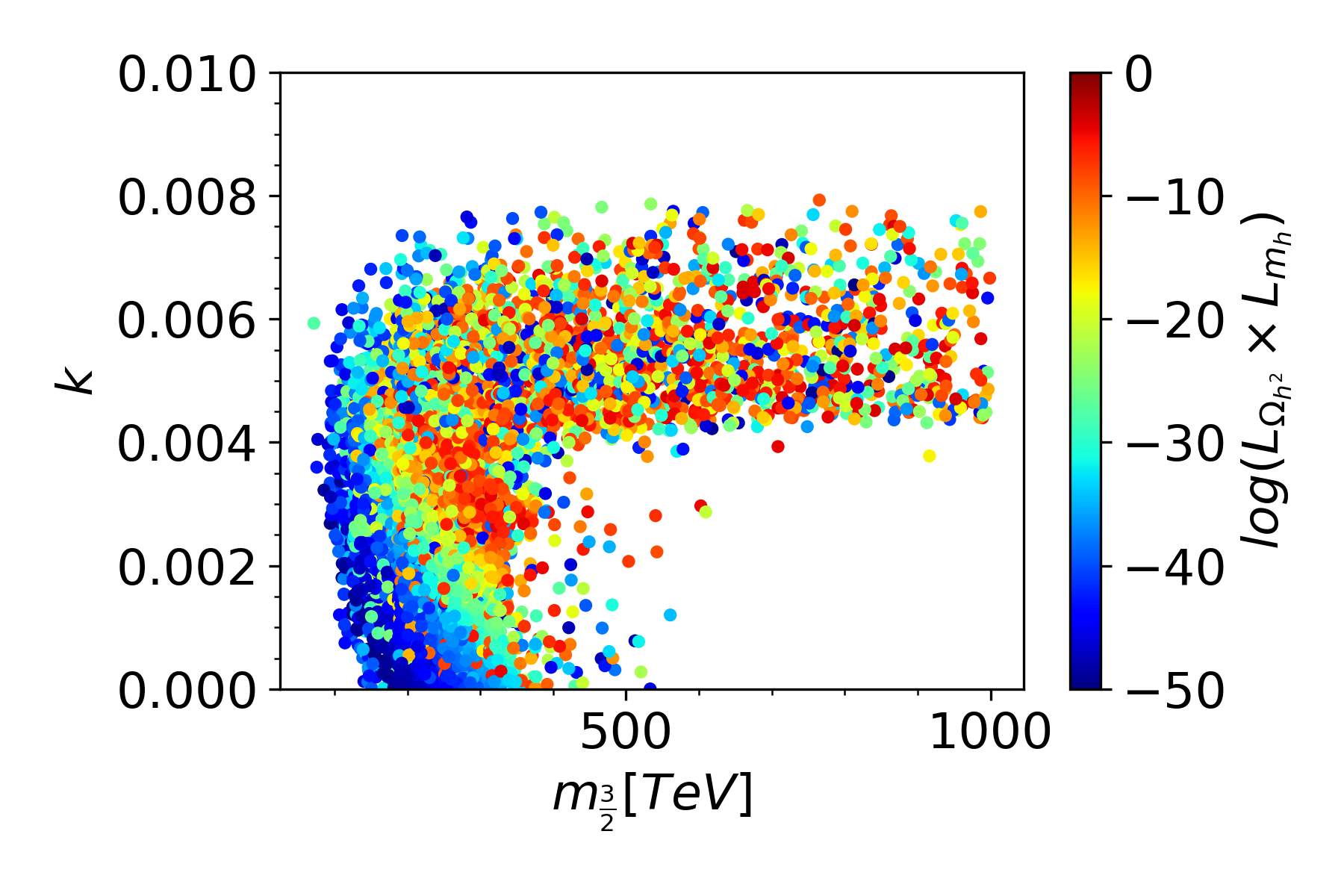}
\caption{$\mu<0$, case II  }
\label{fig:A0_posk/mu-/m32_vs_k_mupos}
\end{subfigure}
\begin{subfigure}{.5\textwidth}
\includegraphics[width=.9\linewidth]{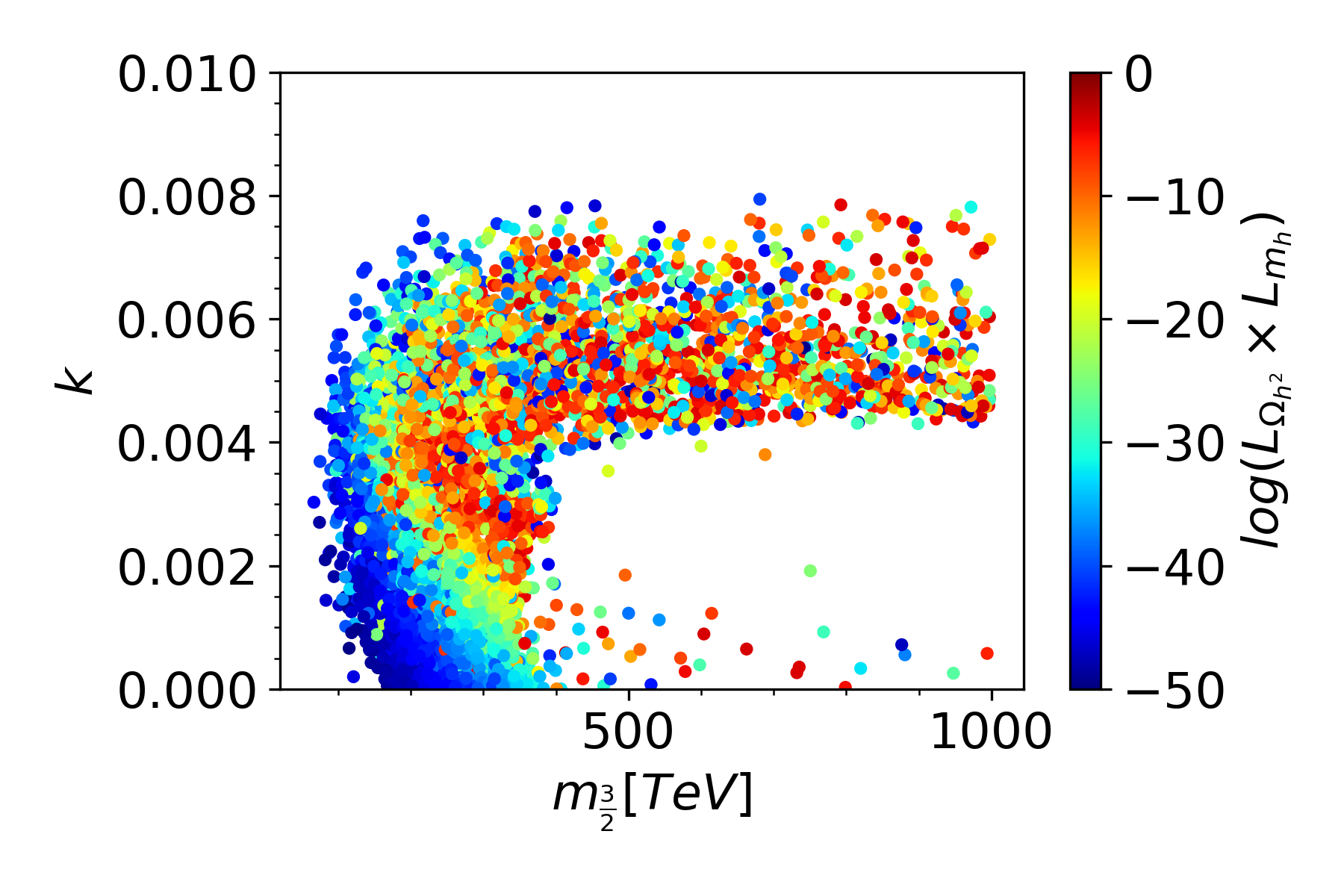}
\caption{$\mu>0$, case II }
\label{fig:A0_posk/mu-/m32_vs_k_mupos}
\end{subfigure}
\caption{Shows $k$ against $m_\frac{3}{2}$ where case II data from a Monte-Carlo scan with parameter ranges ranges $tan(\beta) \in [1.5,50], k \in [0,0.1]$, $\alpha \in [-0.166, 0.166]$, and $m_\frac{3}{2} \in [10^{3}GeV,10^{6}GeV]$. Colour denotes likelihood, with hotter colours corresponding with high likelihoods. As before, the likelihood is dominated by the relic density calculation.  }
\label{fig:A0_posk/m32_vs_k_mupos}
\end{figure}

Fig \ref{fig:A0_posk/A0} shows the distribution of $A_0$ against $k$ for both signs of $\mu$. Interestingly, the shape of the two plots for the two signs of $\mu$ is quite different, with the $\mu<0$ sign yielding a mushroom shaped distribution of points, while the $\mu>0$ sign results in a wigwam shaped plot. However, there are some key features of both that are shared. In neither do we see any particular likelihood increase for non-0 $A_0$. This suggests that moving to case II will not give significant improvement compared to the previous case. Furthermore, in both we observe a preferred region of $k\approx 0.003$.  

Fig \ref{fig:A0_posk/m32_vs_k_mupos} closely mirrors the structure of Fig \ref{fig:scaleless_k/mu+/m32_vs_k_mupos}; with $A_0$ variation causing some spread in the likelihoods. The structure of LSP make-up as well as spectrum is generally similar. For this reason we refer you to the previous for detailed analysis. Furthermore, the density of points is greatly reduced as variation in $A_0$ produces many low-likelihood points. This does support claims of naturalness for the exclusively scaleless model previously presented. Although this scan was not truly exhaustive, no gains were made in terms of the overall likelihood. 

As in case I with positive $k$, we see that no point can satisfy the higgs mass  and $g-2$ simultaneously. The same is true of $g-2$ and the relic density. In order to achieve the correct higgs mass, $m_\frac{3}{2}$ must be at least 200TeV. This has the effect of increasing the slepton masses and thus decreasing $\Delta a_\mu$. We therefore, turn our attention to $k<0$ values.

\subsection{Negative $k$ (Scan 4) \label{sec:s4}}

\begin{figure}[h!]
\begin{subfigure}{.5\textwidth}
\includegraphics[width=.9\linewidth]{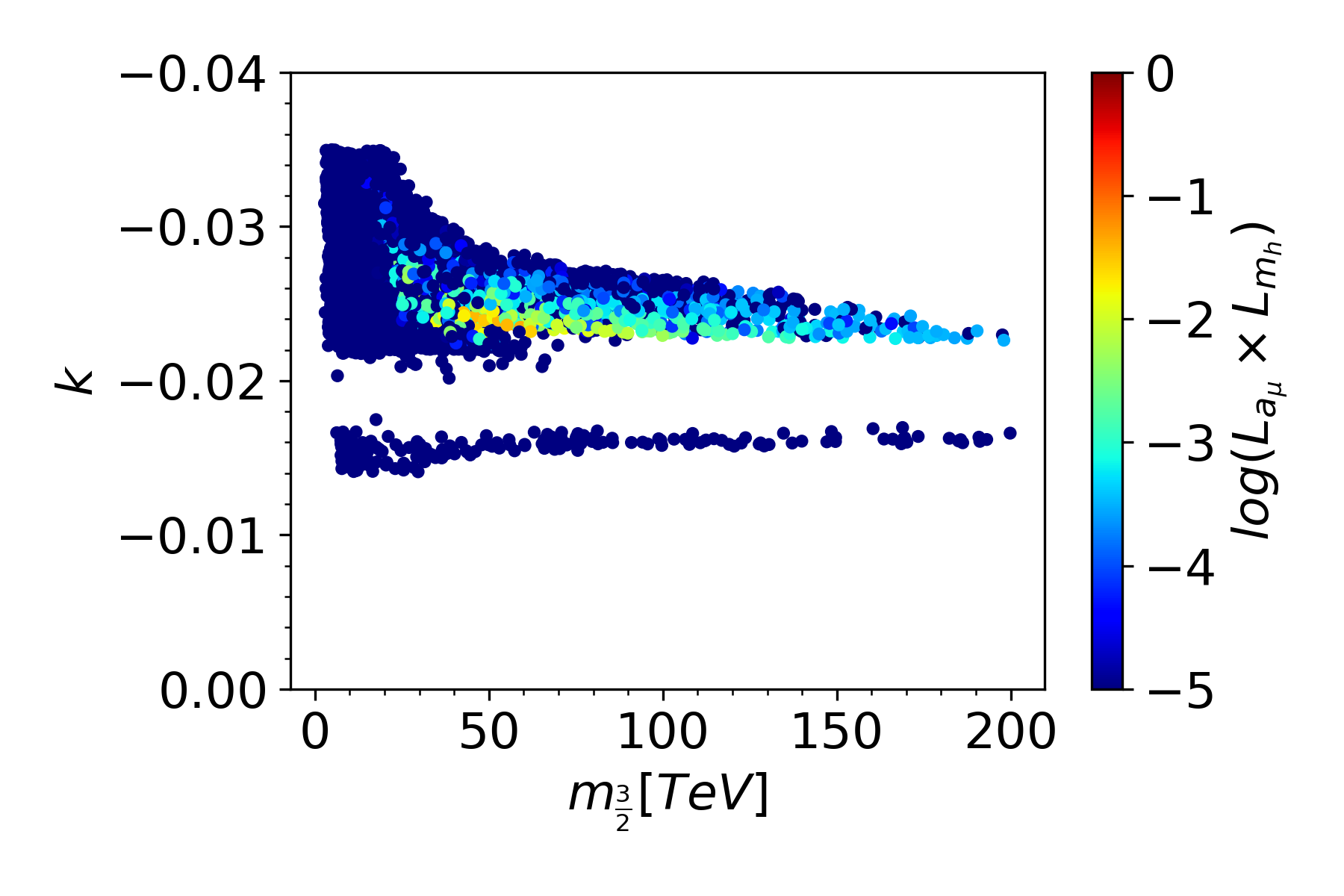}
\caption{$\mu<0$, case II }
\label{fig:A0/gmin2/mu-/m32_vs_k_mupos_amu_mh_like_constrained}
\end{subfigure}
\begin{subfigure}{.5\textwidth}
\includegraphics[width=.9\linewidth]{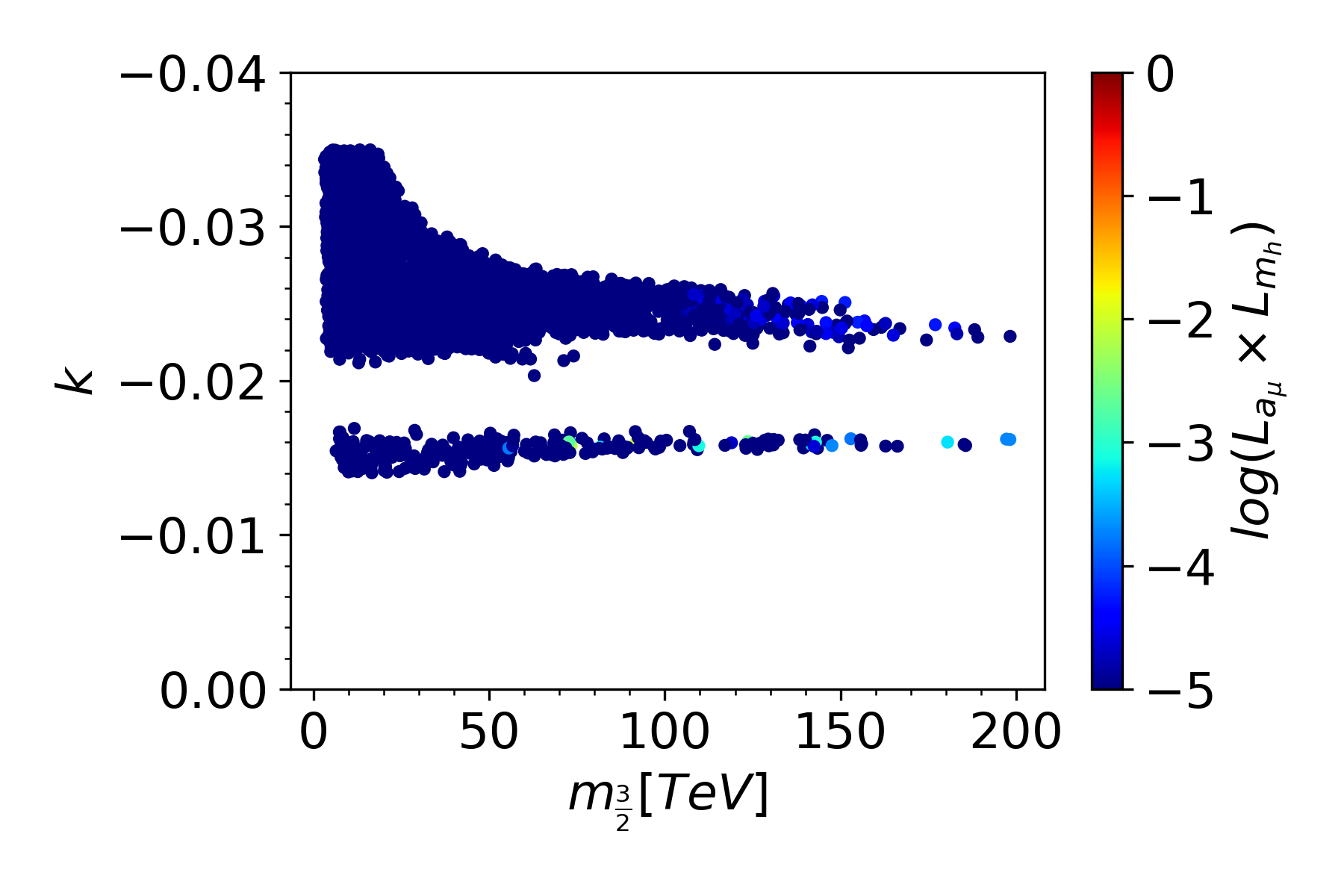}
\caption{$\mu>0$, case II }
\label{fig:A0/gmin2/mu+/m32_vs_k_mupos_amu_mh_like_constrained}
\end{subfigure}
\caption{Shows the distribution of $k$ against $m_\frac{3}{2}$ with $a_\mu$ and $m_h$ log likelihood for $m_{\frac{3}{2}}\in[0TeV,200TeV]$, $\alpha \in [-0.005, 0.005]$ and $k\in[-0.035,-0.014]$. }
\label{fig:A0/gmin2/m32_vs_k_mupos_amu_mh_like_constrained}
\end{figure}

\begin{figure}[h!]
\begin{subfigure}{.5\textwidth}
\includegraphics[width=.9\linewidth]{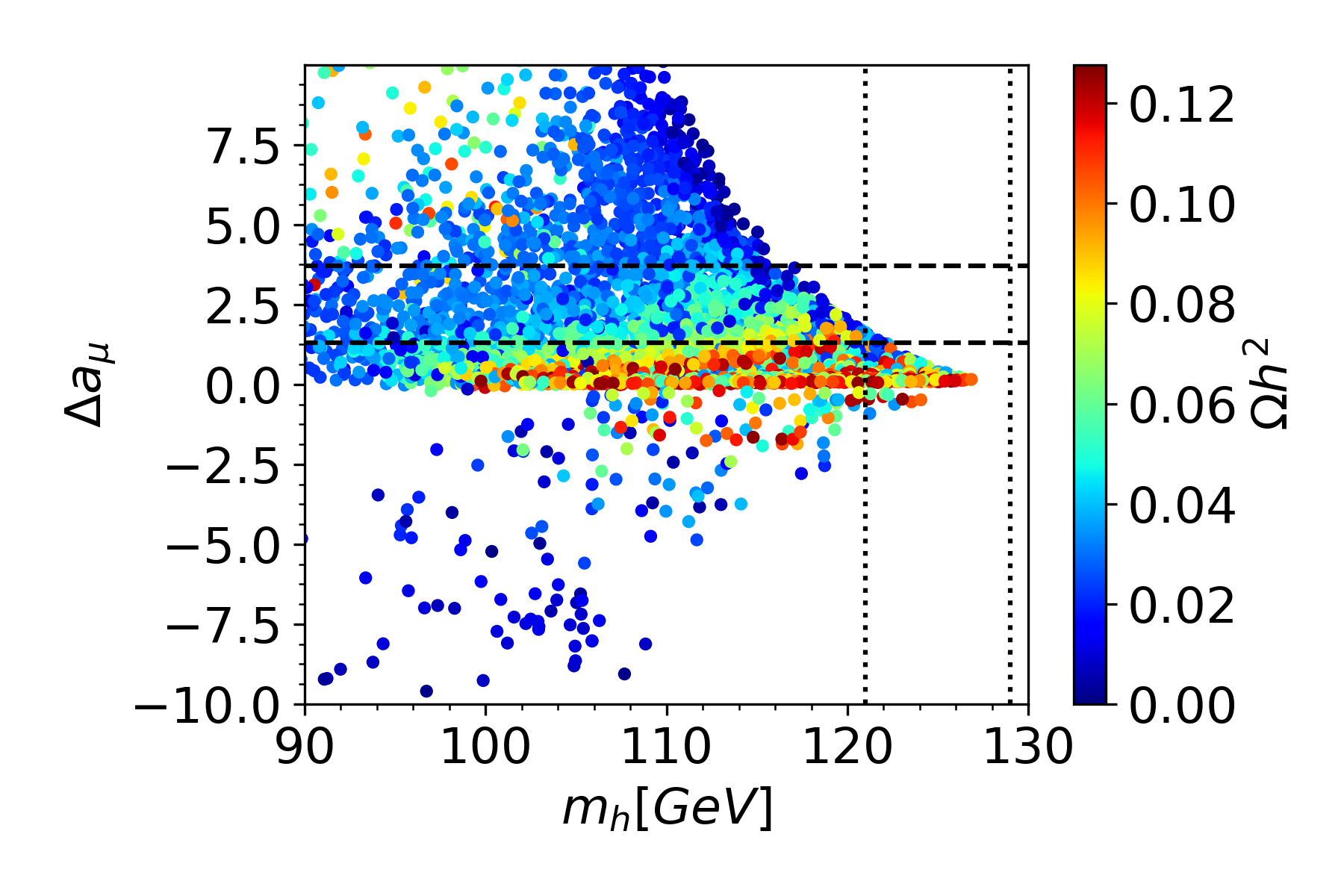}
\caption{$\mu<0$, case II  }
\label{fig:A0/gmin2/mu-/mh_vs_amu_mupos_constrained}
\end{subfigure}
\begin{subfigure}{.5\textwidth}
\includegraphics[width=.9\linewidth]{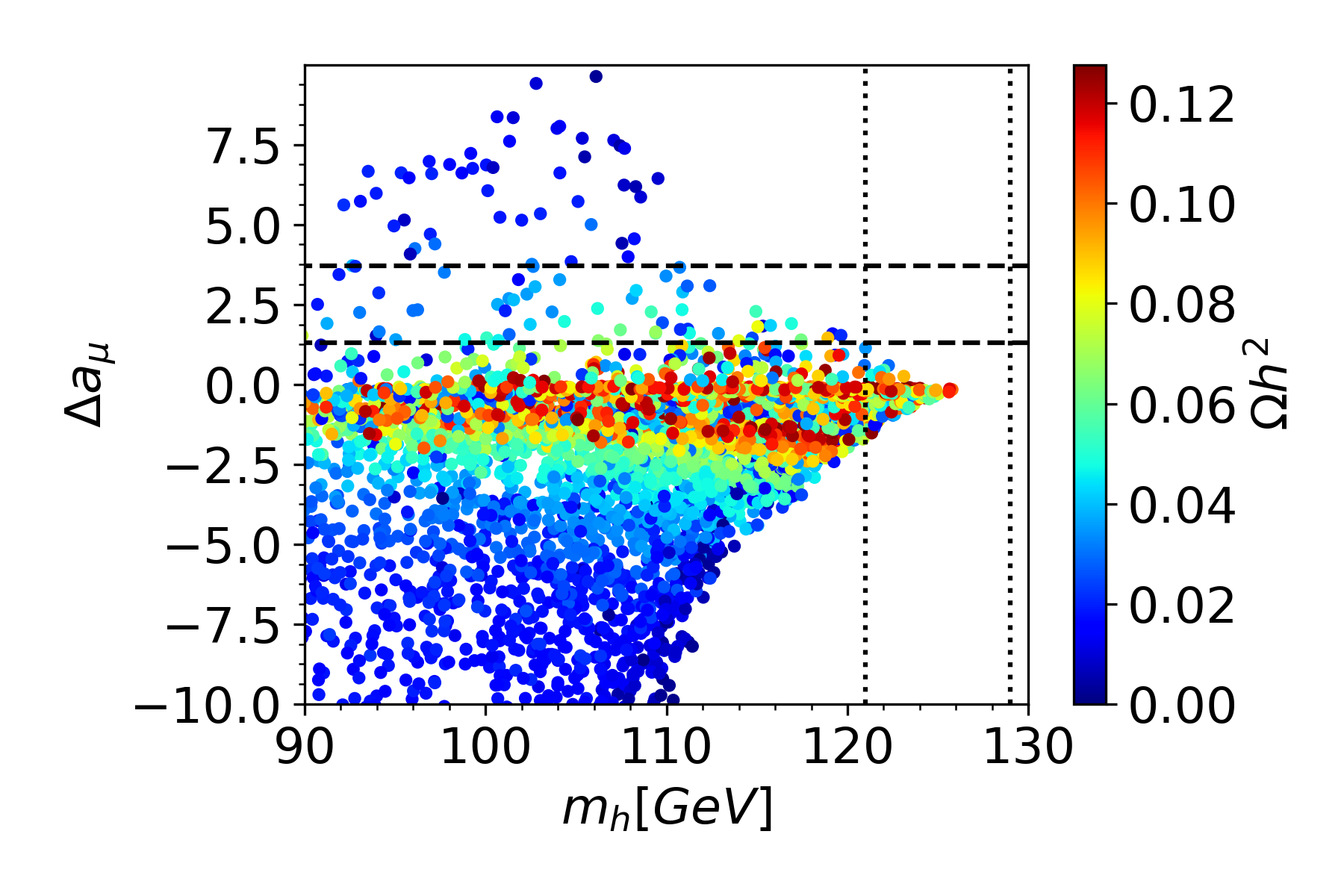}
\caption{$\mu>0$, case II }
\label{fig:A0/gmin2/mu_/mh_vs_amu_mupos_constrained}
\end{subfigure}
\caption{Shows a scatter plot of the higgs mass against the anomalous muon magnetic moment for a scan with ranges $m_{\frac{3}{2}}\in[0TeV,200TeV]$, $\alpha \in [-0.005, 0.005]$ and $k\in[-0.035,-0.014]$. The $2\sigma$ region of $a_\mu$ is marked with dotted lines while the $2\sigma$ region of $m_h$ is marked with dashed lines. The colour denotes the relic density.}
\label{fig:A0/gmin2}
\end{figure}

\begin{figure}[h!]
\begin{subfigure}{.5\textwidth}
\includegraphics[width=.9\linewidth]{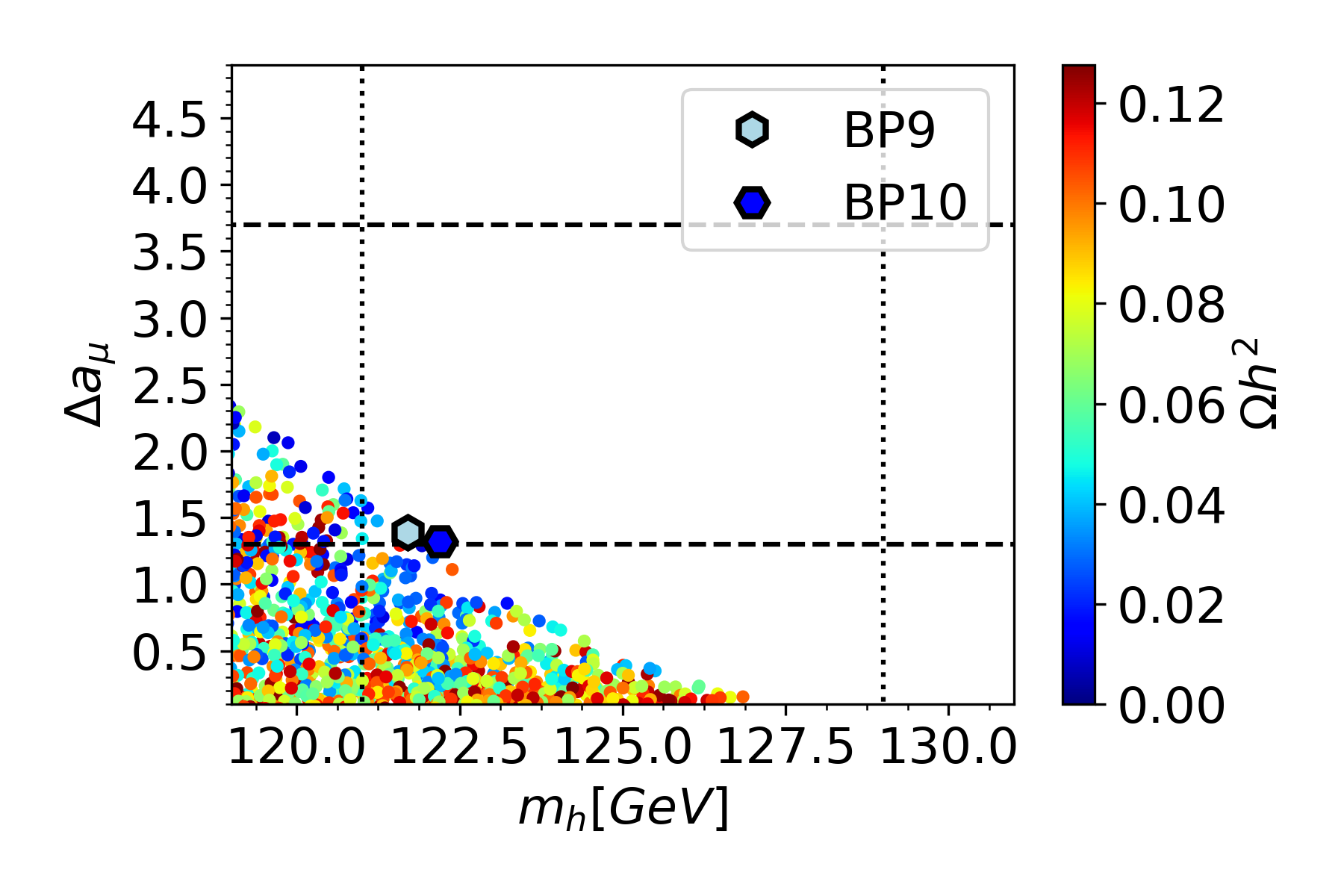}
\caption{$\mu<0$, case II }
\label{fig:A0/gmin2/mu-/mh_vs_amu_mupos_constrained_zoomed}
\end{subfigure}
\begin{subfigure}{.5\textwidth}
\includegraphics[width=.9\linewidth]{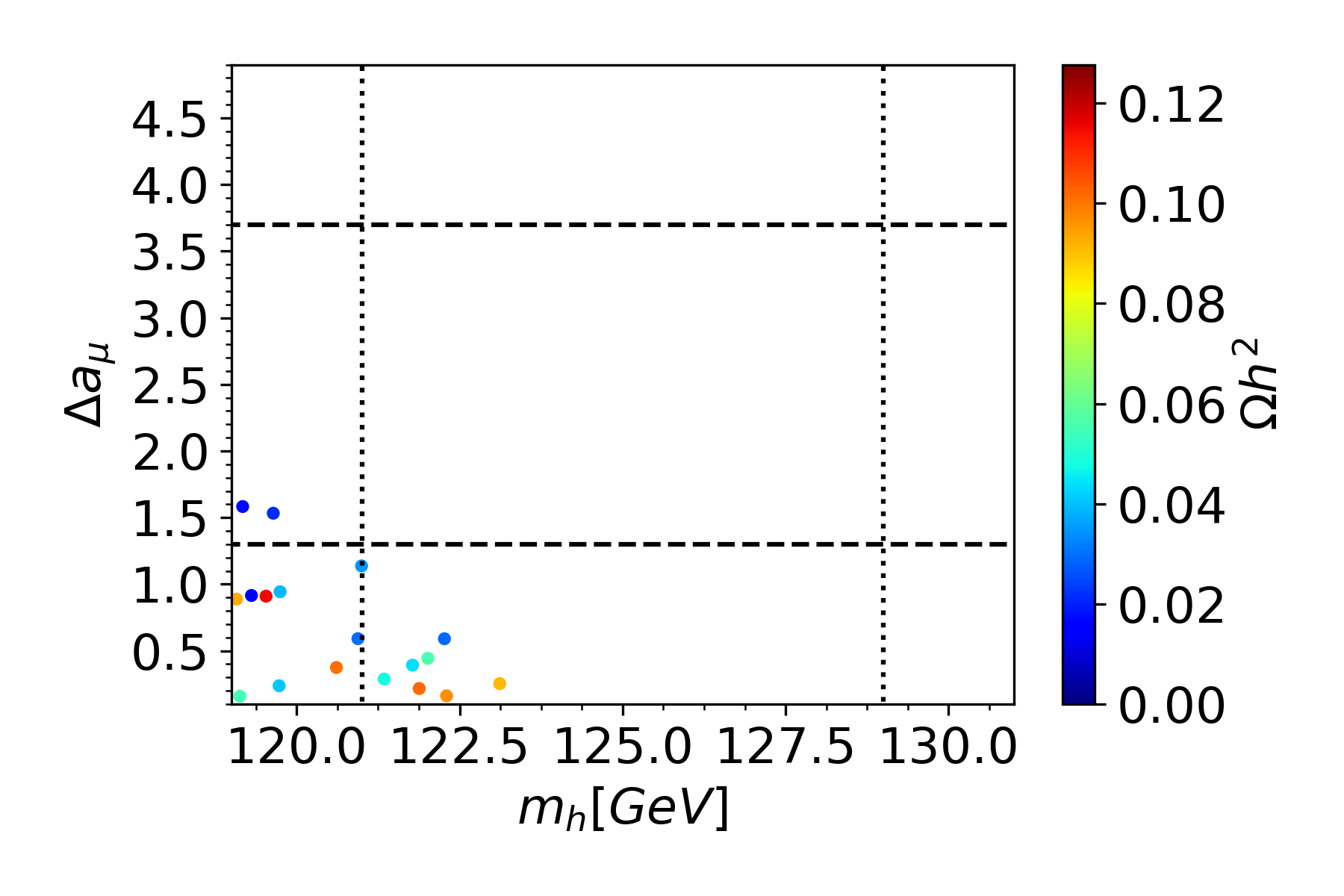}
\caption{$\mu>0$, case II }
\label{fig:A0/gmin2/mu+/mh_vs_amu_mupos_constrained_zoomed}
\end{subfigure}
\caption{As for Fig \ref{fig:A0/gmin2} but zoomed in on the $3\sigma$ region for both $a_\mu$ and $m_h$. Two benchmark points (presented below are marked)}
\label{fig:A0/gmin2_zoomed}
\end{figure}

\begin{figure}[h!]
\begin{subfigure}{.5\textwidth}
\includegraphics[width=.9\linewidth]{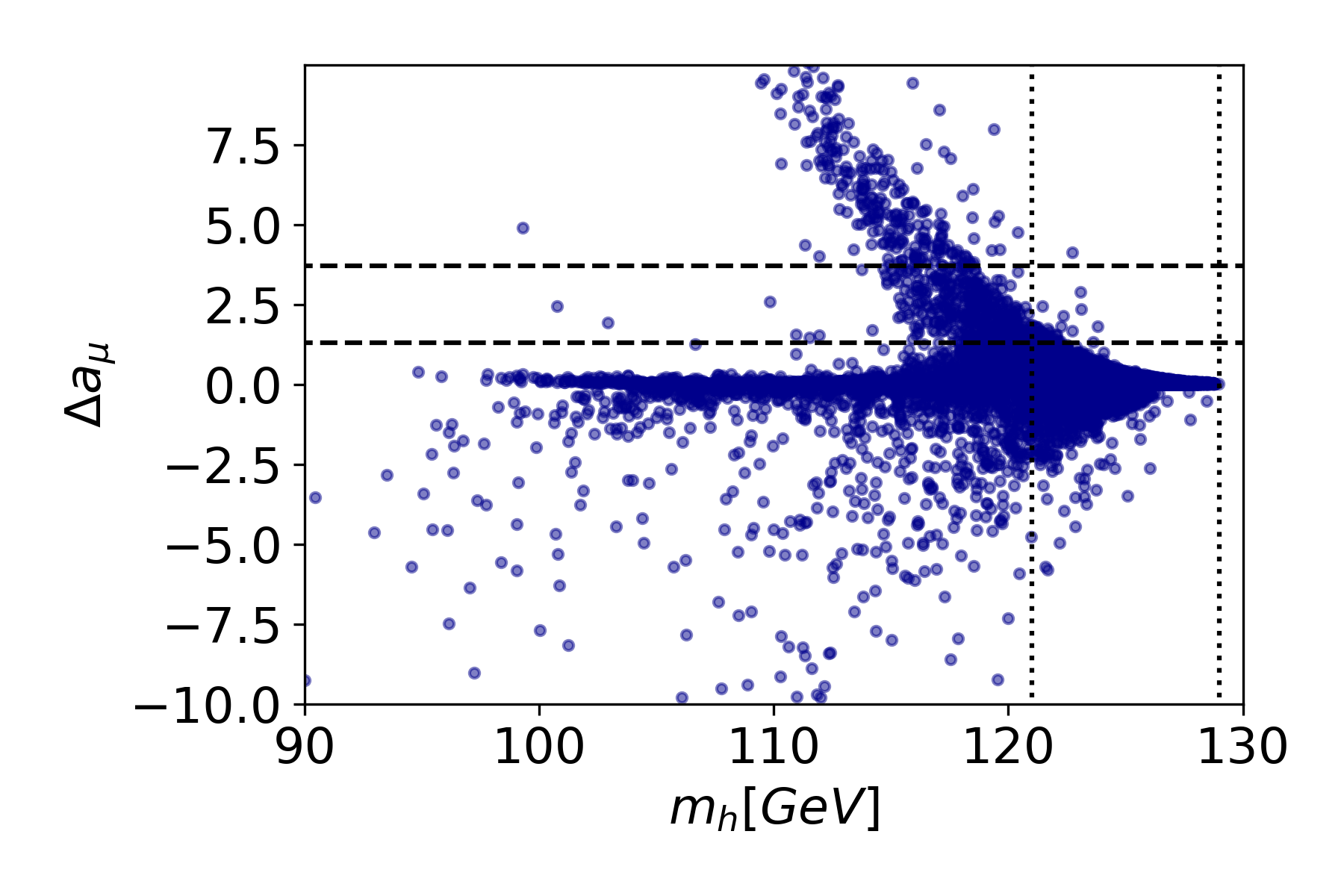}
\caption{$\mu<0$, case II }
\label{fig:A0/gmin2_RPV/mu+/mh_vs_amu_mupos_constrained}
\end{subfigure}
\begin{subfigure}{.5\textwidth}
\includegraphics[width=.9\linewidth]{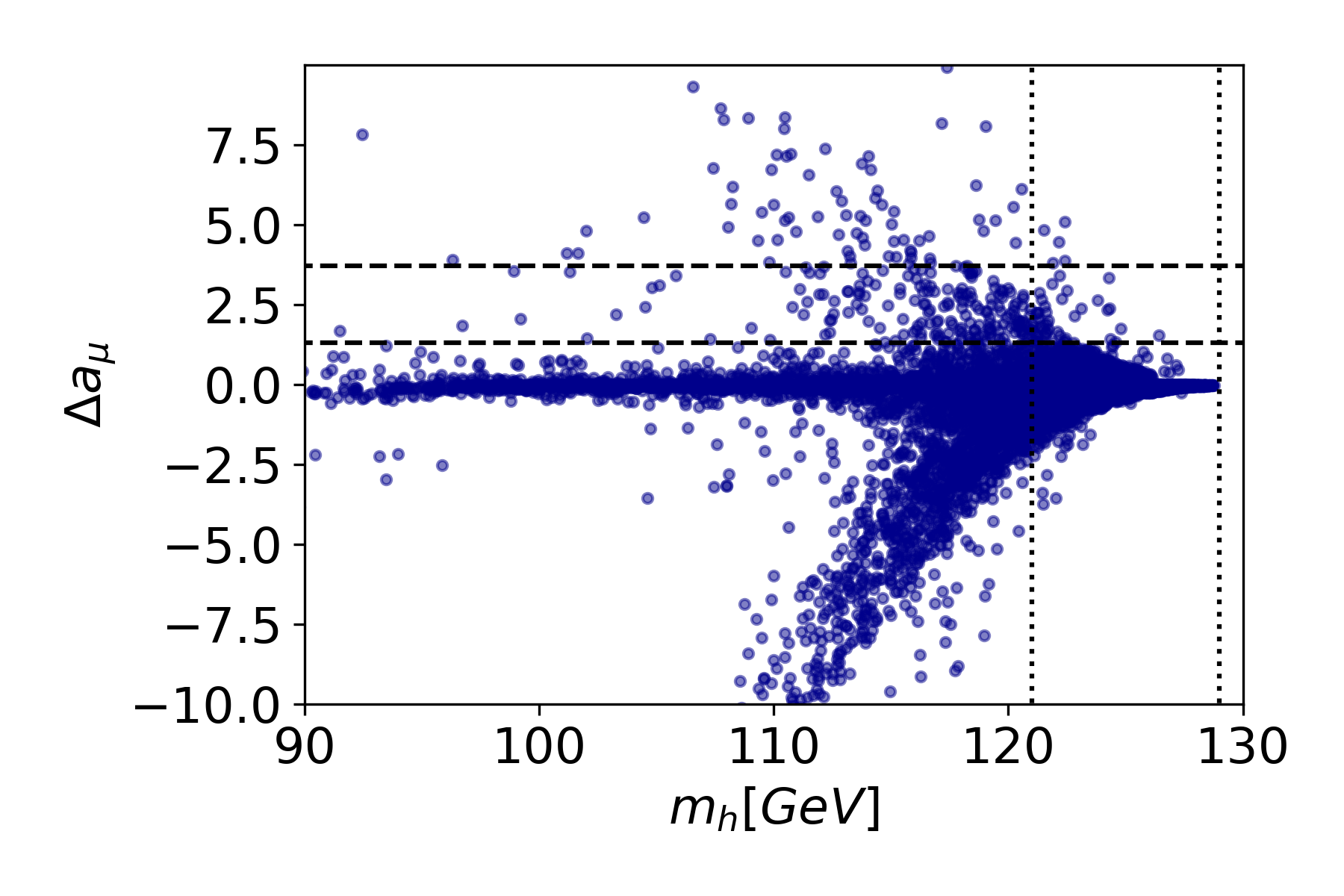}
\caption{$\mu>0$, case II }
\label{fig:A0/gmin2_RPV/mu-/mh_vs_amu_mupos_constrained}
\end{subfigure}
\caption{As for Fig \ref{fig:A0/gmin2} but where points that have charged LSP states are shown. Such states are marked in blue to indicate that no relic density could be calculated.}
\label{fig:A0/gmin2_RPV}
\end{figure}

\begin{figure}[h!]
\begin{subfigure}{.5\textwidth}
\includegraphics[width=.9\linewidth]{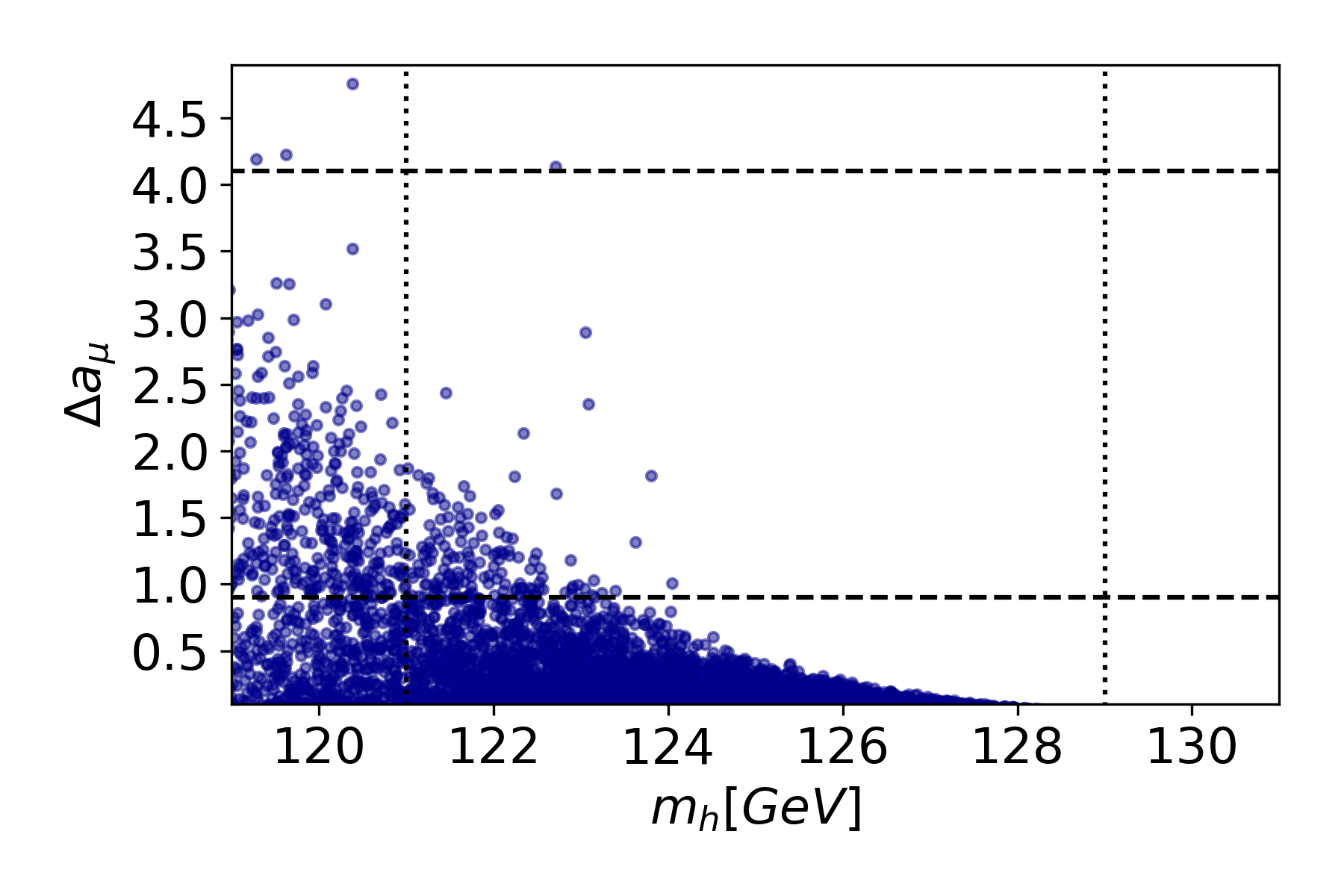}
\caption{$\mu<0$, case II }
\label{fig:A0/gmin2_RPV/mu+/mh_vs_amu_mupos_constrained_zoomed}
\end{subfigure}
\begin{subfigure}{.5\textwidth}
\includegraphics[width=.9\linewidth]{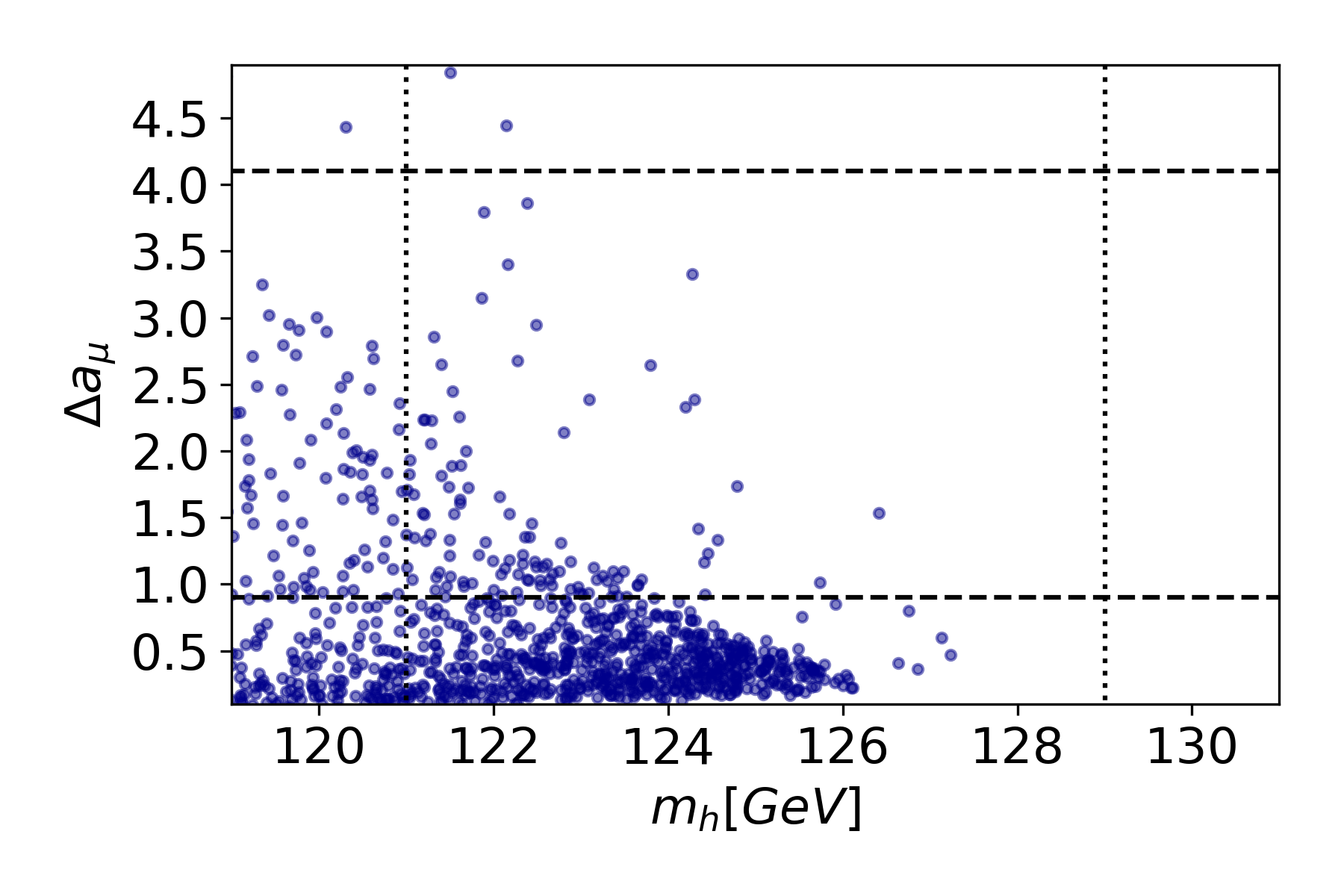}
\caption{$\mu>0$, case II }
\label{fig:A0/gmin2_RPV/mu-/mh_vs_amu_mupos_constrained_zoomed}
\end{subfigure}
\caption{As for Fig \ref{fig:A0/gmin2_RPV} but zoomed in on the $3\sigma$ region for both $a_\mu$ and $m_h$.}
\label{fig:A0/gmin2_RPV_zoomed}
\end{figure}

As was previously argued, the inclusion of negative $k$ values should allow for larger Higgs boson masses while keeping slepton masses low; thus incorporating a mechanism for generating the anomalous muon magnetic moment. As was done previously, we loosen the relic density constraint insisting only that the relic density is sufficiently small so as not to completely rule out the given point. Furthermore, we find large parts of the parameter space give the right handed stau state as the lightest sparticle. In Figures \ref{fig:A0/gmin2/m32_vs_k_mupos_amu_mh_like_constrained} to \ref{fig:A0/gmin2_zoomed} we present results where the neutralino is the LSP. However, it is also interesting to consider potential R-parity violating models in which these light, charged particles decay into standard model particles. As was previously stated, a detailed discussion of the viability of such points is beyond the scope of this paper, but we do include Figures \ref{fig:A0/gmin2_RPV} and \ref{fig:A0/gmin2_RPV_zoomed} as these states can lead to excellent fits for $a_\mu$.

Below we present the results of two scans (over positive and negative $\mu$ values). We limit the range of $m_\frac{3}{2}$ under the upper limit set by the starobinsky-like inflation limits previously discussed. This eliminates particularly massive gaugino states and thus large slepton mass states. 

In both cases, two values of k are favoured by the scan. Analogously to case I, $k\approx -0.016$ and $k\approx -0.0023$ with the latter snf $\mu<0$ representing a greater likelihood region of parameter space. This gives an approximate ratio of the high scale gaugino mass parameters as $|M_1|:|M_2|:|M_3| \approx 1:4:7$.

From Fig.~\ref{fig:A0/gmin2} and  \ref{fig:A0/gmin2_zoomed} we again see that $\mu<0$ is favoured with a proportion of points sitting well within the $2\sigma$ range. We also observe that some points have a remarkably high relic density almost satisfying the $1\sigma$ region. This is a tantalising suggestion that this effectively scaleless model may be able to satisfy the higgs mass, relic density, and $a_\mu$ simultaneously. Having said this, the inclusion of $A_0$ does not give any discernible improvement in the allowed values of the Higgs boson mass due to the previously discussed natural tendency toward a scale-less model. However, these results are still included as the inclusion of a small $A_{0}$ parameter can modify the SUSY parameters sufficiently such that some points pass the CheckMATE collider constraints as presented in benchmark points 9 and 10 (Table \ref{tab:input-table-c1-an0}). 

From Fig \ref{fig:A0/gmin2_RPV} we can see that allowing for these RPV parameter points slightly increases the $a_\mu$ values. Conversely to the previous, $\mu<0$ now represents the higher likelihood points. As the stau is the lightest particle in the cases, no relic density can be calculated, and therefore we mark all points in blue to indicate this issue. Having said this, some parameter points perfectly attain the higgs mass and $a_\mu$. For a discussion of the phenomenology of such points see \cite{Jeong:2021qey}.

\FloatBarrier

We now present the two benchmark points plotted on Fig \ref{fig:A0/gmin2_zoomed}. Both were selected to represent a particularly high value of $a_\mu$ as well as interesting collider results. 

Both benchmarks 9,10 have the same value of k as shown in Fig \ref{fig:A0_posk/A0} where $k=-0.23$ had clear phenomenological advantages. Furthermore, the values of $m_\frac{3}{2}$ and $tan(\beta)$ are also approximately similar. They are both selected from the $\mu<0$ scan due to its more promising phenomenology. In both cases $\beta\approx1$ and $\alpha$ is very small. This is consistent with the no-scale model we are examining. The high scale gaugino mass parameters are also presented. After running to the low scale the ratio becomes far more extreme causing a great hierarchical divergence between the colour charged and the non-colour charged sparticles as can be seen below.

\FloatBarrier
\begin{table}
\begin{tabularx}{\textwidth}{  m{0.11\textwidth}     m{0.35\textwidth} m{0.35\textwidth}     }
\hline
Quantity &    BP9 & BP10    \\
\hline
$\alpha $ &    			  						0.0007	&	0.0003					\\                  
$\beta $ &     									1		&	1						\\                             
$ m_{\frac{3}{2}} $ [TeV] &  						56		&	57						\\	
 $k$ &     				 						-0.023	&	-0.023					\\              
\hline
\underline{SPheno:} &          								&							\\ 
\hline
$m_{0}$ [GeV] &      								0		&	0						\\		                   
$tan(\beta)$ &         								10.5		&	10.0						\\		     
 $sign(\mu)$ &       								-1		&	-1						\\		          
 $A_{0}$ [GeV] &    								-227		&	-109						\\		                      
 $M_{1}$ [GeV] &    								-275		&	-271 						\\		               
 $M_{2}$ [GeV] &      							-1144	&	-1163					\\		               
 $M_{3}$ [GeV] &       							-1841	&	-1877					\\	
\hline
\end{tabularx}
\caption{Shows two benchmark points representing two different areas of interest in the parameter space. We present the model parameters and the resultant SPheno input parameters. BP9 shows a point with high $a_\mu$ and BP10 shows a point with high $\Omega$. Dimensions of the parameters are given where "-" means dimensionless. $ m_{\frac{3}{2}} $ is given in units of TeV, and $m_{0}, A_{0}, M_i $ are given in GeV.\label{tab:input-table-c1-an0}}
\end{table}

Table \ref{tab:table-spectrum-c1-an0} shows the mass spectrum for the two benchmark points BP9, BP10. Large values of $M_3$ give large values of squark and gluon masses and contribute to the relatively large higgs mass. Conversely, the smaller values of the other two gaugino mass parameters, in combination with the absence of $m_0$, gives the requisite small slepton states. Furthermore, the hierarchy between $M_1$ and $M_2$ gives a notable disparity between the right and left handed states. Due to the small value of $tan(\beta)$ and the negative sign of $\mu$, $\widetilde \tau_{1}$ is predominantly right handed. We note the very light smuon states that give key contributions to $a_\mu$. Finally, the small value of $M_1$ leads to predominantly bino-like, light dark matter candidate. 

\begin{table}
\vspace{-1em}
\begin{tabularx}{\textwidth}{  m{0.11\textwidth}   m{0.35\textwidth} m{0.35\textwidth}     }
\hline
Masses &   BP9 & BP10  \\
\hline
$\widetilde e_{L} $&    							712		&	724				\\
$\widetilde e_{R} $&    	 						139 	&	140				\\
$\widetilde \nu_{e L} $&      						708		&	719				\\
$\widetilde \mu_{L} $ &        						712		&	724				\\
$\widetilde \mu_{R} $ &         					139		&	140				\\
$\widetilde \nu_{\mu L} $ &  						708		&	719				\\
$\widetilde \tau_{1} $ &       						107		&	110				 \\
$\widetilde \tau_{2} $ &       						712		&	724				\\
$\widetilde \nu_{\tau L} $ &       					706		&	717				\\
\hline
$\widetilde{d}_{L} $  &        				        3360	&	3420				\\
$\widetilde d_{R} $ &         						3290	&	3350				\\
$\widetilde u_{L} $ &      							3360	&	3420				 \\
$\widetilde u_{R} $  &       						3290	&	3350					\\
$\widetilde s_{L} $ &        						3360	&	3420					\\
$\widetilde s_{R} $ &        						3290	&	3350				\\
$\widetilde c_{L} $ &         						3360	&	3420				 \\
$\widetilde c_{R} $ &          						3290	&	3350				\\  
$\widetilde b_{1} $  &        						3120	&	3180				\\
$\widetilde b_{2} $  &     							3280	&	3340					\\
$\widetilde t_{1} $ &        						2800	&	2840				 \\
$\widetilde t_{2} $ &     							3140	&	3190				\\
\hline
$\widetilde g $ &                   				3850		&	3920				 \\
$\widetilde \chi_{10} $ &         					103			&	103		\\
$\widetilde \chi_{20} $ &               			926			&	942			\\
$\widetilde \chi_{30} $ &             				2000		&	2050				 \\  
$\widetilde \chi_{40} $ &            				2000		&	2060					\\  
$\widetilde \chi_{1+} $ &           				926			&	943			\\
$\widetilde \chi_{2+} $ &            				2009		&	2060				 \\  
\hline
${h_{0}}$  &        								122			&	122			\\                  
${H_{0}}$  &    									2110		&	2160					\\               
${A_{0}}$  &    									2110		&	2160					\\                   
${H_{\pm}}$  & 										2110		&	2170					 \\
$\mu$   &  											-1870		&	-1930				\\
$B_{0}$  &    										-68			&	-46				\\                               
\hline
$\widetilde \tau_1 -\widetilde \chi_{10} $  & 		3.87		&	6.77							\\       
\hline
\end{tabularx}
\caption{Shows the spectrum of SUSY masses for the benchmark points given in Table \ref{tab:input-table-c1-an0}. The difference between the mass of $\widetilde \tau_1$ and $\widetilde \chi_{10}$ is also given as this pertains to the production of dark matter. We also include the high scale bilinear coupling value $B_0$ for its relevance to the high scale parameters of the model. All parameters are given in GeV.  \label{tab:table-spectrum-c1-an0}}
\end{table}

Table \ref{tab:table-omega-c1-an0} shows the key parameters for the relic density calculation for the two points BP9, BP10. As the stau is the nLSP, the mass difference between it and $\widetilde \chi_{10}$ is presented. As its mass gap is so small, the $\widetilde \tau$ plays a critical role in the mechanism for dark matter annihilation. Indeed the dominant decay channels contributing to the relic density calculation are $ \widetilde \tau_1 \: \widetilde \tau_1 \rightarrow \tau \: \tau$ and  $ \widetilde \tau_1 \: \widetilde \chi_{10} \rightarrow \tau \: \gamma$. As the mass gap is especially small in the BP9, dark matter is over-annihilated by these channels leading to a relic density below the desired value. 

\begin{table}
\begin{tabularx}{\textwidth}{  m{0.3\textwidth}     m{0.35\textwidth} m{0.35\textwidth}     }
\hline
Quantity &   BP9 & BP10   \\
\hline     
$\Omega_{DM} h^2$   &						0.0397	&	0.020			\\           
\hline
$\widetilde \chi_{10} $ [GeV] &         					103		&	103				\\
$\widetilde \tau_1 -\widetilde \chi_{10} $ [GeV] & 	3.87	&	6.77				\\     
\hline
  $|\alpha_1|^2$ &							 	1		&	1						\\
  $|\alpha_2|^2$ &								0		&	0						\\
  $|\alpha_3|^2$ &								0		&	0						\\
  $|\alpha_4|^2$ &								0		&	0						\\
\hline
\end{tabularx}

\caption{Shows the relic density of the LSP for each benchmark point. The difference between the LSP and the nLSP is also given. Finally, we give the probability of finding the  LSP in a particular flavour state. That is to say; we give $|\alpha_i|^2$  where $\widetilde \chi_{10} = \alpha_1 \widetilde B + \alpha_2 \widetilde W + \alpha_3 \widetilde H_1 + \alpha_4 \widetilde H_2$ and $\sum |\alpha_i|^2 = 1$. Dimensionful parameters are given in GeV. \label{tab:table-omega-c1-an0} }
\end{table}

Table \ref{tab:decay-BP910} shows the key collider and phenomenological findings resulting from these parameter points BP9, BP10. Both the nnLSP, $\widetilde \mu_R$, and the LSP $\widetilde \tau_1$, have short lifetimes with only one decay channel perhaps suggesting a strong collider signature. Indeed, the strongest signal region for both BP9 and BP10 is one that focuses on dileptonic final states with missing transverse energy in the context of "electroweakinos". Both b-type branching ratios are very well fitted in both cases. Furthermore, $a_\mu$ is within the $2\sigma$ region for BP9 and BP10. We see relatively high values of $r$ calculated by CheckMATE due to the low mass sleptons. 

\begin{table}
\begin{tabularx}{\textwidth}{  m{0.3\textwidth}   m{0.35\textwidth} m{0.35\textwidth}   }
\hline
Quantity &     BP9 & BP10 \\
\hline
$\Gamma \widetilde \tau_1$ [Gev] &									$2.61\times10^{-3}$  	&	$8.18\times10^{-3}$						\\
BR $(\widetilde \tau^-_1 \rightarrow \chi_{10} \tau^-)$ [\%]	&				100					&	100   					\\
\hline
$\Gamma \widetilde \mu_{R}$ [Gev] &  								$1.47\times10^{-1}$ 		&	$1.57\times10^{-1}$				\\
BR $(\widetilde \mu^-_R -\rightarrow \chi_{10} \mu^-)$ [\%]& 				100					&	100							\\
\hline
BR $(b \rightarrow s \: \gamma)$ [\%] & 								0.032 				&	0.032					\\ 
BR $(B_s \rightarrow \mu^+ \: \mu^-)$ [\%] & 							$2.96\times 10^{-7}$		&	$2.95\times10^{-7}$			\\
$\Delta \frac{(g-2)_\mu}{2}$ &  										$1.39\times10^{-9}$ 		&	$1.32\times10^{-9}$			\\
$\Omega_{DM} h^2$  & 											0.0397				&	0.020					\\
$\chi_{10}$ [Gev]& 												103					&	103						\\
$\sigma_{q \overline{q} \rightarrow  \chi_{10} \chi_{10}}$ [pb] &  			$3.982\times10^{-14}$	&	$2.290\times10^{-14}$						\\

\hline
$r_{max}$ &													$0.40$				&	$0.57$						\\
$\sqrt{s}$ [TeV] &												13					&	13						\\
Analysis	&													cms\_sus\_16\_039		&	cms\_sus\_16\_039		\\
Signal Region &												SR\_A44				&	SR\_A44							\\
Ref.			&												\cite{CMS:2017moi}		&	\cite{CMS:2017moi}						\\
$\sigma_{LO}$ [pb]	& 											$6.342\times10^{-11}$	&	$5.893\times10^{-11}$		\\
\end{tabularx}

\caption{Shows branching ratios for lightest supersymmetric particles in the spectrum for BP7 and BP8. Only branching ratio greater than $1\%$ are included. We also include some beyond the standard model observables BR $(b \rightarrow s \: \gamma)$, BR $(B_s \rightarrow \mu^+ \: \mu^-)$, $\Delta \frac{(g-2)_\mu}{2}$, and $\Omega_{DM} h^2$, where $\Delta \frac{(g-2)_\mu}{2}$ is a calculation of the SUSY contribution beyond the standard model. The model successfully predicts the b decays discrepancy and satisfies the anomalous muon magnetic moment to $2/3\sigma$. The relic density is too small and is therefore not ruled out phenomenologically. CheckMATE runs using 13TeV and 8TeV analyses do not rule out these points. Decay widths and masses are given in GeV, branching ratios are given in \%, and cross sections are given in pb.   \label{tab:decay-BP910} }
\end{table}

\section{Conclusion}
\label{conclusion}

Inflation represents a very attractive solution to a number of cosmological problems and, in combination with supersymmetry, a very attractive model for Beyond the Standard Model physics emerges based on no-scale SUGRA.
We have focussed on the particular case where the Polonyi term in the superpotential acts as a slow roll inflaton for Starobinsky inflation, leading to an upper bound on the gravitino mass $m_{3/2}<1000$ TeV.

The recent Fermilab muon $g-2$ result further motivates a
no-scale model, where all the dimensionful parameters the model are zero (except the gaugino masses which arise via mixed modulus and anomaly mediation), naturally leading to light slepton masses for certain gaugino masses. 
For negative universal gaugino masses, $k<0$,
we find relatively light bino/wino masses together with light sleptons, as suggested by the muon $g-2$ measurement. 
In general, such a model is also capable of providing a good dark matter candidate whilst satisfying constraints from collider physics, as well as yielding the correct Higgs boson mass, but it turns out to be non-trivial to achieve this while satisfying the 
desired muon $g-2$ constraints.

We have conducted a Monte Carlo parameter scan over the given model in two conditions: case I and case II,
corresponding to zero or non-zero trilinear soft parameter $A_0$. We show that case I with $k>0$ can give excellent fits for the Higgs boson mass and relic density whilst easily satisfying the collider and flavour constraints. Furthermore, with a reversal in the sign of $k$ we show that $g-2$ can be satisfied to $2\sigma$, as exemplified by BP7 and BP8, where BP8 also satisfies the desired relic density. Since BP7 and BP8 are on the edge of exclusion, with $r_{max}$ values slightly above unity,
this motivates the study of case II where we allow a non-zero $A_0$ trilinear soft parameter.
By including a small $A_0$ parameter at the high scale, we find that muon $g-2$ can be satisfied whilst satisfying the collider constraints that threaten case I. However we find, while the case II benchmark points BP9 and BP10 satisfy the collider constraints and the muon $g-2$, they both predict a relic density which is below the desired value, which does not exclude these points of course, but is somewhat disappointing from the point of view of dark matter. 
Interestingly, we find that, in both cases, a large part of parameter space is dominated by RPV-style points, where the LSP becomes the lightest $\widetilde \tau$ state. Although such states are beyond the scope of this paper, a detailed discussion of these kind of points can be found \cite{Jeong:2021qey}. 

In general, we find that the lightest neutralino $\widetilde \chi_{10}$ should be a purely bino-like state with a mass of around 100 GeV for parameter points BP7-BP10 that satisfy the muon $g-2$ constraint. We highlight in particular the fully no-scale SUGRA point with zero $A_0$, $B_0$ and $m_0$, namely
BP8, which can explain not only the recent Fermilab muon $g-2$ measurement, and has the correct Higgs boson mass, but also yields the desired dark matter relic density, albeit with $r_{max}=1.04$ on the edge of exclusion. We remark that 
current LHC limits are more easily evaded due to right handed sleptons being almost degenerate with the neutralino LSP, while the left handed counterparts are much higher in mass, suppressing chargino decay channels into first and second generation sleptons.
However, even for BP9 and BP10, the $r_{max}$ value is not too far from unity, suggesting that future LHC runs are capable of discovering such SUSY particles for all the interesting benchmarks that satisfy the muon $g-2$. 

In conclusion, no-scale SUGRA is not only well motivated theoretically from string theory and provides an elegant framework for accounting for cosmological Starobinsky inflation, but also has very interesting phenomenological implications as well. 
Ignoring the muon $g-2$ to begin with, and assuming positive universal gaugino mass contributions, in addition to the anomaly mediated contributions, we show that no-scale SUGRA can readily satisfy the dark matter 
and Higgs boson mass requirements, consistently with all other phenomenological constraints.
We then show that the recent Fermilab measurement of the muon $g-2$ may be accommodated,
together with the correct Higgs boson mass, 
for no-scale SUGRA with negative universal gaugino mass contributions in addition to the anomaly mediated contributions.
For the fully no-scale SUGRA case, with all soft parameters equal to zero at the high scale, apart from gaugino masses,
we find that successful points which satisfy the muon $g-2$, and can sometimes yield the desired relic density, although such points tend to be near the edge of LHC collider exclusion.
Analysing no-scale SUGRA with a non-zero $A_0$, we find that the muon $g-2$ can still be explained,
with the collider constraints somewhat relaxed. However, even in this case, light sleptons and charginos are still predicted, with good prospects for discovering these SUSY particles in LHC Run 3.

\section{Acknowledgments }
The authors acknowledge the use of the IRIDIS High Performance Computing Facility, and associated support services at the University of Southampton, in the completion of this work.
SFK acknowledges the STFC Consolidated Grant ST/L000296/1 and the European Union’s Horizon 2020 Research and Innovation programme under Marie Sklodowska-Curie grant agreement HIDDeN European ITN project (H2020-MSCA-ITN-2019//860881-HIDDeN).

\end{document}